\documentclass[fleqn,usenatbib]{mnras}

\usepackage{newtxtext,newtxmath}

\usepackage[T1]{fontenc}
\usepackage{ae,aecompl}

\usepackage{graphicx}	
\usepackage{amsmath}	
\usepackage{amssymb}	
\usepackage{pdflscape}	

\usepackage{subfig}		
\usepackage{afterpage}	
\usepackage{multirow}   

\newcommand*{\thead}[1]{\multicolumn{1}{c}{#1}} 

\title[Mini-halos and AGN feedback in clusters]{On the relation between mini-halos and AGN feedback in clusters of galaxies}

\author[A. Richard-Laferri\`{e}re et al.]{
A. Richard-Laferri\`{e}re,$^{1,2}$\thanks{E-mail: ar999@cam.ac.uk (ARL)}
J. Hlavacek-Larrondo,$^{1}$
R. S. Nemmen,$^{3}$
C. L. Rhea,$^{1}$
\newauthor
G. B. Taylor,$^{4}$
M. Prasow-\'{E}mond,$^{1}$
M. Gendron-Marsolais,$^{5}$
M. Latulippe,$^{1}$
\newauthor
A. C. Edge,$^{6}$
A. C. Fabian,$^{2}$
J. S. Sanders,$^{7}$
M. T. Hogan$^{8,9}$
and G. Demontigny$^{1}$
\\
$^{1}$D\'{e}partement de physique, Universit\'{e} de Montr\'{e}al, C.P. 6128 Succ. Centre-ville, Montr\'{e}al H3C 3J7, Canada\\
$^{2}$Institute of Astronomy, University of Cambridge, Madingley Road, Cambridge CB3 0HA, UK\\
$^{3}$Instituto de Astronomia, Geof\'{i}sica e Ci\^{e}ncias Atmosf\'{e}ricas, Universidade de S\~{a}o Paulo, Rua do Mat\~{a}o 1226, S\~{a}o Paulo 05508-090, Brazil \\
$^{4}$Department of Physics and Astronomy, University of New Mexico, 1919 Lomas Blvd. NE, Albuquerque 87131, USA\\
$^{5}$European Southern Observatory, Alonso de C\'{o}rdova 3107, Vitacura, Casilla 19001, Santiago, Chile\\
$^{6}$Department of Physics, Durham University, South Road, Durham DH1 3LE, UK \\
$^{7}$Max-Planck-Institut f\"{u}r Extraterrestrische Physik, Giessenbachstra{\ss}e 1, Garching 85748, Germany \\
$^{8}$Most recent affiliation: Department of Physics and Astronomy, University of Waterloo, Waterloo, ON, N2L 3G1, Canada \\
$^{9}$Most recent affiliation: Perimeter Institute for Theoretical Physics, Waterloo, ON, N2L 2Y5, Canada
}

\date{Accepted 2020 August 30. Received 2020 August 29; in original form 2019 June 28}

\pubyear{2020}

\begin{document}
\label{firstpage}
\pagerange{\pageref{firstpage}--\pageref{lastpage}}
\maketitle

\begin{abstract}
A variety of large-scale diffuse radio structures have been identified in many clusters with the advent of new state-of-the-art facilities in radio astronomy.  Among these diffuse radio structures, radio mini-halos are found in the central regions of cool core clusters. Their origin is still unknown and they are challenging to discover; less than thirty have been published to date. Based on new VLA observations, we confirmed the mini-halo in the massive strong cool core cluster PKS~0745$-$191 ($z=0.1028$) and discovered one in the massive cool core cluster MACS~J1447.4+0827 ($z=0.3755$). Furthermore, using a detailed analysis of all known mini-halos, we explore the relation between mini-halos and AGN feedback processes from the central galaxy. We find evidence of strong, previously unknown correlations between mini-halo radio power and X-ray cavity power, and between mini-halo and the central galaxy radio power related to the relativistic jets when spectrally decomposing the AGN radio emission into a component for past outbursts and one for on-going accretion. Overall, our study indicates that mini-halos are directly connected to the central AGN in clusters, following previous suppositions. We hypothesize that AGN feedback may be one of the dominant mechanisms giving rise to mini-halos by injecting energy into the intra-cluster medium and reaccelerating an old population of particles, while sloshing motion may drive the overall shape of mini-halos inside cold fronts. AGN feedback may therefore not only play a vital role in offsetting cooling in cool core clusters, but may also play a fundamental role in re-energizing non-thermal particles in clusters.
\end{abstract}
\begin{keywords}
surveys -- galaxies: clusters: general -- galaxies: clusters: individual (PKS~0745$-$191, MACS~J1447.4+0827) -- galaxies: clusters: intracluster medium -- radio continuum: general -- X-ray: galaxies: clusters 
\end{keywords}



\section{Introduction}\label{intro}

Galaxy clusters are extremely large ($\approx10^{14}-10^{15}$ M$_{\sun}$), gravitationally bound objects that consist mainly of dark matter, hot X-ray emitting gas, known as the intra-cluster medium (ICM), and hundreds to thousands of galaxies. However, galaxy clusters also contain vast amounts of relativistic particles, as well as magnetic fields. Combined, these components produce strong synchrotron emission detectable at radio wavelengths that can reach cluster-wide sizes ranging from a few kpc to $\sim$1~Mpc (see \citealt{Feretti2012}; \citealt{Brunetti2014}; \citealt{vanWeeren2019}, for reviews); including structures known as giant radio halos and radio mini-halos. Giant radio halos consist of diffuse emission extending to the size of the cluster ($\sim$Mpc) and they are almost exclusively found in non-cool core clusters, meaning that the cluster was recently involved in a major (cluster-cluster) merger \citep{Giovannini1999,Buote2001,Cassano2010}. On the other hand, mini-halos, which are the central topic of this paper, are smaller and are found in relaxed, cool core clusters (see \citealt{Gitti2015} for a recent overview).

Cool core clusters are relaxed clusters that exhibit highly peaked central X-ray surface brightness distributions, implying that the radiative cooling of the hot X-ray gas must be balanced by a source of heating to follow observations. The candidate of choice is the mechanical feedback from the central Active Galactic Nucleus (AGN) \citep[e.g.][]{Fabian1994,Peterson2006,Fabian2012,McNamara2012}, which comes from the powerful relativistic jets of the central AGN in the Brightest Cluster Galaxy (BCG) that inflate bubbles filled with relativistic plasma displacing the ICM and creating X-ray cavities. The energy of these jets is then thought to be injected into the ICM through turbulence, shocks or sound waves (e.g. \citealt{Birzan2004}; \citealt{Dunn2006}; \citealt{Rafferty2006}; \citealt{Julie2015}).

Radio mini-halos consist of faint, diffuse radio structures that surround the central radio loud BCG (e.g. \citealt{Gitti2002}, \citealt{Govoni2009}, \citealt{Julie2013}, \citealt{ZuHone2013} \citealt{Giacintucci2014}, \citealt{vanWeeren2014}, \citealt{Kale2015}, \citealt{Yuan2015}).  They have amorphous shapes and extend on $\sim50-300~$kpc scales. They are therefore smaller than galaxy clusters, but extend well beyond the host galaxy and to larger scales than the average size of the jetted emission produced by the AGN in the BCG, jets having typical sizes of $\sim30~$kpc (e.g. \citealt{vonDerLinden2007}). Mini-halos also have very low surface brightnesses and steep radio spectra with typical spectral indexes of $\alpha < -1$, where the flux density ($S_{\nu}$) is defined as $S_\nu\propto\nu^\alpha$ and $\nu$ is the frequency \citep{Giacintucci2014}.

Some giant radio halos have been found to extend on Mpc scales in a few cool core clusters (e.g. CL1821+643, \citealt{Bonafede2014}, \citealt{Kale2016}, \citealt{Boschin2018}; A2390 \& A2261, \citealt{Sommer2017}) and some clusters have been found to host both a mini-halo and a radio halo (e.g. PSZ1~G139.61+24.20, \citealt{Savini2018}, \citealt{Giacintucci2019}; RX~J1720.1+2638, \citealt{Giacintucci2014}, \citealt{Savini2019}), pointing to a possible connection between giant radio halos and mini-halos \citep{vanWeeren2019}. However, the usual discrepancy in the size and type of host cluster of radio mini-halos and giant mini-halos has been used as a way to differentiate them from one another. The origin of these diffuse radio structures is therefore still not well understood and remains an active topic of research with about thirty cool core clusters known to host mini-halos \citep{Giacintucci2017,Giacintucci2019}. Extending the sample of radio mini-halos and giant radio halos is therefore one of the ways we can further understand the origin and properties of these structures.

It is thought that the relativistic particles creating the radio emission in mini-halos may originate from the central AGN (e.g. \citealt{Fujita2007,Cassano2008}), but the radiative cooling times are much shorter than the time required for them to reach the extent of the mini-halo (e.g. \citealt{Taylor2002}). This implies that the particles must be produced and/or reaccelerated in situ. 

One possible mechanism proposed in the literature is based on hadronic models, in which new relativistic electrons are produced as secondary products from the interaction between relativistic cosmic-ray protons and thermal protons in the ICM (e.g. \citealt{Pfrommer2004}; \citealt{Fujita2007}; \citealt{Zandanel2014}; \citealt{Ignesti2020}). Such hadronic collisions should however produce detectable $\gamma$-ray emission, yet extended $\gamma$-ray emission has not been detected so far in clusters (e.g. \citealt{Ackermann2010}; \citealt{Ahnen2016}). 

The second mechanism proposed in the literature consists of reaccelerating a pre-existing population of electrons by phenomena such as turbulence in the ICM (e.g. \citealt{Gitti2002, Gitti2004}; \citealt{Eckert2017}). The cause of this turbulence is linked to sloshing motions in the ICM (\citealt{Mazzotta2008}; \citealt{ZuHone2013}; \citealt{Giacintucci2014b}), which are caused by gravitational perturbations arising from minor mergers perturbing the colder gas of the ICM in the central core. This proposed model of reacceleration is supported by observations since many mini-halos appear to be bounded by cold fronts \citep[e.g.][]{Mazzotta2008,Julie2013,Giacintucci2014, Giacintucci2014b}, a classical signature of sloshing motions \citep[e.g.][]{Markevitch2001,Mazzotta2001, Mazzotta2003,Ascasibar2006,Markevitch2007,Owers2009,Ghizzardi2010,Lagana2010}. Furthermore, numerical simulations show that turbulence created by sloshing motions is sufficient to reaccelerate electrons and give rise to the morphology, radio power and spectral indexes of observed mini-halos \citep{ZuHone2013}. Moreover, the \textit{Hitomi} Soft X-ray Spectrometer observations of the Perseus cluster showed that the turbulence of the hot X-ray gas, whatever the caused for this turbulence is, contained enough energy to generate the synchrotron emission of mini-halos via reacceleration of the particles and even to balance the radiative cooling in the core \citep{Hitomi2016}. Although, it was found that the turbulence can only spread $<10~$kpc and act during 4\% of the cooling time, therefore another process would be needed to transport the energy in the cooling core. 
Recently, based on new high-dynamic range Karl G. Jansky Very Large Array (VLA) observations of the mini-halo in the Perseus cluster, it was suggested that the particles creating mini-halos may also be reaccelerated by turbulence being generated from the jetted outflows of the BCG instead of from sloshing motion \citep{Marie-Lou2017}. The same phenomenon was observed in the Phoenix cluster with new VLA observation at 1.52~GHz \citep{McDonald2019,Raja2020}, however the elongated beam could also be responsible for that alignment. This would imply that mini-halos should be connected to the feedback properties of BCGs, a hypothesis already suggested by e.g. \citet{Cassano2008,Gitti2015,Gitti2015b,Bravi2016,Gitti2018}, who stated that the same turbulence could be responsible for the origin of mini-halos and for the heating of the cooling flow. This was linked with the discovery of \citet{Zhuravleva2014}, claiming that the turbulence created by AGN should dissipate into heat, which would be sufficient to offset the radiative cooling. For this reason, in this paper, we further explore the connection between mini-halo properties and AGN properties of the BCG. We also report the confirmation of a radio mini-halo in PKS~0745$-$191 and the discovery of a new mini-halo in MACS~J1447.4+0827 through new VLA and \textit{Chandra} observations. In Section \ref{OBS}, we present the observations and the data reduction of the VLA and \textit{Chandra} X-ray datasets for PKS~0745$-$191 and MACS~J1447.4+0827. In Section \ref{Selection}, we present the sample of clusters analysed in this paper, while the results are presented in Section \ref{Analysis}. The implications of the results are discussed in Section \ref{Discussion} and the summary is presented in Section \ref{Summary}.   

A $\Lambda$CDM cosmology with H$_0 = 69.6$ km s$^{-1}$ Mpc$^{-1}$, $\Omega_\textnormal{M} = 0.286$, and $\Omega_{\Lambda} = 0.714$ \citep{Bennett2014} is adopted throughout this manuscript. 

\section{Observations}\label{OBS}

\subsection{PKS~0745$-$191}\label{PKS}

\subsubsection{\textit{Chandra} X-ray Observations}

\begin{figure*}
\centering
\begin{minipage}[c]{1.0\linewidth}


\begin{minipage}[c]{0.48\textwidth}
\subfloat{\includegraphics[width=1\textwidth]{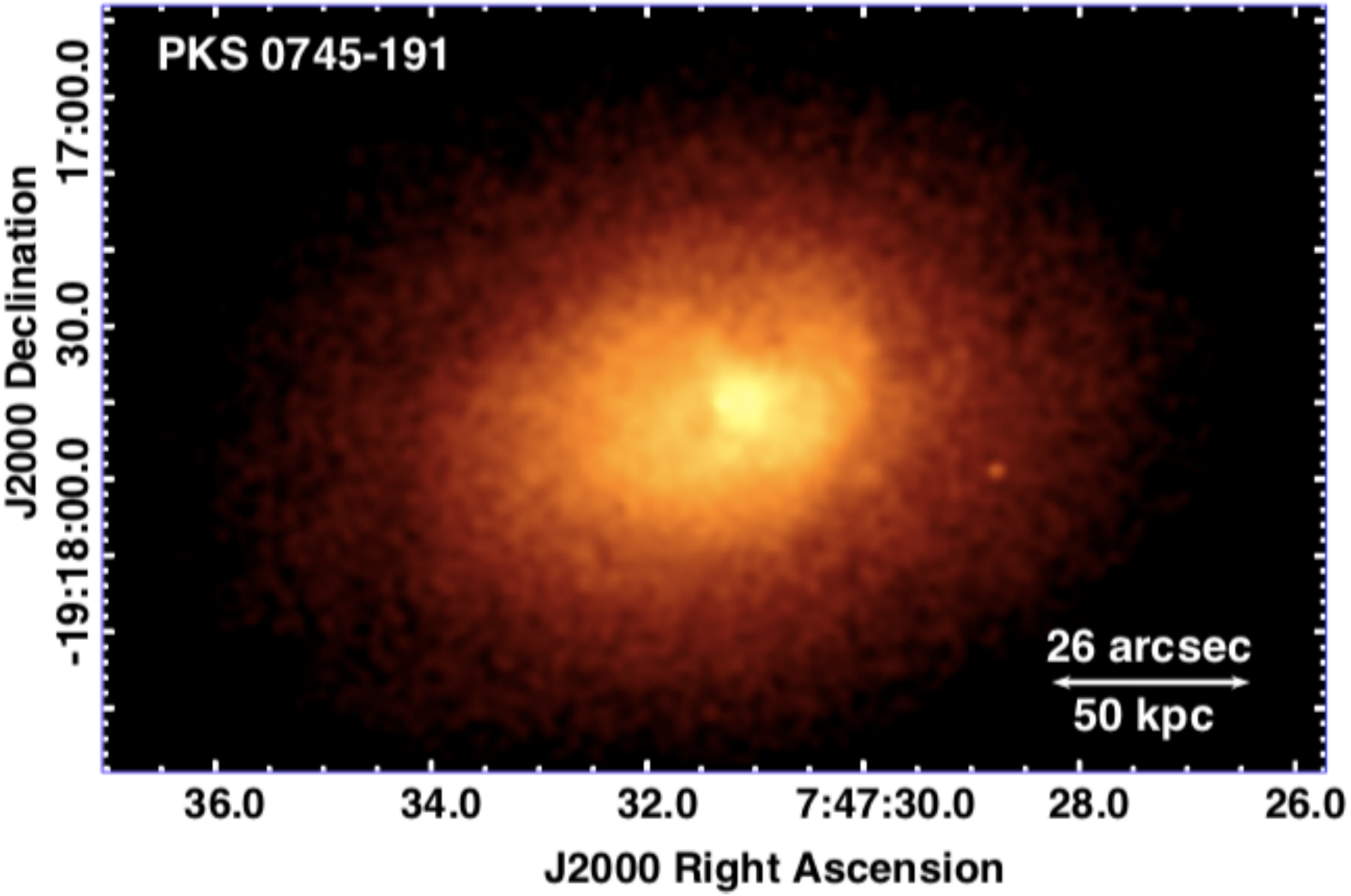}}
\end{minipage}
\hspace{4mm}
\begin{minipage}[c]{0.48\textwidth}
\subfloat{\includegraphics[width=1\textwidth]{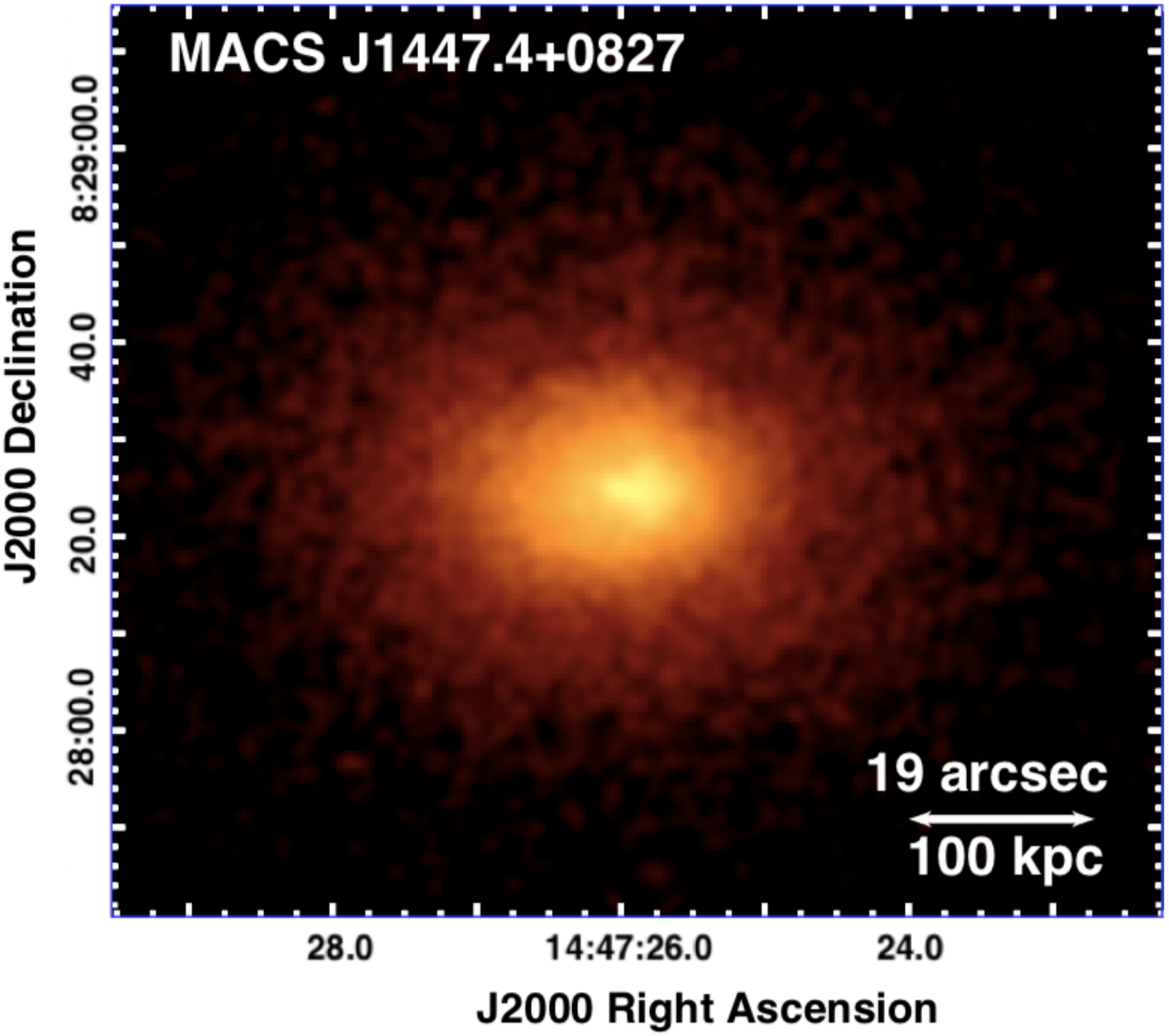}}
\end{minipage}


\end{minipage}
\caption[]{\textbf{Left:} Merged \textit{Chandra}
  X-ray image of PKS~0745$-$191, exposure-corrected, in the $0.5-7$ keV range. \textbf{Right:} Merged \textit{Chandra}
  X-ray image of MACS~J1447.4+0827, exposure-corrected, in the $0.5-7$ keV
  range.}
\label{fig:Xray}
\end{figure*}

\begin{figure*}
\centering
\begin{minipage}[c]{1.0\linewidth}
\subfloat{\includegraphics[width=0.48\textwidth]{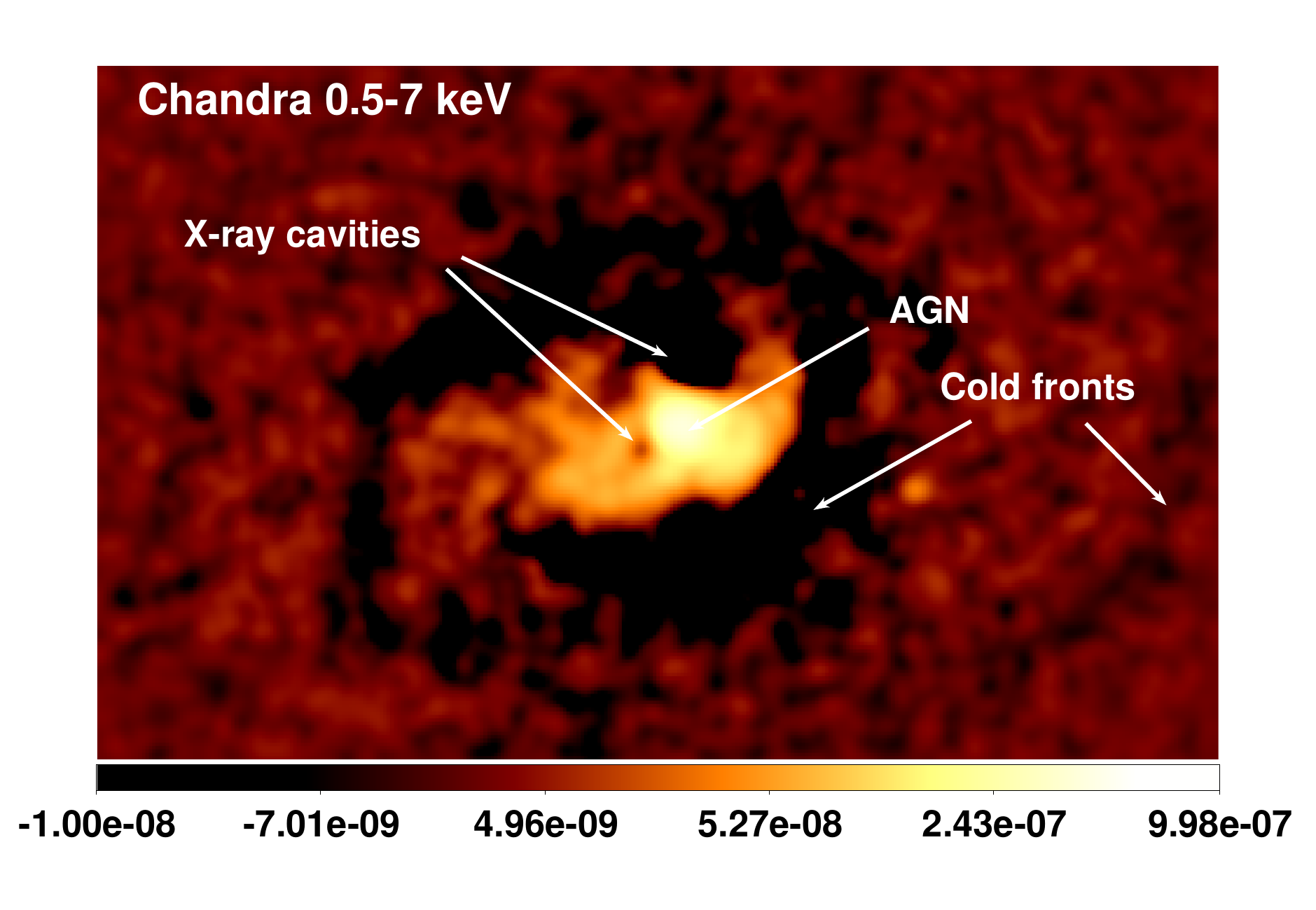}}
\subfloat{\includegraphics[width=0.48\textwidth]{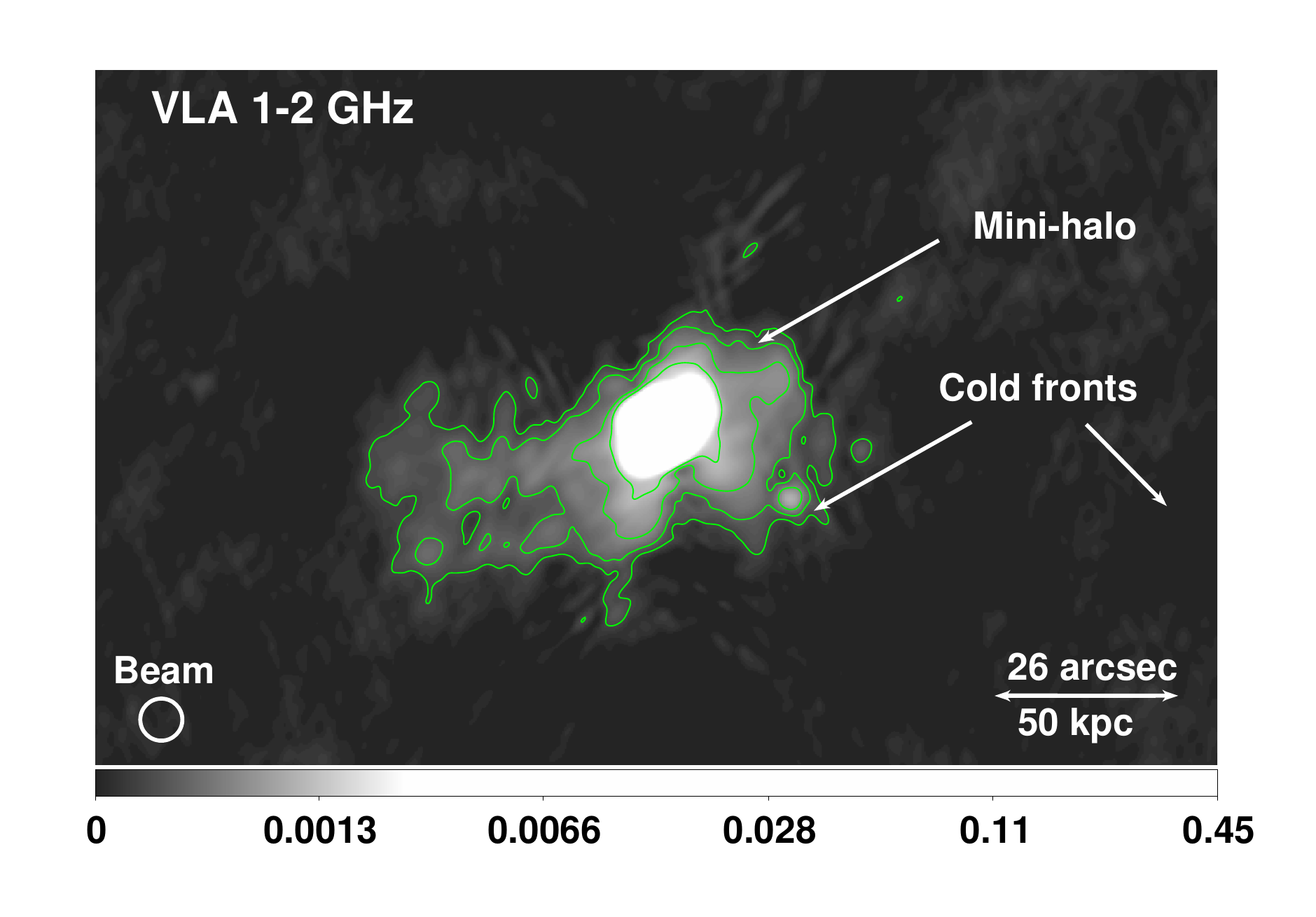}}
\end{minipage}
\caption[]{Images of PKS~0745$-$191, both on the same scale than the left panel of Fig.~\ref{fig:Xray}. \textbf{Left:} Unsharp-masked image, showing fractional residuals between data smoothed by a Gaussian of $\sigma$ = 2 and 16 pixels (1\arcsec and 8\arcsec) of the \textit{Chandra} X-ray image of PKS~0745$-$191. The positions of the X-ray cavities, cold fronts and AGN have been identified. \textbf{Right:} VLA image obtained at $1-2$~GHz with a $\sigma_{\rm rms}=50~$\textmu Jy beam$^{-1}$ and a beam size of $3.0\arcsec\times3.0\arcsec$ (shown with the white circle). The positions of the cold fronts have been identified. Four radio contours are overlaid on the radio image: $3\sigma_{\rm rms}$, $6\sigma_{\rm rms}$, $12\sigma_{\rm rms}$ and $24\sigma_{\rm rms}$.}
\label{fig1}
\end{figure*}

Advanced CCD Imaging Spectrometer (ACIS) \textit{Chandra} observations were obtained in 2011 with ACIS-S in VFAINT mode for PKS~0745$-$191 (ObsID 12881, PI Sanders), totaling an exposure time of 118.1 ks. The data are presented in \citet{Sanders2014} and were reduced following the standard procedure, and then combined with the existing 17.9 ks ACIS-S observations (ObsID 2427, PI Fabian). These observations are presented in Table~\ref{tab:X-ray}. The ObsID 508 observations were not used because of extensive flare contamination. The resulting exposure time of the observations is $135.5~$ks, and the resulting exposure-corrected image is shown in the left panel of Fig.~\ref{fig:Xray}. We refer the reader to \citet{Sanders2014} for a more detailed description of the data reduction and flare removal procedure. Due to the bright central peak, the features in the X-ray emission are seen more clearly using an unsharp-masked image to remove the large scale emission. In the left panel of Fig.~\ref{fig1}, we highlight some of the interesting features discussed in \citet{Sanders2014}, including two X-ray cavities and the western edges associated with cold fronts.


 \begin{table}
  \caption{Details of the \textit{Chandra} X-ray observations. The columns are: 1. Cluster Name; 2. Observation identification number (ObsID); 3. Observation date 4. \textit{Chandra} detector; 5. Total clean exposure time.}
  \label{tab:X-ray}
  \begin{tabular}{ccccc}
    \hline
    \thead{Name} & \thead{ObsID} & \thead{Date} & \thead{Detector} & \thead{Exp} \\
    \thead{} & \thead{} & \thead{} & \thead{} & \thead{[ks]} \\
    \hline
  PKS~0745$-$191 & 2427 & 2001 June 16  & ACIS-S & 17.9 \\
  & 12881 & 2011 Jan 27 & ACIS-S & 118.1 \\
  MACS J1447.4 & 10481 & 2008 Dec 14 & ACIS-S & 11.1 \\
  \quad \; \; \; +0827& 17233 & 2016 Apr 5 & ACIS-I & 41.0 \\
  & 18825 & 2016 Apr 6 & ACIS-I & 24.2 \\
        \hline
  \end{tabular}
 \end{table}
 

\subsubsection{VLA Observations}\label{sec:PKS_VLA}

We obtained a total of 5 hours in L-band ($1-2$~GHz) with the VLA (PI Sanders) for PKS~0745$-$191. The observations were taken in 2012 in A-configuration with 27 operational antennas. This observation is summarised in Table~\ref{tab:radio}. 3C 286 was used as the flux calibrator and observed at the end of the observing run, while J0735-1735 was selected as the phase calibrator.  The data were reduced using the standard Astronomical Image Processing System (AIPS, \citealt{Greisen2003}). A significant fraction of the observations were affected by radio frequency interference (RFI). The \textsc{flagr} task was used to remove most of this RFI, but some minor artefacts remained in the final image. To obtain the final image, the \textsc{imagr} task was used, and included multi-field imaging with 35 facets. The resulting rms and beam size are presented in Table~\ref{tab:radio}. This image is presented in the right panel of Fig.~\ref{fig1}, along with radio contours starting at $3\sigma_{\rm rms}$. 


 \begin{table*}
  \caption{Details of the VLA radio observations. The columns are: 1. Cluster Name; 2. Array configuration; 3. Observation frequency; 4. Observation bandwidth; 5. Number of intermediate frequency (IF) or spectral window (spw); 6. Number of channel per IF; 7. Date when the observation was taken; 8. Total time on source; 9. Beam size at full width half (FWHM) of the beam; 10. Image rms level.}
  \label{tab:radio}
  \resizebox{17.7cm}{!}{%
  \begin{tabular}{cccccccccc}
    \hline
    \thead{Name} & \thead{Conf.} & \thead{Frequency} & \thead{Bandwidth} & \thead{IF} & \thead{channels/IF} & \thead{Date} & \thead{Duration} & \thead{FWHM} & \thead{$\sigma_{\rm rms}$} \\
    \thead{} & \thead{} & \thead{[GHz]} & \thead{[MHz]} & \thead{} & \thead{} & \thead{} & \thead{[min]} & \thead{[$\arcsec \times\arcsec$]} & \thead{[$\mu$Jy beam$^{-1}$]} \\
    \hline
  PKS~0745$-$191 & A & $1-2$ & 64 & 16 & 64 & 2012 Oct 21 & 300 & $3.0\times3.0$ & 50 \\
  MACS~J1447.4+0827 & A & $1-2$ & 64 & 16 & 64 & 2015 Aug 14 & 180 & $0.9\times1.1$ & 11 \\
  & B & $1-2$ & 64 & 16 & 64 & 2016 May 25 & 180 & $3.1\times3.6$ & 16 \\
  & C & $1-2$ & 64 & 16 & 64 & 2016 Feb 2 & 180 & $9.6\times12.5$ & 15 \\
        \hline
  \end{tabular}%
  }
 \end{table*}
 

PKS~0745$-$191 harbours a bright central AGN, making the current VLA image dynamic range limited (in this case, $30,000:1$). Furthermore, antennae and side lobes are cast across our source due to a strong distant source outside the field of view. However, even if some minor artefacts remain, Fig.~\ref{fig1} reveals a faint radio component that extends out to $\sim42\arcsec \sim 80~$kpc in the eastern direction. The component is diffuse and has a $1-2$~GHz spectral index of $\alpha\approx-2$. 

The diffuse component in our observations can either be interpreted as a mini-halo, which is expected considering that PKS~0745$-$191 is a massive cool core cluster \citep[e.g.][]{Giacintucci2017}, or it could represent radio lobes seen end-on, meaning that the radio lobes are seen at a small angle with the line-of-sight. However, there is no sign of an established jet on either side of the nucleus at this or higher frequencies (see \citealt{Baum1991}; \citealt{Taylor1994}), and the core is not significantly brighter compared to the other inner radio structures. If the lobes were seen end-on then the core should be more highly beamed. The diffuse component also does not appear to connect directly to the central AGN as would be expected in radio lobes. The radio structure is also very faint (\textmu Jy level) and appears to be strongly bounded by the inner cold front to the west found by \citet{Sanders2014}, as is seen in several other clusters of galaxies hosting mini-halos (e.g. \citealt{Mazzotta2008}). We therefore interpret this structure as a mini-halo, confirming the uncertain classification by \citet{Gitti2004}, and discuss the implications of this discovery in the following sections. 

\subsection{MACS~J1447.4+0827}\label{MACS}

\subsubsection{\textit{Chandra} X-ray Observations}

\begin{figure*}
\centering
\begin{minipage}[c]{1.0\linewidth}
\subfloat{\includegraphics[width=0.48\textwidth]{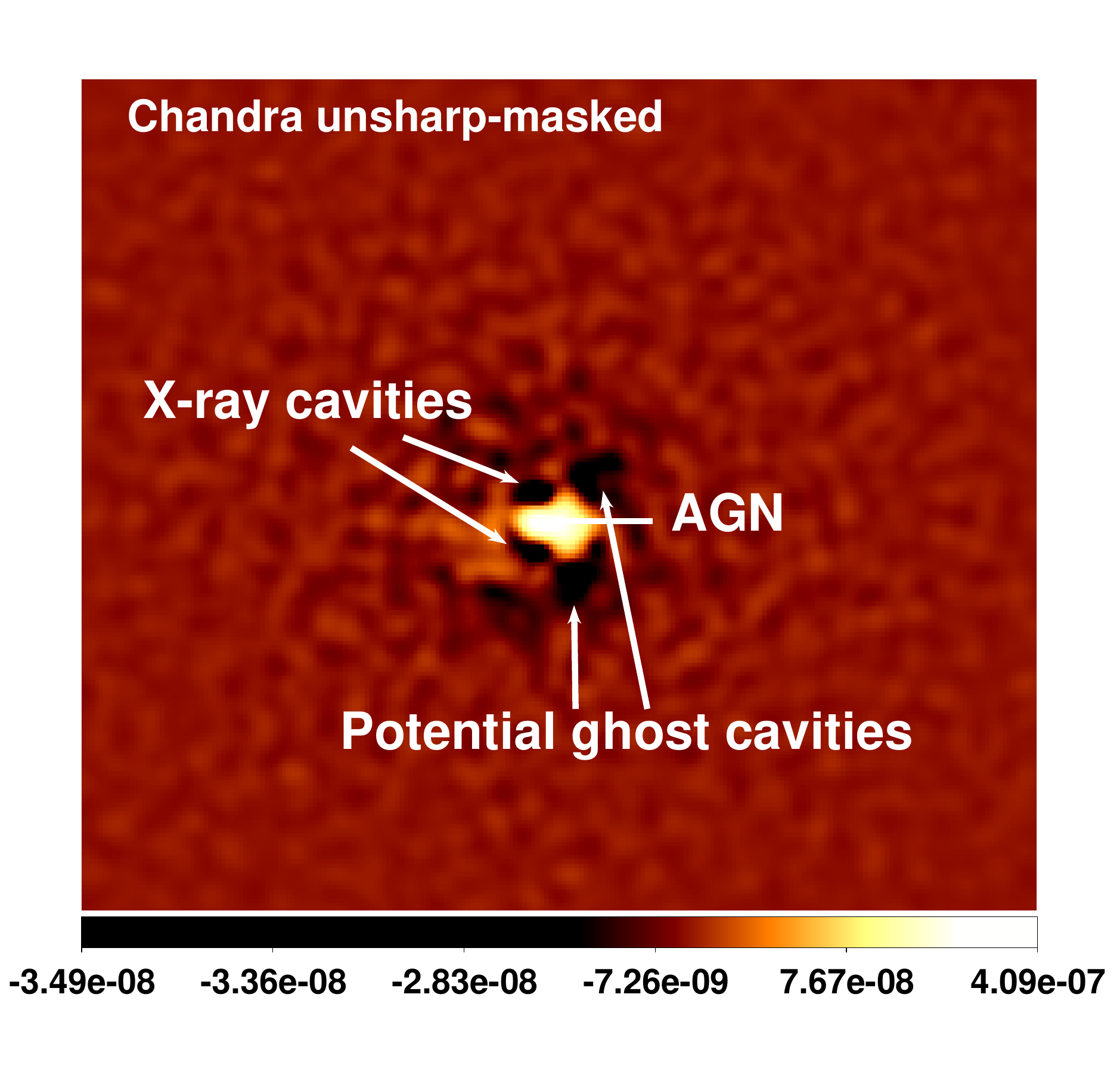}}
\subfloat{\includegraphics[width=0.48\textwidth]{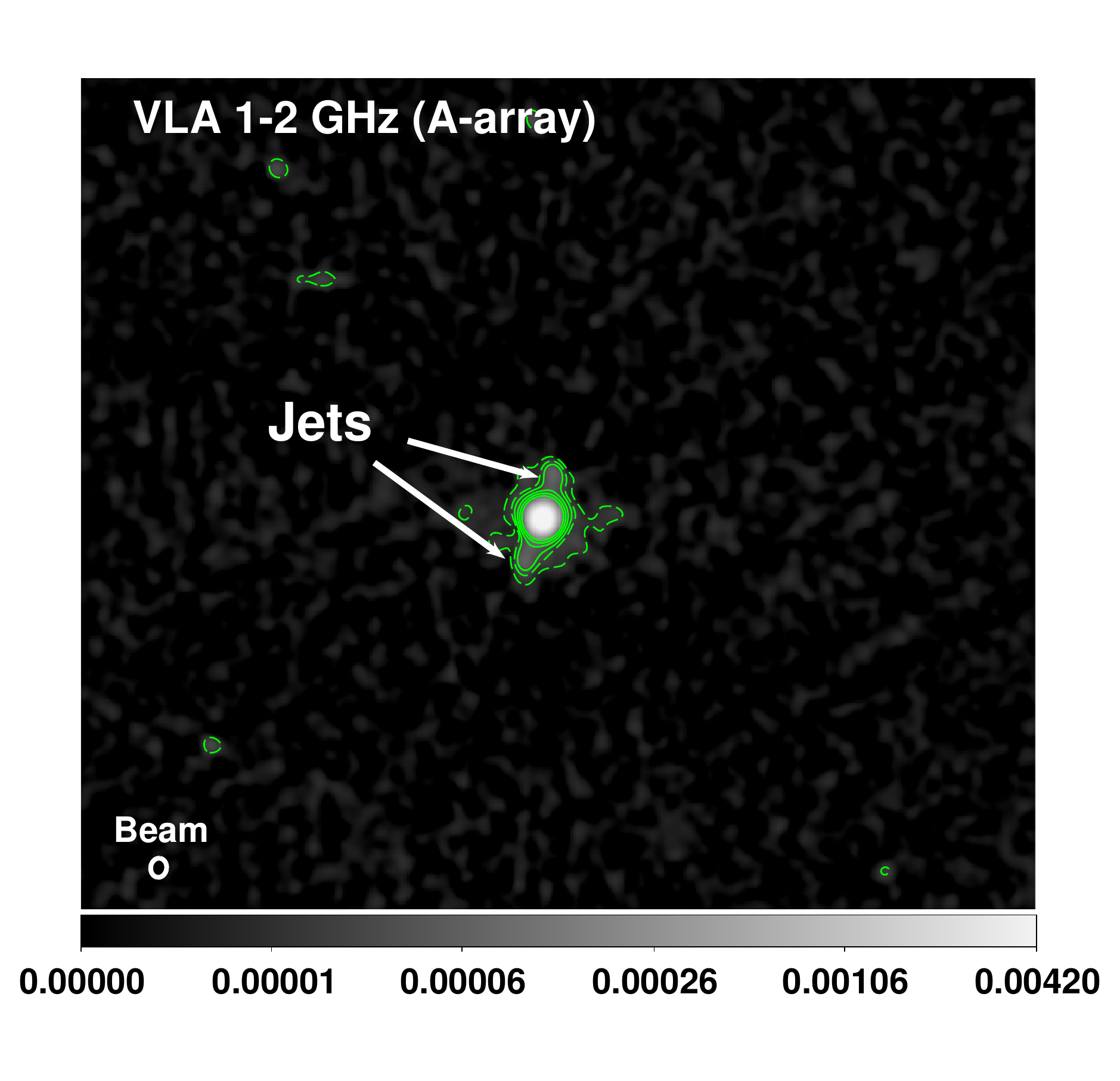}}
\end{minipage}
\begin{minipage}[c]{1.0\linewidth}
\subfloat{\includegraphics[width=0.48\textwidth]{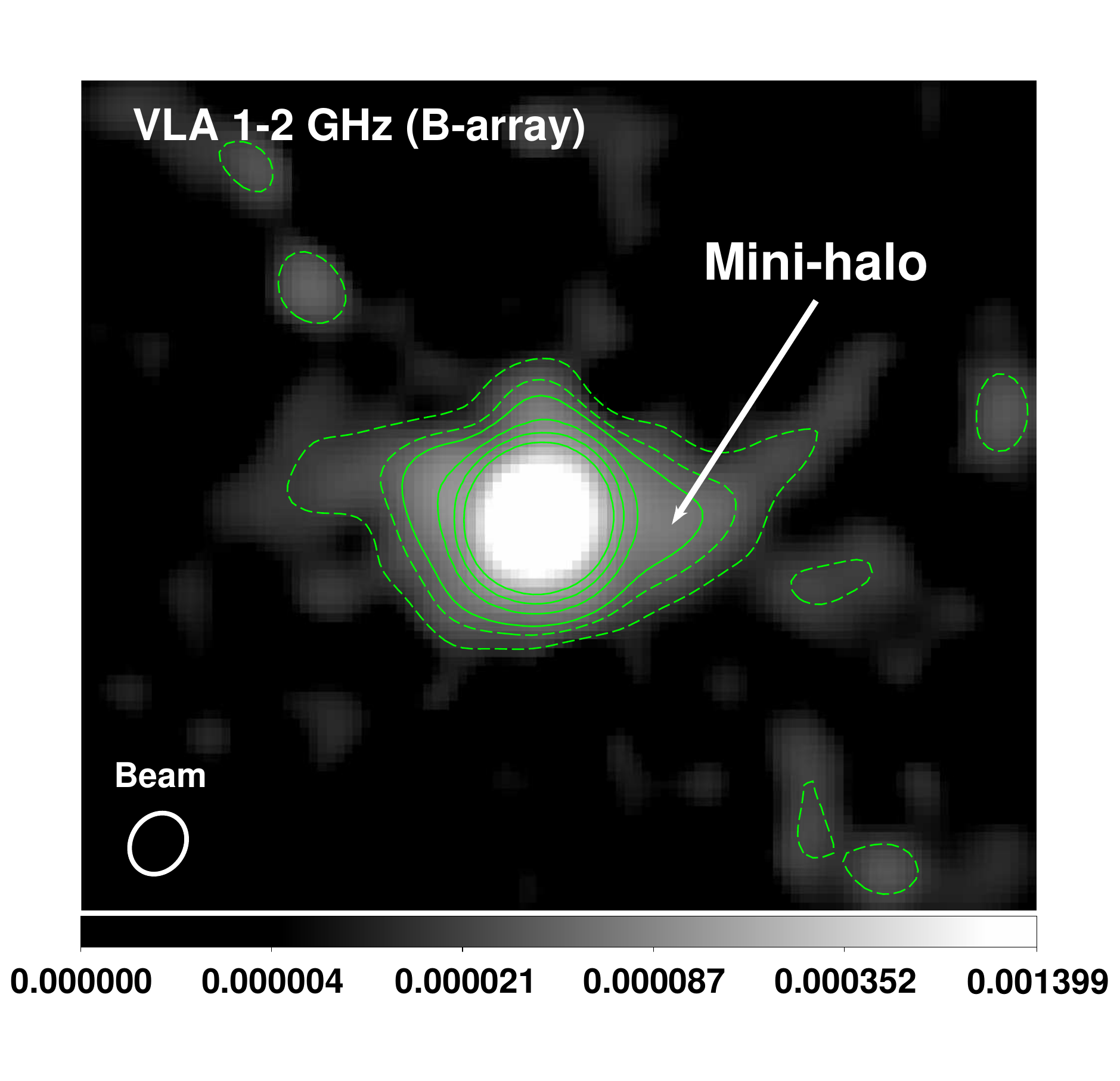}}
\subfloat{\includegraphics[width=0.48\textwidth]{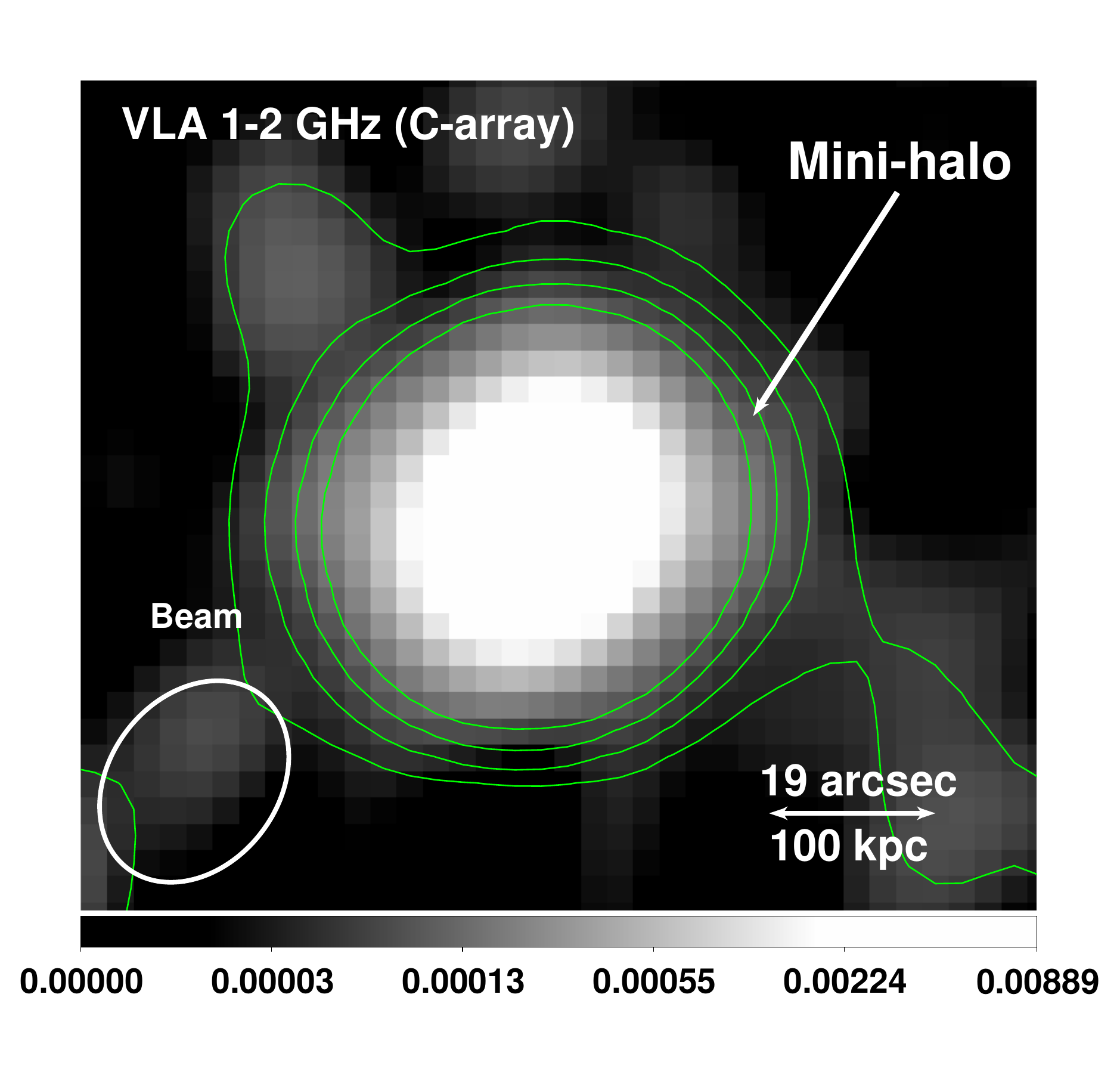}}
\end{minipage}
\caption[]{Images of MACS~J1447.4+0827, the four of them on the same scale than the right panel of Fig.~\ref{fig:Xray}. \textbf{Top-Left:} Unsharp-masked \textit{Chandra} image where a 2D gaussian image ($\sigma=6$) was subtracted
from a less smoothed image ($\sigma=1$). The positions of the X-ray cavities, potential ghost cavities and AGN have been identified. \textbf{Top-right:} VLA A-array image obtained at $1-2$~GHz with a
  $\sigma_{\rm rms}=11~$\textmu Jy beam$^{-1}$ and a beam size of $0.9\arcsec\times1.1\arcsec$ (shown with the white circle). The positions of the jets have been identified. \textbf{Bottom-Left:} VLA B-array image obtained at $1-2$~GHz with a
  $\sigma_{\rm rms}=16~$\textmu Jy beam$^{-1}$ and a beam size of $3.1\arcsec\times3.6\arcsec$. The image was smoothed with a Gaussian function with radius = 5 and $\sigma = 3$. \textbf{Bottom-Right:} VLA C-array image obtained at $1-2$~GHz with a
  $\sigma_{\rm rms}=15~$\textmu Jy beam$^{-1}$ and a beam size of $9.6\arcsec\times12.5\arcsec$. The image was smoothed with a Gaussian function with radius = 3 and $\sigma = 1.5$. Four radio contours are overlaid on each of the radio images: $3\sigma_{\rm rms}$, $6\sigma_{\rm rms}$, $12\sigma_{\rm rms}$ and $24\sigma_{\rm rms}$. Two additional contours (dash lines) are overlaid on the A- and B-array images: $1\sigma_{\rm rms}$ and $2\sigma_{\rm rms}$.}
\label{fig2}
\end{figure*}

New \textit{Chandra} ACIS observations of MACS~J1447.4+0827 were obtained in 2016 with ACIS-I in VFAINT mode (ObsID 17233 and 18825, PI Hlavacek-Larrondo), totalling an exposure time of 68 ks. These were combined with the existing 12 ks of observations taken with ACIS-S in VFAINT in 2008 (ObsID 10481, PI Hicks). These observations are presented in Table~\ref{tab:X-ray}. Data reduction was done following the standard procedure using \textsc{ciao} version 4.7. After reprocessing the datasets with \textsc{chandra\_repro}, point sources were removed from the image. Flares were then identified and removed using \textsc{deflare} script and \textsc{lc\_clean} routine. To combine the images, we used the \textsc{merge\_obs} task, which gave us the resulting 0.5-7 keV \textit{Chandra} exposure-corrected image shown in the right panel of Fig.~\ref{fig:Xray}. The total exposure time is 76.3~ks, and a detailed description and analysis of the X-ray image is presented in \citet{Myriam2020}. Again, we performed an unsharp-masked image, shown in the top-left panel of Fig.~\ref{fig2}, to highlight the X-ray cavities, the potential ghost cavities and the AGN. For the unsharp-masked image, we substracted a 2D gaussian smoothed image ($\sigma=6~$pixels) from a less smoothed image ($\sigma=1~$pixel) using \textsc{fgauss} and \textsc{fairith} from \textsc{ciao}.

\subsubsection{VLA Observations}

We also obtained 9 hours in L-band ($1-2$~GHz) with the VLA (PI Hlavacek-Larrondo) for MACS~J1447.4+0827: 3 hours in A-configuration (August 2015), 3 hours in C-configuration (February 2016) and 3 hours in B-configuration (May 2016). The observations were taken with 26 operational antennas. These observations are summarised in Table~\ref{tab:radio}. J1445+0958 was used as the phase calibrator while 3C286 was used as the flux and bandpass calibrator. The data were reduced following the standard procedure using the Common Astronomy Software Applications (CASA, \citealt{McMullin2007}; version 4.6). Between 35\% and 48\% of the observations (depending on the configuration) were affected by RFI, thus the \textsc{flagdata} task in \textsc{rflag} mode was used to remove the majority of the affected data. The rest was flagged manually by examining the amplitude in function of frequency plot with \textsc{plotms} to identify bad channels, baselines and time ranges. The data were then calibrated and the \textsc{clean} task was used to obtain an image. The \textsc{clean} task also creates a model that can be used with the \textsc{gaincal} and \textsc{applycal} tasks to calibrate the data, this is called self-calibration. The \textsc{clean} task is then used again to produce a new image, and a second round of self-calibration is performed on that new image. A third round of self-calibration was applied before obtaining the final images shown in the top-right and bottom panels of Fig.~\ref{fig2}. The resulting rms and beam size are presented in Table~\ref{tab:radio}. 

The images clearly reveal the presence of two relativistic jets (A-configuration data; top-right panel of Fig.~\ref{fig2}), as well as a faint and distinct component extending well beyond the inner jets, out to a radius of $\sim 36.2\arcsec \sim190~$kpc (B and C-configuration data; bottom panels of Fig.~\ref{fig2}). This diffuse component has a spectral index calculated between 1 and 2~GHz of $\alpha\sim-1.2 \pm 1.0$. The large uncertainty is due to the small frequency interval used to measure this spectral index. A more detailed analysis of the radio images in conjunction with the X-ray observations is presented in \citet{Myriam2020}. Yet, due to its size, its faint diffuse emission, its shape and its spectral index of $\alpha < -1$, we interpret this component as a mini-halo.

\section{Cluster Selection}\label{Selection}

\subsection{Cluster Sample}

One of the goals of this paper is to determine if mini-halos are connected to the feedback properties of BCGs. Therefore, in addition to the new  mini-halo and the confirmed mini-halo reported in the previous section, we compiled all known mini-halos from the literature. This was accomplished using more than thirty papers including those focusing on the Extended Giant Metrewave Radio Telescope (GMRT) Radio Halo Survey by \citet{Kale2013, Kale2015, Kale2015b}, the study of 75 galaxy clusters by \citet{Yuan2015} and the analysis by \citet{Giacintucci2014,Giacintucci2017,Giacintucci2019} using \textit{Chandra} X-ray data and radio observations from the VLA and the GMRT. A total of 31 mini-halos were found, in addition to the new mini-halo and the confirmed mini-halo reported in Section \ref{OBS} of this paper (see Table~\ref{tab:ClusterProperties}).

We note that our list of mini-halos does not include the one in A2390 even if it was included in \citet{Giacintucci2014}. As mentioned in \citet{Giacintucci2017}, it was recently found to extend to 800~kpc in size, well beyond the cold fronts. Therefore, this source has been classified as a giant radio halo \citep{Sommer2017}.



We also note that according to \citet{Giacintucci2014} and \citet{Giacintucci2017}, six of the mini-halos reported in those papers were classified as candidate mini-halos (A1068, A1413, A1795, MACS~J0159.8$-$0849, MACS~J0329.6$-$0211 and RXC~J1115.8+0129). A1068 and A1413 were considered candidate mini-halos as \citet{Govoni2009} argued that there is only low significance indications of diffuse emission in them, thus a clear classification would ideally require further investigation and deeper data. However, \citet{Savini2019} confirm the presence of a centrally located diffuse emission in the cluster A1413 using a high-resolution 144~MHz Low Frequency Array (LOFAR) image, and we therefore now consider this source as a confirmed mini-halo. We retain the total flux density from \citet{Govoni2009} as \citet{Savini2019} found the total flux density at 144~MHz and not at 1.4~GHz. In the case of A1795, \citet{Giacintucci2014} argued that deeper radio observations were required to confirm the existence of a mini-halo as the radio surface brightness distribution has an unusual filamentary shape. However, in \citet{Giacintucci2019}, the authors confirmed the nature as mini-halos of MACS~J0159.8$-$0849 and MACS~J0329.6$-$0211 using new GMRT observations. RXC~J1115.8+0129 was discovered by \citet{Pandey-Pommier2016}, but not enough parameters were measured to unambiguous classify it as a mini-halo. We therefore follow this classification (see last column in Table~\ref{tab:ClusterProperties}).

\citet{Giacintucci2014} also argued that an additional three clusters contained uncertain mini-halos (A2626, MACS~J1931.8$-$2634 and ZwCl~1742.1+3306; see last column in Table~\ref{tab:ClusterProperties}). For A2626, the extended emission seen in the core may be instead associated with the central radio galaxy \citep{Gitti2004} and recent studies complicated the classification by revealing a complex 'kite-like' radio structure \citep{Gitti2013,Ignesti2017, Kale2017,Ignesti2018}. For ZwCl~1742.1+3306, the observations were not sufficient to be sure of its nature. However, \citet{Paul2019} confirmed the presence of a significant extra diffuse emission surrounding the BCG at the centre of the cluster MACS~J1931.8$-$2634 using GMRT radio observations at 235 and 610~MHz, and we therefore now consider this source as a confirmed mini-halo. We retain the total flux density from \citet{Giacintucci2014} as \citet{Paul2019} found the total flux density at 235 and 610~MHz and not at 1.4~GHz. Finally, \citet{Giacintucci2017} included the mini-halo in RXJ1720.2+3536, referring to Macario et al. (private communication). There is unfortunately not enough information (e.g. mini-halo radio power, etc.) available on this last mini-halo to use it in our statistical analysis.


In summary, considering these potential caveats, the sample analysed in this paper comprises 33 mini-halos. These are presented in Table~\ref{tab:ClusterProperties}, along with the name of the host galaxy cluster, their redshift and eight additional parameters described in the following subsections.

\subsection{Cluster and Mini-Halo Properties}\label{sec:clusterProperties}


\afterpage{
\begin{landscape}
 \begin{table}
  \caption{Cluster and BCG properties for the 33 galaxy clusters with mini-halos. The columns are: 1. Cluster Name; 2. Redshift ($z$); 3. 1.4~GHz radio power of the mini-halo ($P_{1.4\textrm{~GHz}}$), corrected for our cosmology; 4. Reference for $P_{1.4\textrm{~GHz}}$; 5. Mass of the cluster inside the radius for which the density is 500 times the critical density of the universe at this redshift ($M_{500}$), from \citet{Planck2014} unless specified otherwise; 6. Cluster X-ray luminosity ($L_{\rm X}$) derived from observations in the $0.1-2.4$ keV band inside a radius of 600~kpc, from this work; 7. Cooling radius, defined as the radius at which the cooling time ($t_{\rm cool}$) is 3 Gyrs, from \citet{Bravi2016} unless specified otherwise; 8. Average radius of the mini-halo which corresponds to the square root of the maximum radius times the square root of the minimum radius which are derived from the $+3\sigma_{\rm rms}$ isocontours of their radio images; 9. Reference for the average radius of the mini-halos; 10. X-ray cavity power estimated using the buoyancy rise time, except for MACS~J1447.4+0827 where the sound crossing time is used.; 11. Reference for X-ray cavity power; 12. Steep BCG radio luminosity measured at 1~GHz ($\textrm{BCG}_{\textrm{steep}}$), derived from the fluxes in \citet{Hogan2015} using our cosmology unless specified otherwise; 13. Core BCG radio luminosity at 10~GHz ($\textrm{BCG}_{\textrm{core}}$), derived from the fluxes in \citet{Hogan2015} using our cosmology unless specified otherwise; 14. Notes: Uncertain (U) and candidate (C) mini-halos from \citet{Giacintucci2014,Giacintucci2017,Giacintucci2019}. \textbf{Reference code:} (1) \citet{Giacintucci2014}, (2) \citet{Knowles2019}, (3) \citet{Giacintucci2019}, (4) \citet{Govoni2009}, (5) \citet{Gitti2013}, (6) \citet{Kale2015}, (7) This work, (8) \citet{Sijbring1993}, (9) \citet{Raja2020}, (10) \citet{Kale2015b}, (11) \citet{Giacintucci2011}, (12) \citet{Giacintucci2014b}, (13) \citet{Yuan2015}, (14) \citet{Savini2019}, (15) \citet{Paul2019}, (16) \citet{Rafferty2006}, (17) \citet{Sanders2009a}, (18) \citet{Julie2012}, (19) \citet{Myriam2020}, (20) \citet{McDonald2015}, (21) \citet{Julie2013}.}
  \label{tab:ClusterProperties}
  \resizebox{24cm}{!} {
  \begin{tabular}{llllllllllllll}
    \hline
   \thead{Name} & \thead{$z$} & \thead{$P_{1.4\textrm{~GHz}}$} & \thead{Ref.} & \thead{$M_{500}$} & \thead{$L_{\rm X, 600~kpc}$} & \thead{Cooling Radius} & \thead{Average Size} & \thead{Ref.} & \thead{Cavity Power} & \thead{Ref.} & \thead{BCG steep} &  \thead{BCG core} & \thead{Notes}  \\
      &   & \thead{[$10^{24}$~W~Hz$^{-1}$]} & & \thead{[$10^{14}$  M$_{\sun}$]} & \thead{[$10^{44}$~erg~s$^{-1}$]} &  \thead{[kpc]} & \thead{[kpc]} & & \thead{[$10^{42}$~erg~s$^{-1}$]} &  & \thead{[$10^{24}$~W~Hz$^{-1}$]} & \thead{[$10^{24}$~W~Hz$^{-1}$]} & \\
    \hline
    
2A~0335+096&	0.060&	0.060 $\pm$ 0.006 &(1)&	2.27 $\substack{+0.24 \\ -0.25}$& 1.176 $\pm$ 0.011 & 46 $\pm$ 1&	70 & (1) &	24& (16) &	0.162 $\pm$ 0.002&	< 0.0089&\\ 

ACT$-$CL~J0022.2$-$0036	& 0.805	 & 8.4 $\pm$ 2.7 $^a$ & (2)	& 5.9 $\substack{+2.3 \\ -2.1}$ $^c$ & 11.3 $\substack{+2.0  \\ -1.5}$ & $<100$ $^h$	& 85 $^j$ & (2) & NA	  &	&NA	&NA&\\
    
A478&	0.088&	0.33 $\pm$ 0.06 &(1)&	7.1 $\substack{+0.3 \\ -0.4}$& 5.89  $\pm$ 0.12 & 45 $\pm$ 1&	160& (1) &	100 & (16) &	0.56 $\pm$ 0.06&	0.082 $\pm$ 0.005&\\

A907	&	0.153 & 0.89 $\pm$ 0.19 & (3)	& 5.2 $\pm$ 0.5  &  3.940 $\substack{+0.065  \\ 0.045}$ & $28.6 \substack{+6.2 \\ -1.2}$ $^h$	& 65 & (3) & NA & &6.6 $\pm$ 0.4	& $<2.50$	&\\

A1068& 0.137& 0.18 $\pm$ 0.06 &(4) & 3.55 $\substack{+0.38 \\ -0.41}$ & 4.073 $\substack{+0.056  \\ -0.051}$ & $51.77 \substack{+0.19 \\ -0.72}$ $^h$ & 85 $^j$& (7) & NA & & 0.5 $\pm$ 0.3 & $< 0.052$ & C \\

A1413& 0.143& 0.11 $\pm$ 0.04 &(4) & 5.98 $\substack{+0.38 \\ -0.40}$ & 5.78 $\substack{+0.22  \\ -0.25}$ & 34.5 $\substack{+17.0 \\ -9.0}$ $^h$ & 105& (14) & NA & & < 0.085 $^h$& $< 0.047$ $^h$& \\ 
	
A1795&	0.062&	0.80 $\pm$ 0.05 &(1)&	4.5 $\pm$ 0.2& 4.552 $\substack{+0.047  \\ -0.056}$ & 39 $\pm$ 2&	> 100& (1) &	160& (16) &	10.79 $\pm$ 0.17&	0.23 $\pm$ 0.07&	C\\
	
A1835&	0.2532&	1.3 $\pm$ 0.3 &(1)&	8.5 $\substack{+0.5 \\ -0.6}$& 20.1 $\substack{+0.6  \\ -0.3}$ & 57 $\pm$ 1&	240& (1) &	1800& (16)&	6.3 $\pm$ 0.9&	$< 0.55$&\\	
	
A2029 &	0.0765&	0.29 $\pm$ 0.04 &(1)&	6.8 $\pm$  0.2 & 7.366 $\substack{+0.029  \\ -0.044}$ & 35 $\pm$ 1&	270& (1) &	87& (16)&	9.35 $\pm$ 0.09&	0.0360 $\pm$ 0.0014&\\	
		
A2204&	0.152&	0.56 $\pm$ 0.06 &(1)&	8.0 $\pm$ 0.4 & 12.21 $\substack{+0.19  \\ -0.25}$ & 	50 $\pm$ 1&	50& (1) &	50000& (17)&	3.6 $\pm$ 1.1&	0.50 $\pm$ 0.15&\\
			
A2626&	0.055&	0.132 $\pm$ 0.013 &(5)&	2.4 $\pm$ 0.5 $^d$& 0.825 $\substack{+0.029  \\ -0.037}$ & 	17 $\pm$ 2&	30& (1) &	NA& &	0.57 $\pm$ 0.04&	0.037 $\pm$ 0.003&	U\\

A2667	& 0.230	 & 1.17 $\pm$ 0.11 & (3)	&  6.8 $\pm$ 0.5 & 11.19 $\substack{+0.34  \\ -0.37}$ & $<70$ $^h$	& 70	 & (3) & NA	   & & $<2.97$	& 0.9 $\pm$ 0.3	&\\
			
A3444&	0.254& 	2.5 $\pm$ 0.2	 &(3)&	7.6 $\substack{+0.5 \\ -0.6}$& 11.2 $\substack{+3.0  \\ -1.8}$ & 	103.9 $\substack{+38.1 \\ -16.9}$ $^h$ &	120& (3) &	NA& &	1.68 $\pm$ 0.08 &	$< 0.040$&\\

AS780&	0.236&	2.0 $\pm$ 0.5 &(6)&	7.7 $\pm$ 0.6 & 7.36  $\pm$ 0.15  & 	111.1 $\substack{+7.2 \\ -7.3}$ $^h$ &	80 $^j$ & (7) &	NA& &	4.89 $^h$&	19.71 $^h$&\\	
			
MACS~J0159.8$-$0849 &	0.405&	1.50 $\pm$ 0.15 &(1)&	6.88 $\substack{+0.90 \\ -0.98}$& 15.12 $\substack{+0.65  \\ -0.59}$ & 	53 $\pm$ 1&	90& (1) &	477& (18)&	1.40 $^h$&	27.59 $^h$&	\\
			
MACS~J0329.6$-$0211 &	0.450&	3.1 $\pm$ 0.4 &(1)&	4.9 $\pm$ 0.7 $^d$& 11.3  $\pm$ 0.7  & 	54 $\pm$ 1& > 70& (1) &	NA& &	2.02 $^h$&	0.49 $^h$&	\\

MACS~J1447.4+0827 & 0.3755 & 3.0 $\pm$ 0.3 &(7) & 7.46$\substack{+0.80 \\ -0.86}$ $^e$ & 24.0  $\pm$ 0.4 &  71 $\pm$ 5 $^{i}$ & 190 $^k$ & (7) & 570 &(19) & 2.10 $^h$& 3.38 $^h$ \\
			
MACS~J1931.8$-$2634&	0.352&	21.4 $\pm$ 1.6 &(1)&	6.19 $\substack{+0.77 \\ -0.83}$& 15.0  $\pm$ 0.2  & 	61 $\pm$ 1&	150& (15) &	8760& (18)&	27.58 $^h$&	17.80 $^h$&	\\
			
MS~1455.0+2232&	0.2578&	1.8 $\pm$ 0.2 &(1)&	3.5 $\pm$ 0.4 $^d$& 9.18 $\substack{+0.15  \\ -0.10}$ & 	55 $\pm$ 1&	120& (1) &	NA& &	2.6 $\pm$ 0.3&	0.22 $\pm$ 0.09&\\	
			
Ophiuchus&	0.028& 0.113 $\pm$ 0.016	 &(3)&	12.4 $\pm$ 1.2 $^d$& 2.035 $\substack{+0.055  \\ -0.036}$ & 	41.1 $\substack{+3.6 \\ -2.7}$ $^h$ &	250& (1) &	NA& &	0.052 $^h$&	$< 0.0088$ $^h$&\\	
			
Perseus&	0.0179&	2.21 $\pm$ 0.11 &(8)&	6.1 $\pm$ 0.6 $^d$& 1.9599 $\substack{+0.0024  \\ -0.0025}$ & 	78 $\pm$ 2 $^i$ &	130& (1) &	150& (16)&	2.8 $\pm$ 0.3&	7.9 $\substack{+4.7 \\ -4.0}$&\\	
			
Phoenix&	0.596&	 15 $\pm$ 3 $^b$  & (9) &	12.6 $\substack{+2.0 \\ -1.5}$ $^f$& 59.7 $\substack{+1.8  \\ -2.0}$ & 	73 $\pm$ 1&	310& (9) &	4450& (20)&	26.83 $^h$&	2.14 $^h$&\\	
			
PKS~0745$-$191 &	0.1028&	1.39 $\pm$ 0.03 &(7)&	7.3 $\pm$ 0.8 $^g$& 6.27 $\substack{+0.08  \\ -0.11}$ & 	59 $\pm$ 6 $^i$&	35 $^k$& (7) &	1700 & (16)&	83.8 $\pm$ 1.3&	$< 1.82$&\\

PSZ1~G139.61+24.20	& 0.267	 & 0.14 $\pm$ 0.02 & (3)	&  7.1 $\pm$ 0.6  & 6.94 $\substack{+0.31  \\ -0.29}$ & $<35$ $^h$	& 50	 & (3) & NA &	& NA	 & NA	&\\
			
RBS~797&	0.35&	2.3 $\pm$ 0.3 &(1)&	6.3 $\substack{+0.6 \\ -0.7}$& 18.3 $\substack{+0.6  \\ -0.5}$ & 	69 $\pm$ 1&	120& (1) &	1200& (16)&	4.34 $^h$&	0.88 $^h$&\\	

RXC~J1115.8+0129& 0.350 & 7.9 $\pm$ 0.4 & (10) & 6.4 $\pm$ 0.7 & 12.1  $\pm$ 0.5  & 101.4 $\substack{+27.2 \\ -8.6}$ $^h$ & NA $^l$ & & NA & & NA & NA & C \\

RX~J1347.5$-$1145&	0.4516& 29 $\pm$ 3	 &(3)&	10.61 $\substack{+0.74 \\ -0.77}$& 40.1 $\substack{+0.6  \\ -0.5}$ & 	62 $\pm$ 2&	320& (1) &	NA& &	6.05 $^h$&	9.83 $^h$&\\	
			
RXC~J1504.1$-$0248&	0.2153&	2.84 $\pm$ 0.16 &(11)&	7.0 $\pm$ 0.6 & 22.5 $\substack{+0.3  \\ -0.1}$ & 	64 $\pm$ 1&	140& (1) &	NA& &	5.13 $^h$&	3.27 $^h$&\\
				
RX~J1532.9+3021&	0.3621&	3.6 $\pm$ 0.3 &(1)&	4.7 $\pm$ 0.6 $^d$& 18.71 $\substack{+0.18  \\ -0.23}$ & 	67 $\pm$ 1&	100& (1) &	2220& (21)&	7.5 $\pm$ 0.8&	$< 1.26$&\\	
				
RX~J1720.1+2638&	0.1644&	5.6 $\pm$ 0.4 &(12)&	6.3 $\pm$ 0.4 & 7.17 $\substack{+0.12  \\ -0.11}$ & 46 $\pm$ 2&	140& (1) &	NA& &	5.6 $\pm$ 0.5&	0.21 $\pm$ 0.11&\\	
				
RX~J2129.6+0005&	0.235&	0.42 $\pm$ 0.07 &(6,13)&	4.24 $\substack{+0.55 \\ -0.59}$& 7.42 $\substack{+0.40  \\ -0.39}$ & 70.9 $\substack{+4.7 \\ -4.9}$ $^h$ &	100 $^j$& (7) &	NA& &	2.9 $\pm$ 0.4&	0.46 $\pm$ 0.12&\\	
				
Z3146&	0.290&	1.5 $\pm$ 0.2 &(1,13)&	6.7 $\pm$ 0.8 $^d$& 17.67 $\substack{+0.33  \\ -0.30}$ & 	60 $\pm$ 1&	90& (1) &	5800& (16)&	1.35 $\pm$ 0.19&	0.11 $\pm$ 0.06&\\
					
ZwCl~1742.1+3306&	0.076&	0.200 $\pm$ 0.012 &(1)&	2.63 $\substack{+0.27 \\ -0.29}$& 2.2  $\pm$ 0.3 & 	32 $\pm$ 1&	40& (1) &	NA& &	0.55 $\pm$ 0.11&	0.78 $\pm$ 0.16&	U\\

    \hline

\end{tabular}}

$^a$ Found flux at 1.4~GHz using $\alpha=1.15 \pm 0.15$ from the flux at 610~MHz in \citet{Knowles2019}. $^b$ Found flux using \citet{Raja2020} $P_{1.4\textrm{~GHz}}$ and their spectral index between 610~MHz and 1.52~GHz of $\alpha = -0.98 \pm 0.16$. $^c$ From \citet{Miyatake2013}. \textbf{$^d$} Estimated by \citet{Giacintucci2017} from the $M_{500} - T_X$ relation of \citet{Vikhlinin2009} using the core-excised temperatures. \textbf{$^e$} From \citet{Planck2015}. \textbf{$^f$} Estimated by \citet{McDonald2012}  using the $M_{500} - T_X$ relation of \citet{Vikhlinin2009}. \textbf{$^g$} Derived by \citet{Arnaud2005} from an NFW fit to the observed mass profile. \textbf{$^h$} From this work. $^i$ Estimated from the radial profile of the cooling time by \citet{Myriam2020} for MACS~J1447.4+0827, \citet{Sanders2014} for PKS~0745$-$191 and by \citet{Fabian2007} for Perseus. $^j$ Estimated from the GMRT 610~MHz image of \citet{Knowles2019} for ACT$-$CL~J0022.2$-$0036, from the VLA 1.4~GHz image of \citet{Govoni2009} for A1068 and from the GMRT 610~MHz image of \citet{Kale2015} for AS780 and RX~J2129.6+0005. $^k$ Estimated from the VLA 1-2~GHz image (see Fig. \ref{fig2}, bottom-right panel for MACS~J1447.4+0827 \& Fig. \ref{fig1}, right panel for PKS~0745$-$191). \textbf{$^l$} NA means that the value is not available. 

 \end{table}
\end{landscape} 
}


First, we consider the mini-halo's radio power at 1.4~GHz ($P_{1.4\textrm{~GHz}}$). It should be noted that the steep and negative spectral index of the mini-halo emission means that lower observing frequencies are preferred but the intrinsic sensitivity, radio frequency interference and angular resolution are more challenging below $1~$GHz. Thus, the literature often considers the radio power at 1.4~GHz since it provides optimal values for mini-halo detection. 


For the 31 mini-halos identified in the literature, we use the $P_{1.4\textrm{~GHz}}$ reported in those manuscripts (see Table~\ref{tab:ClusterProperties}). These have been corrected for our cosmology from the fluxes reported (see Table~\ref{tab:flux} for the fluxes). For the mini-halo in ACT$-$CL~J0022.2$-$0036, the flux density was only estimated at 610~MHz \citep{Knowles2019}. Therefore, we calculated the equivalent flux at 1.4~GHz using a spectral index of $\alpha=-1.15 \pm 0.15$, before calculating its $P_{1.4\textrm{~GHz}}$ value. We estimated the radio flux for PKS~0745$-$191 using the $4\sigma_{\rm rms}$ contours of Fig.~\ref{fig1}, which contained 99.5\% of the flux. To estimate the error on the mini-halo flux ($S_{\rm MH}$), we used the following equation from \citet{Cassano2013}:

\begin{equation}\label{Eq:Flux}
\sigma_{\rm S_{\rm MH}}=\sqrt{(\sigma_{\rm cal} {\rm S}_{\rm MH})^2 + ({\rm rms}\sqrt{N_{\rm beam}})^2 + \sigma_{\rm sub}^2},
\end{equation}

\noindent where $\sigma_{\rm cal}$ is the percentage of the flux outside of the extraction region, ${\rm rms}$ is the image noise, $N_{\rm beam}$ is the number of independent beams in the region and $\sigma_{\rm sub}$ is the error due to the uncertainty in the subtraction of the flux density of discrete radio sources in the mini-halo's region. $\sigma_{\rm sub}$ is estimated following \citet{Giacintucci2014}:

\begin{equation}\label{Eq:Sub}
\sigma^2_{\rm sub}=\sum_{i=1}^{N}(I_{{\rm MH},i} \times A_{s,i})^2,
\end{equation}

\noindent where $A_{s,i}$ is the area of the $i$-th radio source and $I_{{\rm MH},i}$ is the flux density of the mini-halo in that region. The mini-halo's flux density was obtained by subtracting the flux from the point sources as well as from the central AGN and was measured to be $S_{\rm MH}= 50.2 \pm 0.7~$mJy. The corresponding k-corrected $1.4~$GHz radio power was found using the equation:

\begin{equation}\label{Eq:Power}
P_{\rm MH} = 4\pi {\rm S}_{\rm MH} D_L^2 (1+z)^{-(\alpha + 1)},
\end{equation}

\noindent which takes into account the cluster redshift and applies a k-correction \citep[e.g.][]{vanWeeren2014}, using the spectral index ($\alpha$) and the luminosity distance ($D_L$). For the mini-halos spectral index, we adopted a value of $\alpha = -1.15 \pm 0.15$ \citep[e.g.][]{Giacintucci2014,Giacintucci2014b,Giacintucci2019}.


For MACS~J1447.4+0827, we used the $4\sigma_{\rm rms}$ contours on the C-configuration observations, which contained 99\% of the flux. The error on the mini-halo flux was estimated using Equation~\eqref{Eq:Flux}. The AGN's flux was found using an ellipsoidal region of 10~kpc with the A-configuration observations. The error was again found using Equation~\eqref{Eq:Flux}. This flux was subtracted from the C-configuration flux to obtain the mini-halo's flux density of $S_{\rm MH}= 5.7 \pm 0.7~$mJy, and then the corresponding radio power was found using Equation~\eqref{Eq:Power}.

Several authors have reported a statistically significant correlation between $P_{1.4\textrm{~GHz}}$ and the cluster X-ray luminosity in the $0.1-2.4$ keV band ($L_{\rm X}$) for clusters with known mini-halos \citep[e.g.][]{Cassano2008,Kale2013,Kale2015,Gitti2015b,Yuan2015}. However, we note that the X-ray luminosities are often found using different methods and computed for different radii and energy bands. For this reason, in this paper, we not only used the X-ray luminosity taken from the literature, but we also computed the X-ray luminosity in a uniform matter using two different methods, namely measuring the cluster X-ray luminosity inside a circular region of 600~kpc in radius and inside a radius of $R_{500}$ (where $R_{500}$ is the radius for which the density is 500 times the critical density of the universe at the cluster redshift). We chose a radius of 600~kpc as there is no consensus in the literature over which radius to use and papers often do not state which radius they are using, therefore we arbitrary took this one. Furthermore, we think that there will be a stronger correlation between $P_{1.4\textrm{~GHz}}$ and $L_{\rm X}$ in the $0.1-2.4$ keV band if $L_{\rm X}$ represents the central region of the clusters as it is the gas in this region that would interact with mini-halos particles. Moreover, a radius of $R_{500}$ is considerably bigger than mini-halos, considering that \citet{Giacintucci2017} adopted  a physically-motivated definition of mini-halos as having a radius between 50~kpc and $0.2R_{500}$, the radius delimiting the inner region where cooling, AGN feedback and sloshing become more important. The methods will be compared and discussed in Section~\ref{sec:mHClusterProperties} and the values of all $L_{\rm X}$ are reported in Table~\ref{tab:LxMethods}, however only the method using a radius of 600~kpc will be used in the shown correlations and plots (see Table~\ref{tab:ClusterProperties}). 

For the $L_{\rm X}$ values using a uniform method, we identified and downloaded the relevant observations from the \textit{Chandra} archive and followed the standard \textsc{ciao} (version 4.11) reduction pipeline after selecting a good time interval clear of background flaring events detected with \textsc{lc\_sigma\_clip}. Following the classic diffuse emission spectrum extraction procedure, we simultaneously fit, using \textsc{xspec} (version 12.10.1), an absorbed thermal emission model (\textsc{phabs}$*$\textsc{apec}) to the source spectrum and a standard background model compensating for both extragalactic and galactic absorption (see \S2.1 of \citealt{McDonald2015}). For each of them, we used a circular region with a radius of $\sim600~$kpc and of $R_{500}$. Table~\ref{tab:LxMethods} gives the observation IDs from \textit{Chandra} of the different clusters, the parameters used in the fits done with \textsc{xspec} and the values of $R_{500}$. For the values from the literature, the $0.1-2.4$ keV X-ray luminosities were taken from \citet{Kale2015}, \citet{Ettori2013} and \citealt{Yuan2015}, and based on the \textit{ROSAT} All-Sky Survey. All these have also been corrected for our cosmology using the fluxes reported. The X-ray luminosities for PKS~0745$-$191 and MACS~J1447.4+0827 were found using an elliptical region of $\sim600~$kpc radius, and an energy band of $0.1-2.4$ keV. The spectrum was extracted using the \textsc{specextract} task in \textsc{CIAO} and then \textsc{xspec} (v12.9.1b) was used to fit the data with \textsc{phabs}, \textsc{apec} and \textsc{clumin} models. 
We also decided to explore the relation between $P_{1.4\textrm{~GHz}}$ and $M_{500}$, as it was done in past studies \citep[e.g.][]{Giacintucci2014,Yuan2015,Kale2019,Giacintucci2019,Paul2019}. $M_{500}$ provides more systematic results to measure the cluster mass than $L_{\rm X}$ since the Planck collaboration recently reported $M_{500}$ values for hundreds of galaxy clusters \citep{Planck2014}. Here, $M_{500}$ is defined as the mass of the cluster inside the radius $R_{500}$ (see Table~\ref{tab:ClusterProperties}). We note that there is a known correlation between $M_{500}$ and the cluster X-ray luminosity (e.g. \citealt{Pratt2009}). We therefore expect that $P_{1.4\textrm{~GHz}}$ will correlate with $M_{500}$, since it correlates with cluster X-ray luminosity. We also note that while we were able to obtain estimates of $M_{500}$ based on the Planck database for the majority of our clusters, 10 of our clusters were not included in the catalog. For these clusters, we therefore used the most up-to-date estimates available in the literature (see column 5 in Table~\ref{tab:ClusterProperties} for references).

Another important parameter to consider is the cluster cooling radius, a property based on the cooling time ($t_{\rm cool}$). Here, we define the cooling radius as the radius at which $t_{\rm cool}=3$ Gyrs \citep{Bravi2016}, where $t_{\rm cool}$ is chosen such that it represents the average time since the last merger event. Within this radius, AGN feedback must be powerful enough to offset cooling and is therefore a useful parameter to better understand if mini-halos are connected to the feedback properties of BCGs. For most of the clusters in Table~\ref{tab:ClusterProperties}, the values and uncertainties come from the study of \citet{Bravi2016}. The values and uncertainties for MACS~J1447.4+0827, PKS~0745$-$191 and for the Perseus cluster were estimated from the radial profile of the cooling time by \citet{Myriam2020}, \citet{Sanders2014} and \citet{Fabian2007} respectively. In order to calculate the cooling radius for clusters which did not have values listed in the literature, we resorted to X-ray spectral analysis. We used the same method employed to find the uniform $L_{\rm X}$ values with the addition of deprojecting the observations using the \textsc{dsdeproj} software. Using the best fit parameters from the fit in \textsc{xspec}, we calculated the cooling time as a function of radius from the centre of the X-ray emission by the following relation:

\begin{equation}
t_{cool} = \frac{5}{2}\frac{1.9~n_{\rm e}~kT~V}{L},
\end{equation}

\noindent where $n_{\rm e}$ is the electron density, $kT$ is the gas temperature, $V$ is the gas volume contained within each spherical shell and $L$ is the gas X-ray bolometric luminosity. The radii were chosen such that each annulus contained $\sim5000~$counts. Table~\ref{tab:ObsID} gives the observation IDs from \textit{Chandra} of the different clusters and the parameters used in the fit done with \textsc{xspec}. For ACT$-$CL~J0022.2$-$0036, A2667 and PSZ1~G139.61+24.20, we found only an upper limit as the observations in the \textit{Chandra} archive of those clusters did not contain enough counts to have a precise measurements of the cooling radius for $t_{\rm cool}=3$ Gyrs. For all the other clusters, the errors on the cooling radii were found by interpolating the errors of the other points in the plot of the cooling time as a function of radius.

Finally, Table~\ref{tab:ClusterProperties} includes the mini-halo average radius. This radius corresponds to the square root of the maximum radius times the square root of the minimum radius. \citet{Giacintucci2014,Giacintucci2019}, \citet{Paul2019}, \citet{Savini2019} and \citet{Raja2020} report these radii using the $+3\sigma_{\rm rms}$ isocontours of their images, except for four mini-halos: ACT$-$CL~J0022.2$-$0036, A1068, AS780 and RX~J2129.6+0005. For these, we followed the same technique to obtain their average radius, using the GMRT 610~MHz image of \citet{Knowles2019} for the first mini-halo, the VLA 1.4~GHz image of \citet{Govoni2009} for the second mini-halo and the GMRT 610~MHz images of \citet{Kale2015} for the latter two. For our confirmed and newly identified mini-halos in PKS~0745$-$191 and MACS~J1447.4+0827, we applied a similar procedure to our radio images presented in Figs.~\ref{fig1} and~\ref{fig2}. The mini-halos' average radius is an important property as it gives an indication of the distance to the AGN where the particles stop being reaccelerated. At this point, the particles do not have enough energy to produce strong synchrotron emission. We note however that the radii do not have uncertainties in Table~\ref{tab:ClusterProperties} and that they can be affected by the signal-to-noise ratio of the images. Furthermore, the extent of what we can observe of the mini-halo depends greatly on the frequency of the observations due to their extremely steep spectra. Therefore, the radii in Table~\ref{tab:ClusterProperties} should be considered at the very least as lower limits to the true mini-halo radius. Even with this indication, \citet{Giacintucci2014} explicitly state that the mini-halo in A1795 and in MACS~J0329.6$-$0211 may be more extended and this is the reason why they are lower limits in Table~\ref{tab:ClusterProperties}.

\subsection{BCG properties}\label{sec:BCGproperties}

Since our goal is to determine if mini-halos are connected to the feedback properties of BCGs, we outline in this section the parameters used to measure feedback properties in BCGs. Feedback comes from the relativistic jets of the AGN, detected in radio, that create X-ray cavities. Therefore, feedback can be quantified by X-ray cavities in X-ray observations or by the radio emission coming from the jets, two independent methods.

Consequently, we first decided to search the literature to identify which of the 33 mini-halos of Table~\ref{tab:ClusterProperties} harbored known X-ray cavities. We found that a total of 15 clusters have published X-ray cavities (\citealt{Rafferty2006, Sanders2009a, Julie2012, Julie2013, McDonald2015, Myriam2020}). In Table~\ref{tab:ClusterProperties}, we list mechanical power of these X-ray cavities, a quantity that represents the work and internal energy needed to create the cavities divided by the time associated with this formation, the buoyancy rise time. For MACS~J1447.4+0827, the sound crossing time is used. However, the time found using different methods does not vary significantly, and therefore we still use this cluster in the relations. These powers are known to within a factor of a few. For every cluster, only the main central cavities were used to measure the cavity power. Away from the centre of the clusters, it gets harder to detect cavities as the number of counts becomes very small because of the steep central rise in the X-ray profile for cool core clusters, therefore a decrease in counts is difficult to measure. Thus, we would be including a bias in this study by using more than the central cavities, as outer cavities are detectable only in the X-ray brightest clusters. 

Even if only 15 clusters were found with X-ray cavities, it does not mean that the remaining clusters with known mini-halos do not contain X-ray cavities. Following a study of 55 clusters \citep{Dunn2008}, it was found that 95\% of the cool core clusters have X-ray cavities. This means that clusters with mini-halos should also harbour cavities since they are found in strong cool core clusters. We studied the X-ray images of the remaining clusters, and even if some of them showed hints of X-ray cavities, most of them did not due to the lack of good quality X-ray images. \citet{Panagoulia2014b} found that data quality strongly affect the detection of X-ray cavities using a volume- and X-ray luminosity-limited sample of 49 clusters, with a central cooling time of $\leq3~$Gyr. They calculated the number of counts in the central circular region of 20~kpc of radius for all the clusters, 30 of them hosting X-ray cavities. They found that, for most of them, sources with fewer than $\sim20,000~$counts within 20~kpc from their core do not have clearly detected X-ray cavities, and all sources with $\geq30,000~$counts have X-ray cavities. Not enough counts makes it hard to detect the decrease in counts and associate it to X-ray cavities, as due to 3-D effects, the decrease in counts is only of $10-25\%$. Therefore, when deeper images of the clusters without X-ray cavities will be available, X-ray cavities should be detectable. Furthermore, determining upper limits for the cavity power of those clusters is not feasible due to the impossibility of placing meaningful constraints on the possible size of the cavities.

We also explored the properties of the radio emission associated with the AGN in the BCGs. However, studying radio emission from AGNs is challenging since not all of the emission comes from the relativistic jets. There is also radio emission coming from different regions around the AGN. To overcome this problem, observations at different spectral frequencies and spatial resolutions are required. Following the study by \citet{Hogan2015}, it is now possible to decompose the radio spectral energy distributions (SEDs) of BCGs into multiple components, in particular a core component, related to ongoing accretion by the AGN, and a steeper component, related to the lobe emission (jets) linked with the past activity of the AGN. This study focused on the radio properties of over 300 BCGs, using a variety of data from the Australia Telescope Compact Array (ATCA), the VLA and the Very Long Baseline Array (VLBA) telescopes. For each BCG in their sample, \citet{Hogan2015} built radio SEDs and typically decomposed the radio SEDs into a flat spectrum component (if present) with a flatter spectral index ($\alpha > -0.5 $; $\textrm{BCG}_{\textrm{core}}$), and a steeper component ($\alpha < -0.5 $; $\textrm{BCG}_{\textrm{steep}}$). For each BCG, the authors then determined the radio flux of each of these components normalized at a radio frequency of 10~GHz and 1~GHz respectively. Table~\ref{tab:flux} lists these values and their spectral indexes for each of our clusters that had available radio fluxes from \citet{Hogan2015}, the others were found in this work using all the available multi-frequency data on the respective clusters and the same method described above and in \cite{Hogan2015}. The radio powers were derived from the fluxes using our cosmology and applying a k-correction, and are listed in Table~\ref{tab:ClusterProperties}. When the spectral indexes were not available, we used $\alpha_{\rm steep}=1.0$ and $\alpha_{\rm core}=0.2$. According to \citet{Hogan2015}, as mini-halos often have very steep spectral indexes ($\alpha < -1.5$), there may be a link between mini-halos and the emission from persistent AGNs, characterised by an ultra-steep $\textrm{BCG}_{\textrm{steep}}$ component. Finally, it is important to note that if the $\textrm{BCG}_{\textrm{core}}$ is high, the accretion of the AGN is powerful. The same can be said for $\textrm{BCG}_{\textrm{steep}}$ in regard to the lobe emission linked with the past activity of the AGN.


\section{Analysis}\label{Analysis}

In the following section, we search for evidence of correlations between the parameters outlined in Section~\ref{Selection}.

\subsection{Mass $M_{500}$ of the cluster}

\begin{figure*}
\centering
\begin{minipage}[c]{1.0\linewidth}
\centering 
\subfloat{\includegraphics[width=0.48\textwidth]{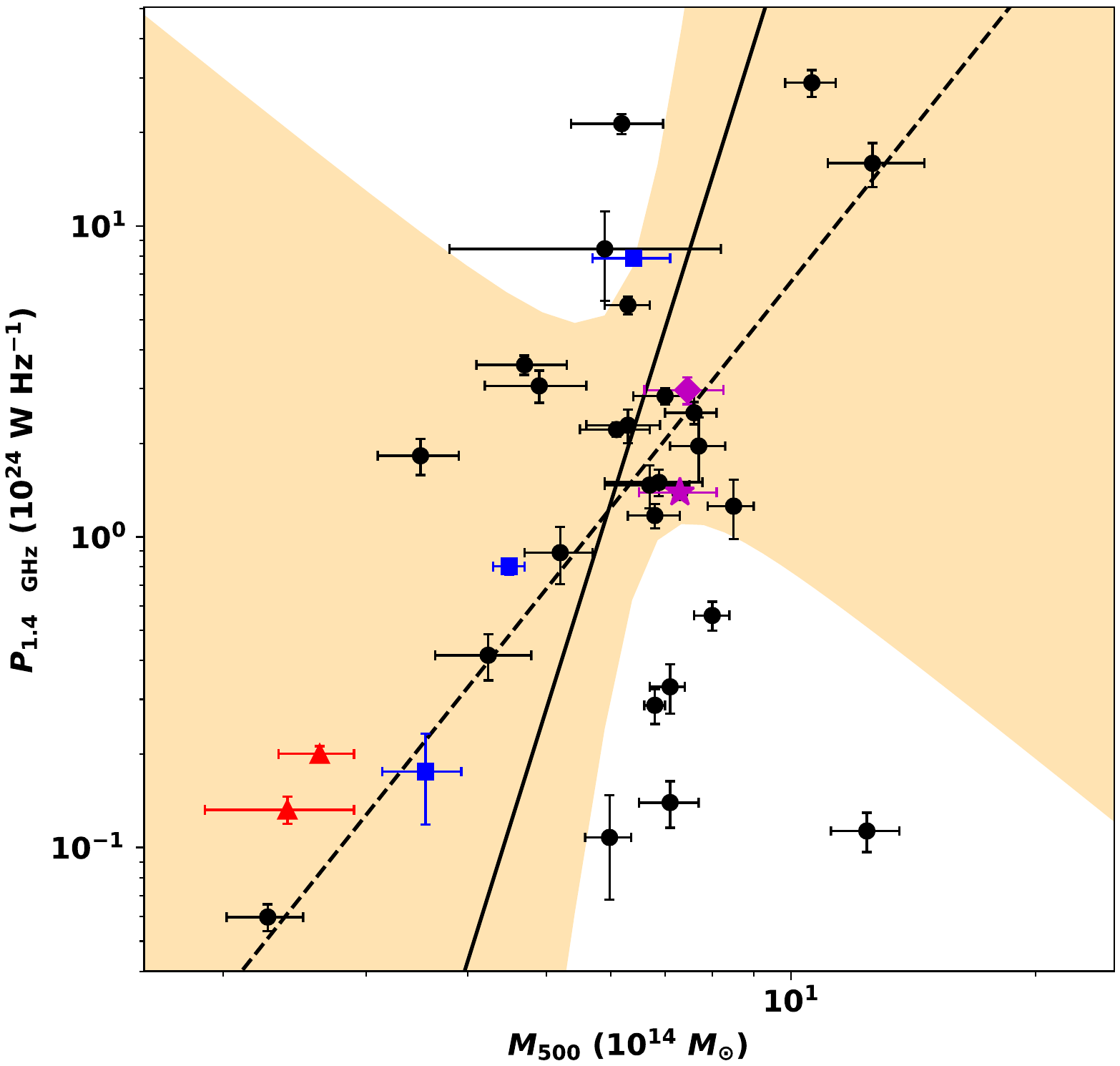}}
\hspace{4mm}
\subfloat{\includegraphics[width=0.48\textwidth]{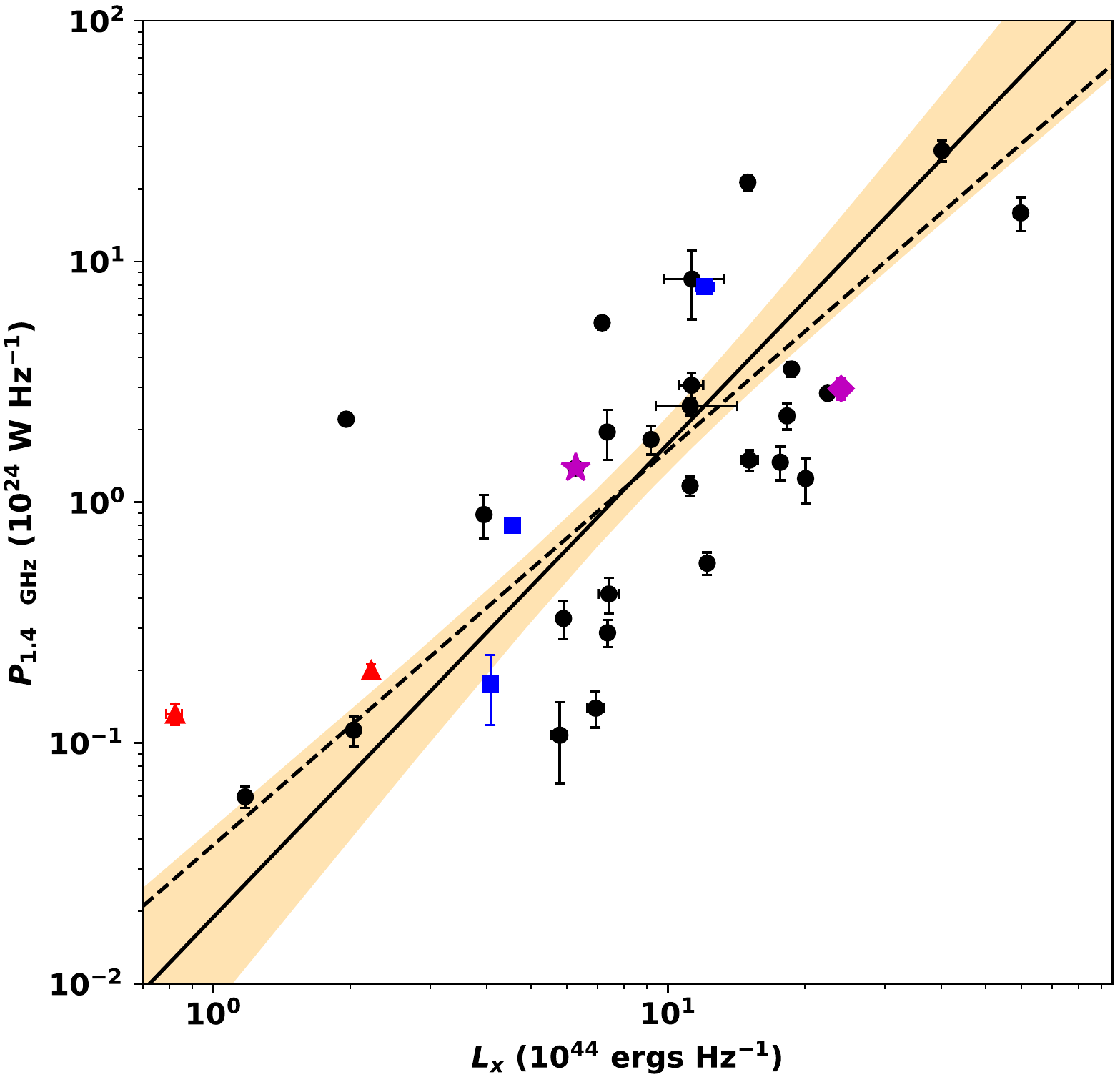}}
\end{minipage}
\caption[]{\textbf{Left:} The 33 mini-halos are shown in the plane of the mini-halo radio power ($P_{1.4\textrm{~GHz}}$) and the cluster mass ($M_{500}$). The candidate (blue squares) and uncertain (red triangles) mini-halos are shown. The best-fit lines using the BCES-orthogonal (solid) and the BCES-bisector (dashed) methods are displayed, as well as the 95\% confidence regions of the best-fit relation for the BCES-orthogonal method (orange region). The best-fit lines are done using every mini-halo, including the candidate and uncertain mini-halos. The magenta star represents PKS~0745$-$191, while the magenta diamond represents MACS~J1447.4+0827. \textbf{Right:} Same but in the plane of $P_{1.4\textrm{~GHz}}$ and the X-ray luminosity inside a radius of 600~kpc ($L_{\rm X}$), for 32 mini-halos.}
\label{fig:PowerM500}
\end{figure*}

We first explored the relation between $P_{1.4\textrm{~GHz}}$ and $M_{500}$. We show the resulting relation in Fig.~\ref{fig:PowerM500} (left panel) with the fit following a power-law relation:

\begin{equation}\label{Eq:Log}
\log(P_{1.4\textrm{~GHz}}) = A_{\rm Eq.\ref{Eq:Log}}~\log(X_{P}) + B_{\rm Eq.\ref{Eq:Log}},
\end{equation}

\noindent
where $X_{P}$ is $M_{500}$. The fit was performed in the log-log space using linear regression by adopting both the BCES-bisector and BCES-orthogonal regression algorithms (\citealt{Akritas1996}; see \citealt{Nemmen2012} as an example of application of the Bivariate Correlated Errors and intrinsic Scatter (BCES) method). On each figure, the best-fit using the BCES-orthogonal method and its $95\%$ confidence region are shown, as well as the linear fit using the BCES-bisector for comparison purposes\footnote{The linear regressions and confidence bands were found using the script on \url{https://github.com/rsnemmen/BCES}}. Each fit was done using every mini-halo, including the candidate and uncertain mini-halos, but excluding the upper limits (see Appendix~\ref{appendix:BCES} for explanation). Furthermore, the results of both algorithms with and without $10,000$ bootstrap resamples are shown in Table~\ref{tab:Correlations}. See Appendix~\ref{appendix:BCES} to understand what is the BCES method, why we use it and why we chose one regression algorithm over the other. 

 \begin{table*}
  \caption{Best-fit correlation parameters. The columns are: 1. Method used to do the linear fit (BCES-orthogonal, BCES-orthogonal using $10,000$ bootstrap resamples, BCES-bisector and BCES-bisector using $10,000$ bootstrap resamples); 2. \& 3. Slope and intercept of the power-law relation, following Equation~\eqref{Eq:Log} for the log-log relations; 4. The Pearson rank correlation coefficient $r_p$; 5. The related probability of no correlation $\rho_p$; 6. The Spearman rank correlation coefficient $r_s$; 7. The related probability of no correlation $\rho_s$; 8. The strength of the correlation based on the Pearson coefficients, which is considered strong if $r_p > 0.60$, moderate if $0.40 < r_p < 0.59$ and weak if $\rho_p > r_p$ (see \citealt{Press1992}, p. 634).}
  \label{tab:Correlations}
  \begin{tabular}{lllllllc}
    \hline
    \thead{Method} & \thead{Slope (A$_{\rm Eq.\ref{Eq:Log}}$)} & \thead{Intercept (B$_{\rm Eq.\ref{Eq:Log}}$)} & \thead{$r_p$} & \thead{$\rho_p$} & \thead{$r_s$} & \thead{$\rho_s$} & \thead{Strength} \\
    \hline 
    \hline
    \multicolumn{8}{|c|}{\boldmath$\log(P_{1.4\textrm{ \textbf{GHz}}}) - \log(M_{500})$}\\
    \hline
    BCES-orthogonal & 8.38 $\pm$ 3.75 & -6.41 $\pm$ 3.04 & \multirow{4}{*}{0.44} & \multirow{4}{*}{0.95\%} & \multirow{4}{*}{0.26} & \multirow{4}{*}{14\%} & \multirow{4}{*}{\textbf{Moderate}}\\
    bootstrap & 10.08 $\pm$ 60.49 & -7.74 $\pm$ 48.07 & &  & & &  \\
    BCES-bisector & 3.28 $\pm$ 0.70 & -2.46 $\pm$ 0.49 &  &  &  & & \\
    bootstrap & 3.07 $\pm$ 1.04 & -2.29 $\pm$ 0.79 &  &  &  & & \\
    
    \hline
    \multicolumn{8}{|c|}{\boldmath$\log(P_{1.4\textrm{ \textbf{GHz}}}) - \log(L_{\rm \textbf{X}})$}\\
    \hline
    BCES-orthogonal & 1.97 $\pm$ 0.31 & -1.72 $\pm$ 0.31 & \multirow{4}{*}{0.73} & \multirow{4}{*}{0.00012\%} & \multirow{4}{*}{0.71} & \multirow{4}{*}{0.00031\%} & \multirow{4}{*}{\textbf{Strong}} \\
    bootstrap & 2.03 $\pm$ 0.39 & -1.79 $\pm$ 0.38 &  &  &  & &  \\
    BCES-bisector & 1.64 $\pm$ 0.17 & -1.42 $\pm$ 0.19 &  &  &  & &  \\
    bootstrap & 1.65 $\pm$ 0.19 & -1.43 $\pm$ 0.21 &  &  &  & &  \\
    
\hline
    \multicolumn{8}{|c|}{\boldmath$\log(P_{1.4\textrm{ \textbf{GHz}}}) - \log(\textrm{\textbf{BCG}}_{\textrm{\textbf{steep}}})$}\\
    \hline
    BCES-orthogonal & 0.99 $\pm$ 0.21 & -0.36 $\pm$ 0.11 & \multirow{4}{*}{0.68} & \multirow{4}{*}{0.0070\%} & \multirow{4}{*}{0.53} & \multirow{4}{*}{0.41\%} & \multirow{4}{*}{\textbf{Strong}} \\
    bootstrap & 1.03 $\pm$ 0.26 & -0.38 $\pm$ 0.15 &  &  &  & &  \\
    BCES-bisector & 0.99 $\pm$ 0.13 & -0.358 $\pm$ 0.096 &  &  &  & &  \\
    bootstrap & 1.01 $\pm$ 0.14 & -0.368 $\pm$ 0.099 &  &  &  & &  \\
    
    \hline
    \multicolumn{8}{|c|}{\boldmath$R_{\textrm{MH}} - \log(\textrm{\textbf{BCG}}_{\textrm{\textbf{steep}}})$}\\
    \hline
     & &  & 0.15 & 47\% & 0.26 & 21\% & \textbf{Not correlated} \\

    \hline
    \multicolumn{8}{|c|}{\boldmath$\log(P_{1.4\textrm{ \textbf{GHz}}}) - \log(\textrm{\textbf{BCG}}_{\textrm{\textbf{core}}})$}\\
    \hline
    BCES-orthogonal & 0.61 $\pm$ 0.16 & 0.24 $\pm$ 0.11 & \multirow{4}{*}{0.63} & \multirow{4}{*}{0.24\%} & \multirow{4}{*}{0.57} & \multirow{4}{*}{0.68\%} & \multirow{4}{*}{\textbf{Strong}}\\
    bootstrap & 0.61 $\pm$ 0.18 & 0.24 $\pm$ 0.12 &  &  &  & &  \\
    BCES-bisector & 0.747 $\pm$ 0.096 & 0.24 $\pm$ 0.12 &  &  &  & & \\     
    bootstrap & 0.76 $\pm$ 0.10 & 0.24 $\pm$ 0.13 &  &  &  & & \\
    
    \hline
    \multicolumn{8}{|c|}{\boldmath$R_{\textrm{MH}} - \log(\textrm{\textbf{BCG}}_{\textrm{\textbf{core}}})$}\\
    \hline
     & &  & 0.13 & 60\% & 0.12 & 62\% & \textbf{Not correlated} \\
    
    \hline
    \multicolumn{8}{|c|}{\boldmath$\log(P_{1.4\textrm{ \textbf{GHz}}}) - \log$(\textbf{Cavity Power) excluding A2204}}\\
    \hline
    BCES-orthogonal & 0.82 $\pm$ 0.11 & -2.12 $\pm$ 0.30 & \multirow{4}{*}{0.83} & \multirow{4}{*}{0.022\%} & \multirow{4}{*}{0.73} & \multirow{4}{*}{0.29\%} & \multirow{4}{*}{\textbf{Strong}} \\
    bootstrap & 0.79 $\pm$ 0.16 & -2.05 $\pm$ 0.46 &  &  &  & & \\
    BCES-bisector & 0.847 $\pm$ 0.082 & -2.21 $\pm$ 0.21 &  &  &  & & \\
    bootstrap & 0.838 $\pm$ 0.094 & -2.18 $\pm$ 0.25 &  &  &  & &  \\
    
    
    \hline
    \multicolumn{8}{|c|}{\boldmath$R_{\textrm{MH}} - \log($\textbf{Cavity Power) excluding A2204}}\\
    \hline
     &  &  & 0.07 & 82\% & 0.05 & 87\% & \textbf{Not correlated} \\
    
    
    \hline
    \multicolumn{8}{|c|}{\boldmath$\log(P_{1.4\textrm{ \textbf{GHz}}}) - \rm{\textbf{Radius}}$}\\
    \hline
     &  & & 0.39 & 3.2\% & 0.44 & 1.6\% & \textbf{Weak} \\
    
        \hline
  \end{tabular}
 \end{table*}
 
 
To evaluate the strength of the correlation we used the Pearson test using every mini-halo and quote the Spearman test for comparison (see Appendix~\ref{appendix:tests} for an explanation of those tests). The coefficients of the Pearson test are $r_p$ and $\rho_p$, and the strength of the correlation is confirmed if $r_p$ is close to 1 or -1 and $\rho_p$, the probability of no correlation, is close to 0 (see Appendix~\ref{appendix:tests} for their definition). The relation is considered strong if $r_p > 0.60$, moderate if $0.40 < r_p < 0.59$ and weak if $\rho_p > r_p$ (see \citealt{Press1992}, p. 634). Similarly for the coefficients $r_s$ and $\rho_s$ of the Spearman test.

With values of $r_p$ = 0.44 and a probability of no correlation of 0.95\% for $P_{1.4\textrm{~GHz}}$ as a function of $M_{500}$, the relation presents a moderate linear correlation. However, when looking at the BCES-orthogonal fit in the left panel of Fig.~\ref{fig:PowerM500}, it can be realized that this fit is almost a vertical line. Since the range of variation of $M_{500}$ is limited and there seems to be considerable intrinsic scatter, a constant $M_{500}$ would probably give a fit almost as good as the BCES-orthogonal algorithm. Therefore, Fig.~\ref{fig:PowerM500} (left panel) should be considered more as a general trend. 


\subsection{Cluster X-ray luminosity}

We also reproduced the known correlation between $P_{1.4\textrm{~GHz}}$ and the X-ray luminosity, this time including our confirmed mini-halo and our newly detected mini-halo in PKS~0745$-$191 and MACS~J1447.4+0827, and a uniform way of finding the X-ray luminosity, namely inside a radius of 600~kpc. This correlation can be seen in the right panel of Fig.~\ref{fig:PowerM500} and the best-fit regression line has a slope of $1.97 \pm 0.31$ and $1.64 \pm 0.17$ for the BCES-orthogonal and BCES-bisector methods respectively (see Table~\ref{tab:Correlations}). The fit was done following the power-law of Equation~\eqref{Eq:Log} with $L_{\rm X}$ instead of $X_{P}$. Those values intersect or are close to the ones found by  \citet{Kale2013, Kale2015} with slopes of 1.43 $\pm$ 0.52, 3.37 $\pm$ 0.70 (BCES-orthogonal method) and 2.49 $\pm$ 0.30 (BCES-bisector method). The Pearson coefficients also point to a strong correlation (see Table~\ref{tab:Correlations}), with $r_p$ = 0.73 and a probability of no correlation of 0.00012\%. This correlation is further supported by the fact that the two BCES methods give a very similar fit. This confirms the previously known correlation from e.g. \citet{Cassano2008,Kale2013,Kale2015,Gitti2015b,Yuan2015,Gitti2015,Gitti2018,Giacintucci2019}.  

\subsection{Steep BCG radio luminosity measured at 1~GHz}

\begin{figure*}
\centering
\begin{minipage}[c]{1.0\linewidth}
\centering 
\subfloat{\includegraphics[width=0.48\textwidth]{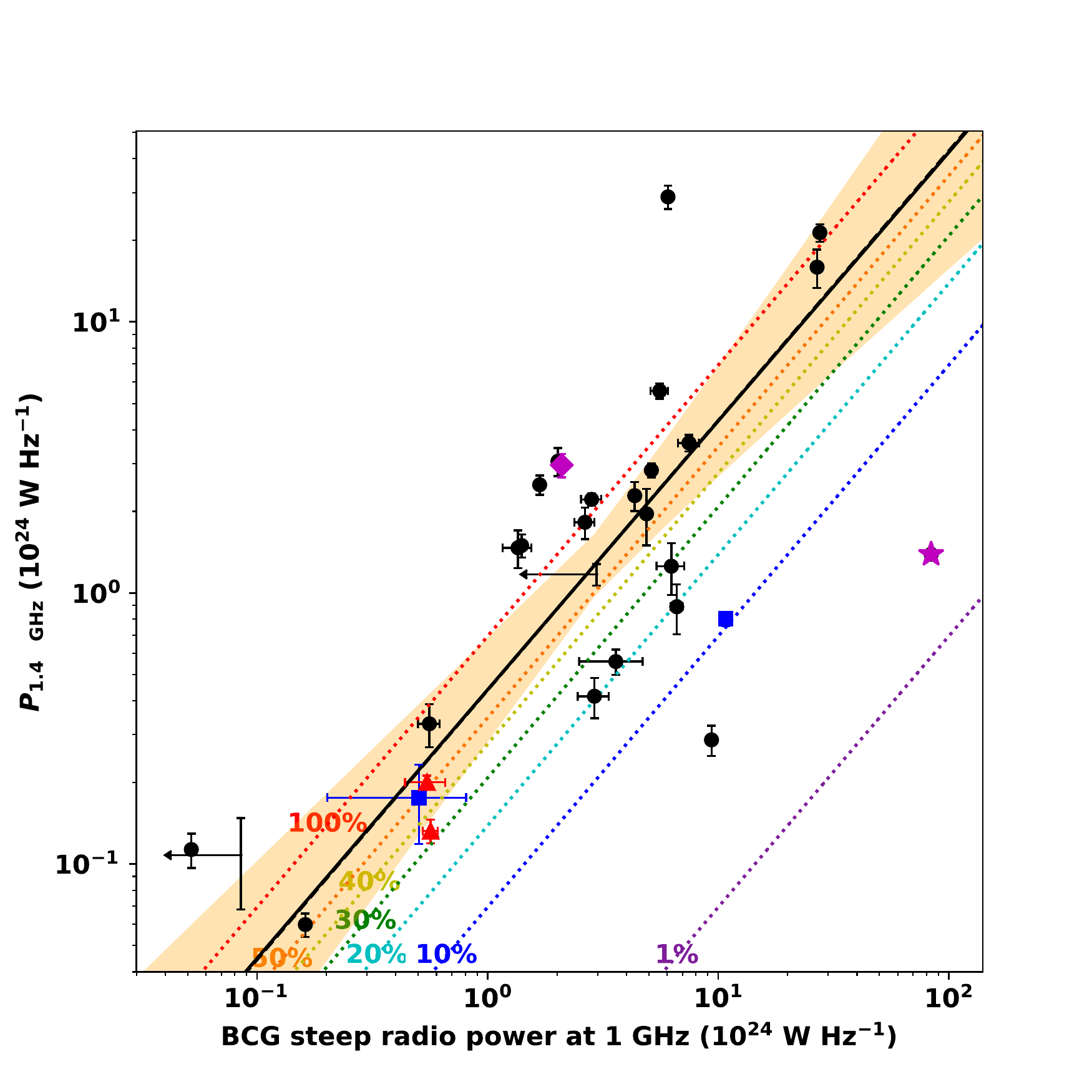}}
\hspace{4mm}
\subfloat{\includegraphics[width=0.48\textwidth]{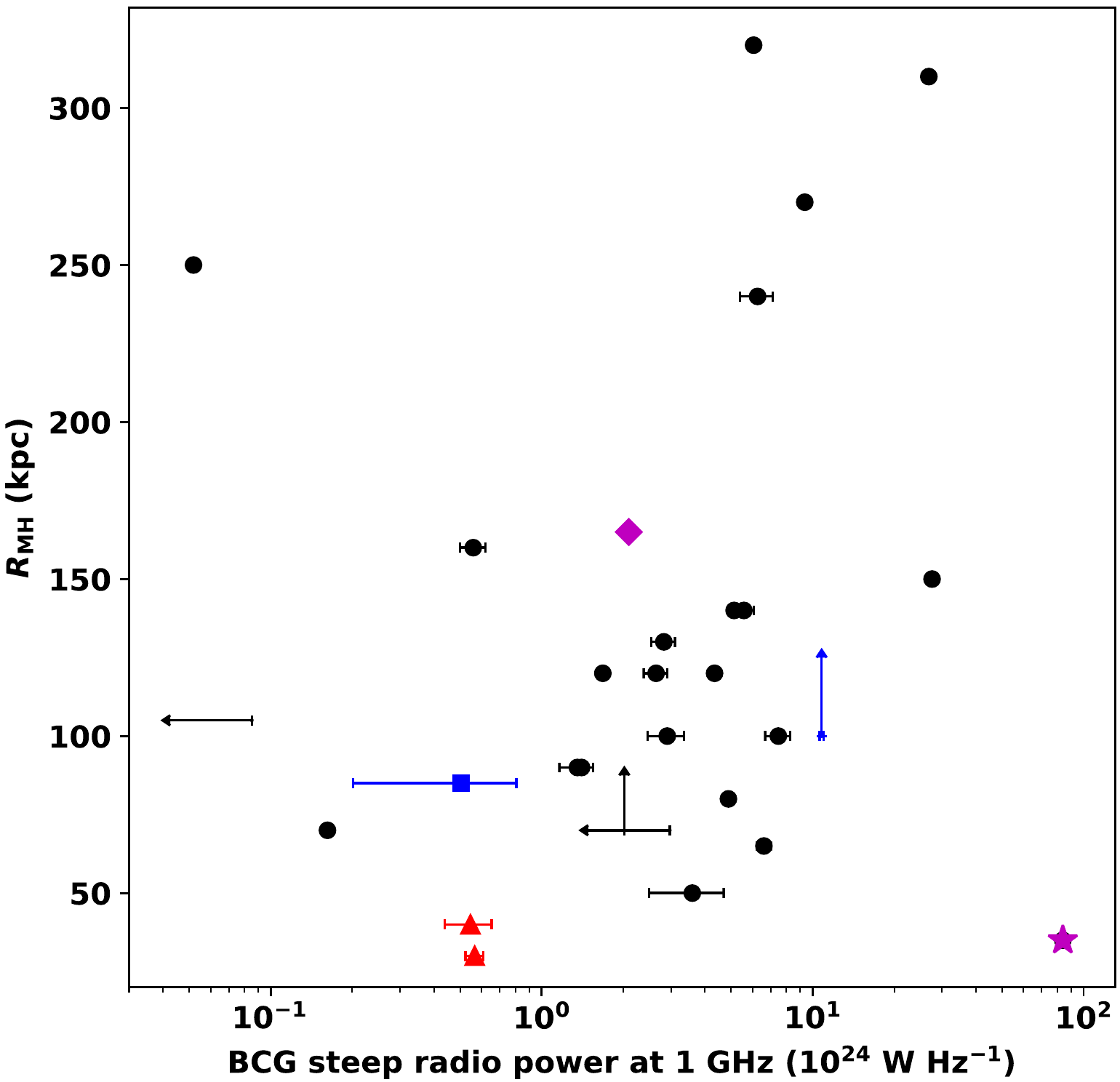}}
\end{minipage}
\caption[]{\textbf{Left:} Mini-halo radio power at 1.4~GHz ($P_{1.4\textrm{~GHz}}$) as a function of the BCG steep radio power at 1~GHz (latter from \citealt{Hogan2015} and this work), for 28 mini-halos and two upper limits (arrow). The candidate (blue squares) and uncertain (red triangles) mini-halos are shown. The best-fit lines using the BCES-orthogonal (solid) and the BCES-bisector (dashed) methods are displayed, as well as the 95\% confidence regions of the best-fit relation for the BCES-orthogonal method (orange region).  The best-fit lines are done using every mini-halo with a value, including the candidate and uncertain mini-halos, but excluding the upper limits. The magenta star represents PKS~0745$-$191, while the magenta diamond represents MACS~J1447.4+0827. The dotted lines represent where $P_{1.4\textrm{~GHz}}$ is 1\% (purple), 10\% (blue), 20\% (cyan), 30\% (green), 40\% (light green), 50\% (orange) and 100\% (red) of the BCG steep radio power at 1.4~GHz, which we extrapolated from the BCG steep radio power at 1~GHz using a spectral index of $\alpha \sim -1.09$, the median spectral index from our clusters' spectral index values in \citet{Hogan2015} and found in this work). \textbf{Right:} Same but in the plane of the average radius of the mini-halo ($R_{\textrm{MH}}$) and the BCG radio steep power at 1~GHz, for 26 mini-halos and four upper/lower limits (arrows). Fits are not shown here as there is no general trend.}
\label{fig:Steep}
\end{figure*}

Looking at Fig.~\ref{fig:Steep} and Table~\ref{tab:Correlations}, we see that $P_{1.4\textrm{~GHz}}$ as a function of the BCG steep radio power at 1~GHz is strongly correlated with a probability of no correlation of 0.0070\% (see Table~\ref{tab:Correlations} for the slopes found following the power law of Equation~\eqref{Eq:Log} with $\textrm{BCG}_{\textrm{steep}}$ instead of $X_{P}$). The strong correlation is further supported by the fact that the two fitting methods, BCES-orthogonal and BCES-bisector, give almost exactly the same fit, the two fits behind one on top of the other (see left panel of Fig.~\ref{fig:Steep}). In Fig.~\ref{fig:Steep} (left panel), we added dotted lines representing where $P_{1.4\textrm{~GHz}}$ is 1\%, 10\%, 20\%, 30\%, 40\%, 50\% and 100\% of the BCG steep radio power at 1.4~GHz, which we extrapolated from the BCG steep radio power at 1~GHz using a spectral index of $\alpha \sim -1.09$, the median spectral index from our clusters' spectral index values in \citet{Hogan2015} and found in this work. The average mini-halo radius and the $\textrm{BCG}_{\textrm{steep}}$ parameters however reveal no clear correlation (with a probability of no correlation of 47\%) as depending on the method used, the fits are not consistent due to the huge scatter of the data points. The study of this linear regression was done in the linear-log plane. The parameters of the fit (slope and intercept) are not shown in Table~\ref{tab:Correlations} as there is no correlation. Those two relations are studied for the first time in this work.



\subsection{Core BCG radio luminosity at 10~GHz}

\begin{figure*}
\centering
\begin{minipage}[c]{1.0\linewidth}
\centering 
\subfloat{\includegraphics[width=0.48\textwidth]{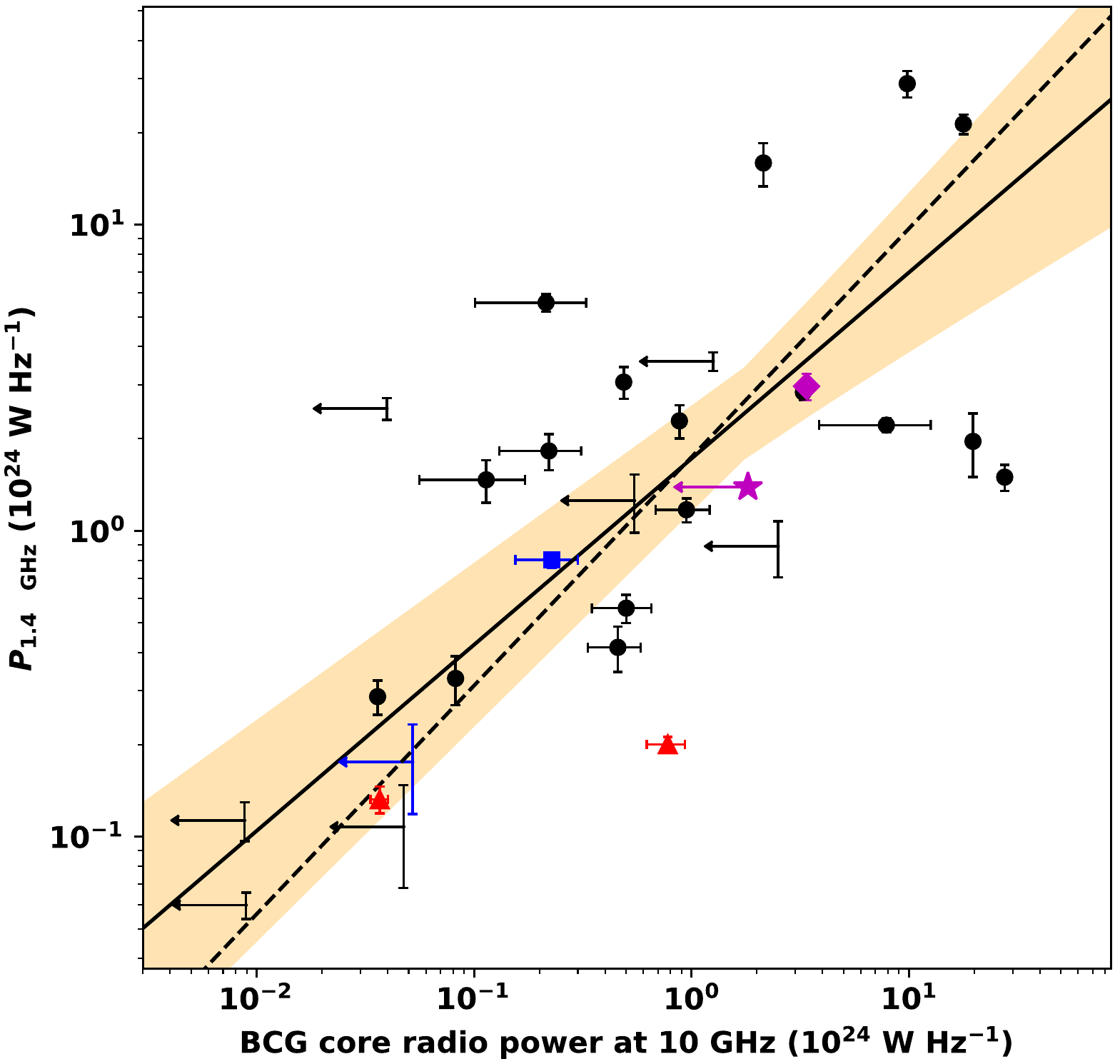}}
\hspace{4mm}
\subfloat{\includegraphics[width=0.48\textwidth]{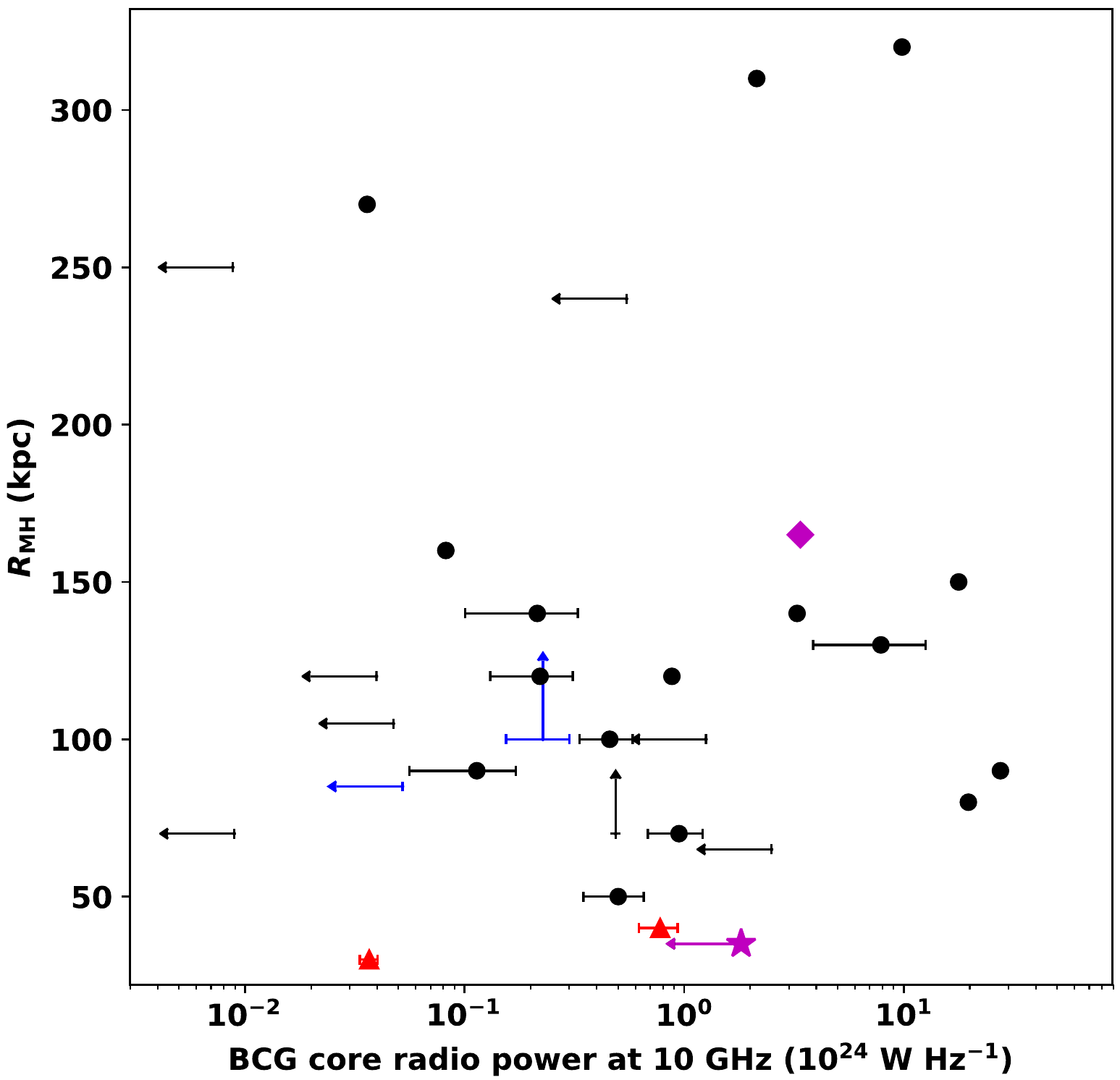}}
\end{minipage}
\caption[]{\textbf{Left:} Mini-halo radio power at 1.4~GHz ($P_{1.4\textrm{~GHz}}$) as a function of the BCG core radio power at 10~GHz (latter from \citealt{Hogan2015} and this work), for 21 mini-halos and nine upper limits (arrows). The candidate (blue squares) and uncertain (red triangles) mini-halos are shown. The best-fit lines using the BCES-orthogonal (solid) and the BCES-bisector (dashed) methods are displayed, as well as the 95\% confidence regions of the best-fit relation for the BCES-orthogonal method (orange region). The best-fit lines are done using every mini-halo with a value, including the candidate and uncertain mini-halos, but excluding the upper limits. The magenta star represents PKS~0745$-$191, while the magenta diamond represents MACS~J1447.4+0827. \textbf{Right:} Same but in the plane of the average radius of the mini-halo ($R_{\textrm{MH}}$) and the BCG radio core power at 10~GHz, for 19 mini-halos and 10 upper/lower limits (arrows). Fits are not shown here as there is no clear general trend.}
\label{fig:Core}
\end{figure*}

On the other hand, for the newly studied relation between $P_{1.4\textrm{~GHz}}$ and $\textrm{BCG}_{\textrm{core}}$ (Fig.~\ref{fig:Core}; left panel), there is a strong correlation ($r_p=0.63$, probability of no correlation of 0.24\%; see Table~\ref{tab:Correlations} for the slopes). The fit was again performed following the power-law of Equation~\eqref{Eq:Log} with $\textrm{BCG}_{\textrm{core}}$ instead of $X_{P}$. The newly studied relation of the mini-halos average radius as a function of the BCG core radio power at 10~GHz (Fig.~\ref{fig:Core}; right panel) does not reveal a statistically significant correlation. This can be confirmed with the Pearson parameters (see Table~\ref{tab:Correlations}). Indeed, $r_p$ is smaller than $\rho_p$ and there is a probability of no linear correlation of 60\% in the linear-log plane. Again, the values of the fits are not shown in Table~\ref{tab:Correlations}.

\subsection{Cavity X-ray power}

\begin{figure*}
\centering
\begin{minipage}[c]{1.0\linewidth}
\centering 
\subfloat{\includegraphics[width=0.48\textwidth]{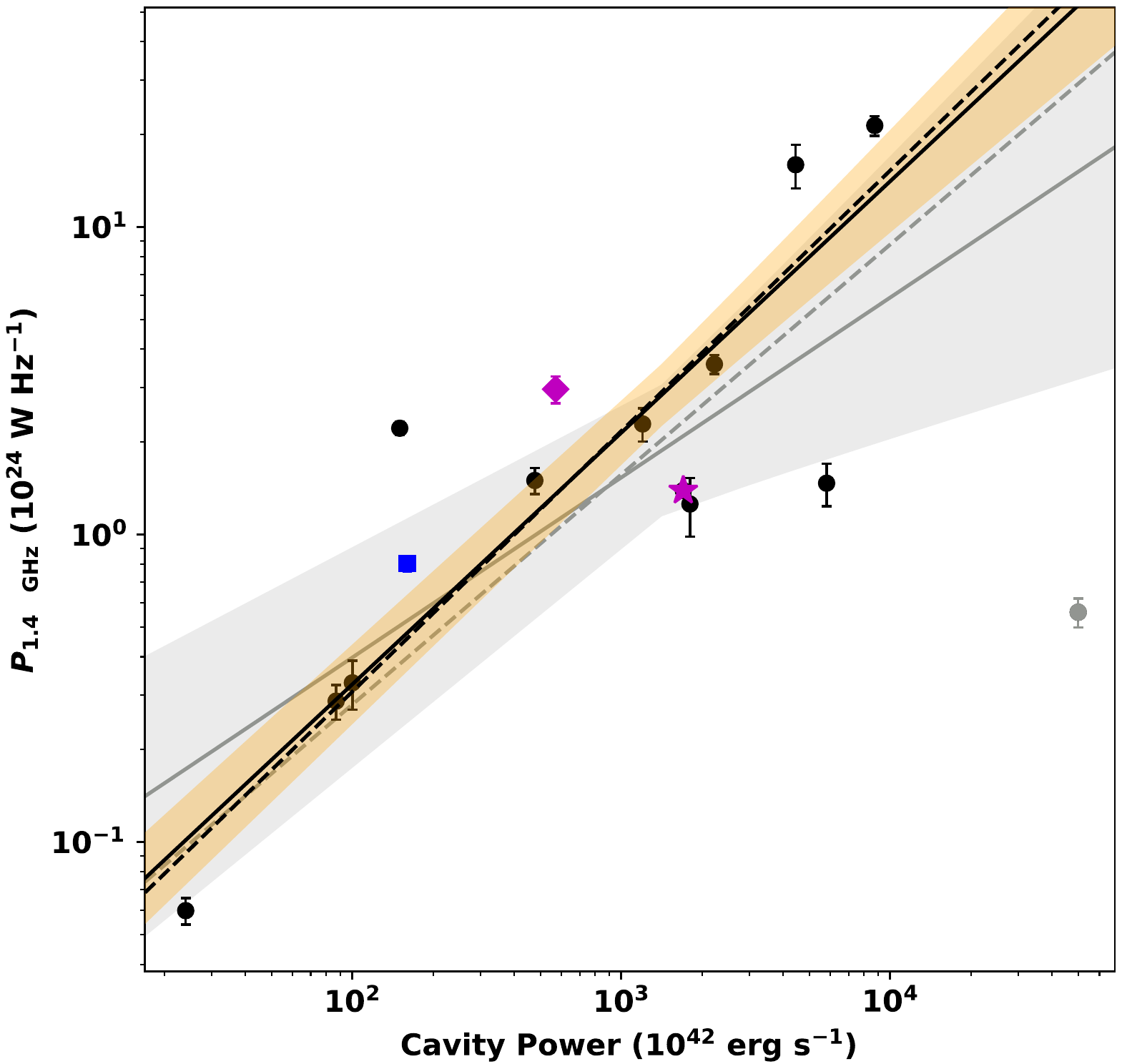}}
\hspace{4mm}
\subfloat{\includegraphics[width=0.48\textwidth]{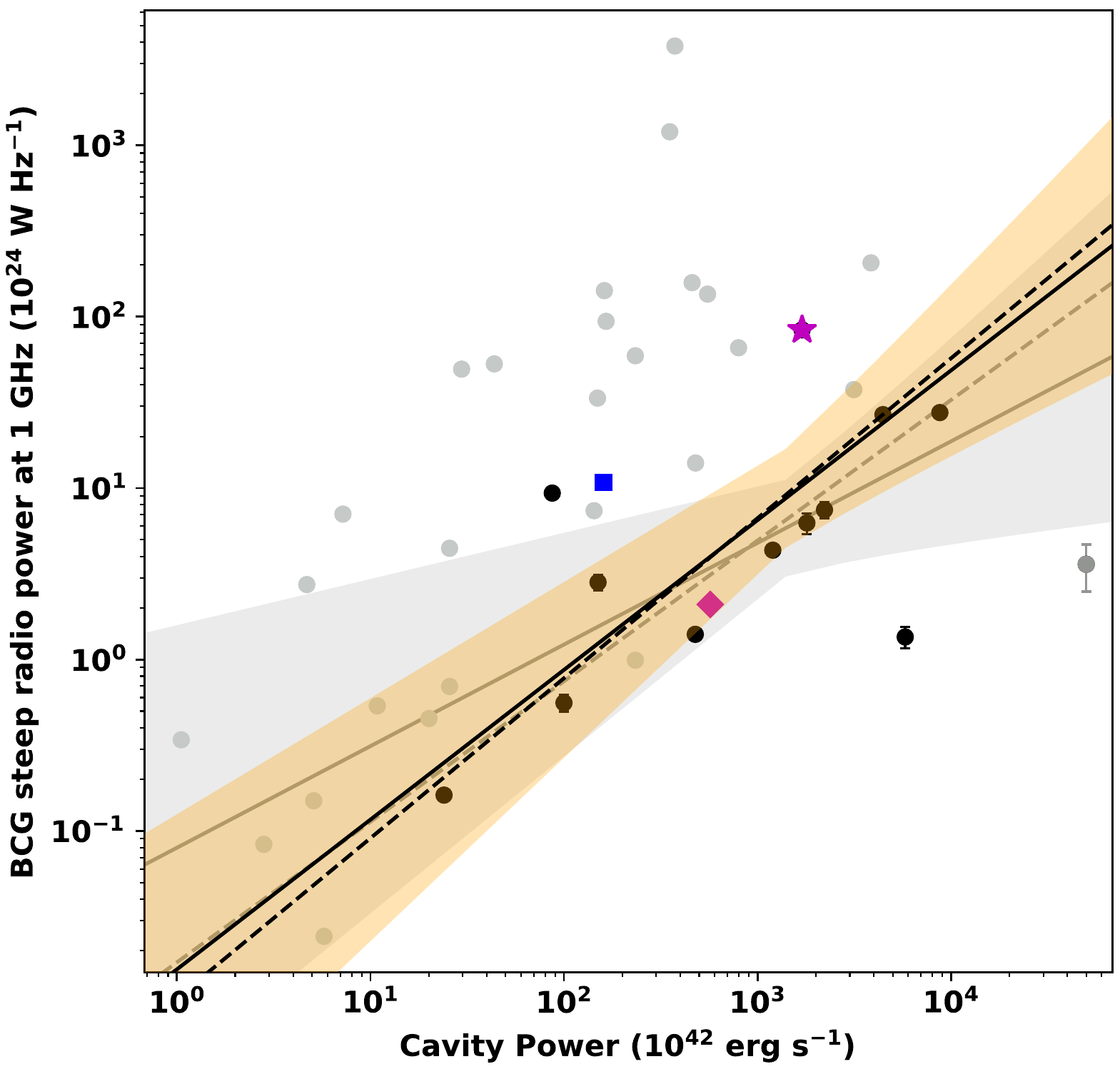}}
\end{minipage}
\caption[]{\textbf{Left:} Mini-halo radio power at 1.4~GHz ($P_{1.4\textrm{~GHz}}$) as a function of the X-ray cavity power, for 15 mini-halos. The candidate (blue squares) mini-halos are shown. The outlier, the anomalously powerful cavities for the mini-halo power, is in the cluster A2204 and is shown in grey. The best-fit lines using the BCES-orthogonal (solid black) and the BCES-bisector (dashed black) methods are displayed, as well as the 95\% confidence regions of the best-fit relation for the BCES-orthogonal method (orange region). Those ones exclude the cluster A2204, while the solid and dashed grey lines in addition to the grey region represent the same thing but including A2204. The best-fit lines are done using every mini-halo with a value, including the candidate and uncertain mini-halos. The magenta star represents PKS~0745$-$191, while the magenta diamond represents MACS~J1447.4+0827. \textbf{Right:} Same but in the plane of the BCG steep radio power at 1~GHz (from \citealt{Hogan2015} and this work) and the X-ray cavity power. The light grey data points are from fig.~13 in \citet{Hogan2015} to compare clusters hosting mini-halos to the general population of clusters.}
\label{fig:Cavity}
\end{figure*}

Finally, in regard to the newly studied relation of $P_{1.4\textrm{~GHz}}$ with the power of the cavities (Fig.~\ref{fig:Cavity}; left panel), a strong correlation was also found ($r_p=0.83$, probability of no correlation of 0.022\%). The fit is done using Equation~\eqref{Eq:Log} with the power of the cavities instead of $X_{P}$ and 14 mini-halos (see Table~\ref{tab:Correlations} for the slopes). 

It is important to note that a small number of the clusters in this paper have multiple system of cavities. For example, the Perseus cluster \citep{Fabian2006} has ghost cavities, and MACS~J1447.4+0827 \citep{Myriam2020}, Phoenix \citep{McDonald2015} and RX~J1532.9+3021 \citep{Julie2013} have potential ghost cavities. However, these are farther away from the cluster center, and therefore the cavity powers listed in Table~\ref{tab:ClusterProperties} are measured only based on the central X-ray cavity pair. 2A~0335+096 contains three smaller X-ray cavities at its centre \citep{Sanders2009b}, however they are distinct from the main cavity pair and therefore only this one is used for the cavity power for consistency in the cavity power measurements. On the other hand, A2204 has anomalously powerful cavities, having the largest bubble heating power known, and the largest by far in Table~\ref{tab:ClusterProperties}. It might be due to the accumulation of repeated outbursts along a single axis \citep{Sanders2009a}, as was seen before in Hydra A \citep{Wise2007}. The relation is therefore studied after excluding the cluster A2204 as those cavities are not well enough understood to be included in the analysis. As mentioned in Section~\ref{sec:BCGproperties}, only the main central cavities were used to determine the cavity power in every cluster to have a uniform approach in this analysis. Therefore, A2204 is not included in the sample, as the most recent outburst cannot be isolated. If A2204 is included, the correlation becomes only moderate, with a $r_p$=0.59 and a probability of no correlation of 1.9\%, and the use of 15 mini-halos.

In Fig.~\ref{fig:Cavity} (right panel), we also find a moderately-strong correlation between the BCG steep radio power at 1~GHz and the power of the cavities, with $r_p=0.59$ and a probability of no correlation of 2.8\% ($r_s=0.45$ and probability of no correlation of 10\% using the Spearman coefficients). We looked at this relation as $\textrm{BCG}_{\textrm{steep}}$ is related to the past activity of the AGN, the lobe emission, which should be linked to the cavities. Here too, the moderately-strong correlation is further supported by the fact that the two fitting methods, BCES-orthogonal and BCES-bisector, give almost exactly the same fit, with slopes of $0.87 \pm 0.22$ and $0.93 \pm 0.11$ (BCES-orthogonal and BCES-bisector respectively). The slopes using bootstrapping are $0.83 \pm 0.97$ and $0.90 \pm 0.30$ (BCES-orthogonal and BCES-bisector respectively). Those values are not shown in Table~\ref{tab:Correlations} as none of the two parameters are associated with mini-halos. Here too, A2204 is excluded; if we include it, $r_p$=0.47 and we have a probability of no correlation of 7.6\%. 

For both relations in Fig.~\ref{fig:Cavity}, as usual, the best-fit using the BCES-orthogonal method and its 95\% confidence regions are shown, as well as the linear fit using the BCES-bisector. For the fit excluding A2204, those are shown with black lines and an orange region, while for the fit including A2204, those are shown with grey lines and region. This clearly state the big deviation if this cluster is included. Furthermore, for the plot of the $\textrm{BCG}_{\textrm{steep}}$ as a function of the power of the cavities, we added light grey points, which are from the fig.~13 of \citet{Hogan2015}, to compare clusters hosting mini-halos to the general population of clusters.

\begin{figure*}
\centering
\begin{minipage}[c]{1.0\linewidth}
\centering 
\subfloat{\includegraphics[width=0.48\textwidth]{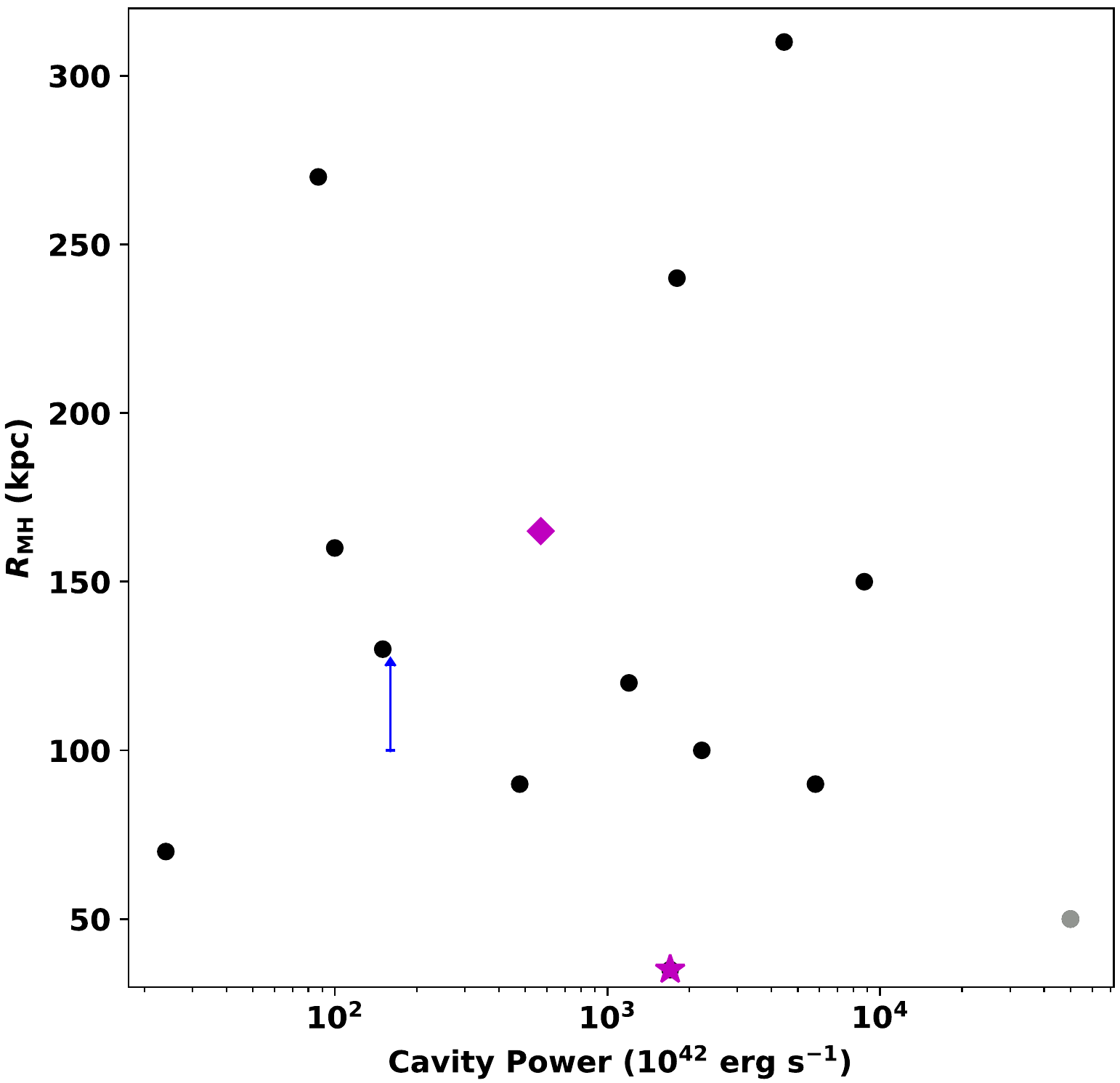}}
\hspace{4mm}
\subfloat{\includegraphics[width=0.48\textwidth]{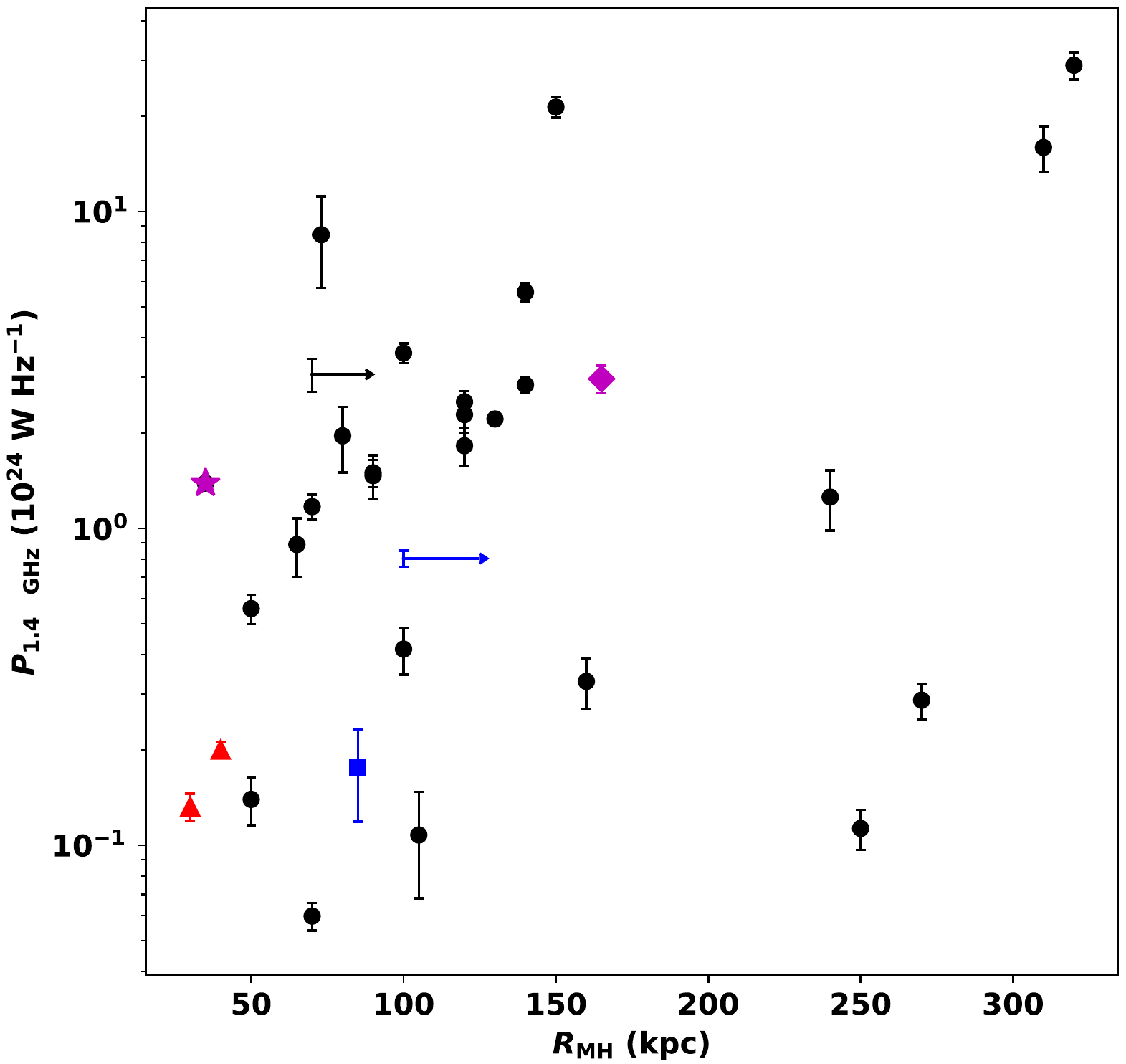}}
\end{minipage}
\caption[]{\textbf{Left:} Average radius of the mini-halo ($R_{\textrm{MH}}$) as a function of the X-ray cavity power, for 14 mini-halos and one lower limit (arrow). The candidate (blue square) mini-halo is shown. The magenta star represents PKS~0745$-$191, while the magenta diamond represents MACS~J1447.4+0827. The outlier, the anomalously powerful cavities for the mini-halo power, is in the cluster A2204 and is shown in grey. Fits are not shown here as there is no clear general trend.  \textbf{Right:} Same but in the plane of the mini-halo radio power at 1.4~GHz ($P_{1.4\textrm{~GHz}}$) and the average radius of the mini-halo ($R_{\textrm{MH}}$), for 30 mini-halos and two lower limits (arrows).}
\label{fig:PvsR}
\end{figure*}

On the other hand, the newly studied relation between the average radius of the mini-halos and the power of the cavities (Fig.~\ref{fig:PvsR}; left panel) does not reveal any general trend as $\rho_p \sim r_p$ and the probability of no correlation is 82\% in the linear-log plane if A2204 is excluded. If A2204 is included, the probability of no correlation becomes 68\%.

\subsection{Mini-halo radio power at 1.4~GHz and average radius}

There is a known trend between the radio power of jets and a proxy of the size of the jets, namely, the mechanical energy being injected into the ICM by the central AGN \citep[e.g.][]{Birzan2004, Birzan2008, Cavagnolo2010, OSullivan2011, Julie2012}. As was explained earlier, the relativistic jets from the central AGN displace the ICM and creates X-ray cavities. The mechanical energy, known as the jet power, is estimated from the energy of the cavities, a value depending on the size of the cavities, and therefore the size of the jets. Consequently, there is a link between the radio power and size of jets and we wanted to explore if it was also the case for mini-halos (Fig.~\ref{fig:PvsR}; right panel).

Similar to the average mini-halo radius and $\textrm{BCG}_{\textrm{steep}}$ plane, there is no clear correlation, only a general trend for the linear regression in the log-linear plane (with a probability of no correlation of 3.2\%), as depending on the method used, the fits are not consistent due to the huge scatter of the data points. This is in accordance to the non-correlation found in \citet{Gitti2018} ($r_s \sim 0.2$ with a probability of no correlation of $\rho_s \sim 40\%$). This relation was also studied in \citet{Cassano2008}, but only six mini-halos were used at the time, therefore a correlation was found. 


\subsection{Potential selectional biases}\label{Warning}

As most of our parameters ($P_{1.4\textrm{~GHz}}$, $L_{\textrm{X}}$, $\textrm{BCG}_{\textrm{steep}}$ and $\textrm{BCG}_{\textrm{core}}$) are distance-dependent quantities, we need to consider potential biases such as the Malmquist bias \citep{Malmquist1922}, which will create a bias towards brighter objects at higher redshifts. It means that for example in Fig.~\ref{fig:PowerM500} (right panel), Fig.~\ref{fig:Steep} (left panel) and Fig.~\ref{fig:Core} (left panel), the bottom right and top left corners will be less occupied. This can generate false correlations due to selection effects. However, as we did not choose our mini-halos based on a minimal brightness value for one of these parameters, but we just use every known mini-halo, the Malmquist bias should not influence strongly our study, we will simply have the brightest mini-halos as they are the easiest to discover.

Another possible bias arises when distances are implicitly included in both axes which can lead to a false correlation. Mini-halo radio power ($P_{1.4\textrm{~GHz}}$), cluster X-ray luminosity ($L_{\textrm{X}}$) and BCG radio luminosity at 1~GHz ($\textrm{BCG}_{\textrm{steep}}$) and at 10~GHz ($\textrm{BCG}_{\textrm{core}}$) are calculated using the distance of the object. Therefore, to verify if the correlations found are real, we did similar figures to Fig.~\ref{fig:PowerM500} (right panel), Fig.~\ref{fig:Steep} (left panel) and Fig.~\ref{fig:Core} (left panel) but using flux values instead of luminosities or power values. We still find that they are strong correlations and Pearson coefficients were either better or similar. For the relation between $P_{1.4\textrm{~GHz}}$ and $L_{\textrm{X}}$ in the flux-flux plane, we have a $r_p$ = 0.70 ($r_p$ = 0.73 in the luminosity-luminosity plane) and a probability of no correlation of 0.00066\% (0.00012\% in the luminosity-luminosity plane). For the relation between $P_{1.4\textrm{~GHz}}$ and $\textrm{BCG}_{\textrm{steep}}$ in the flux-flux plane, we have a $r_p$ = 0.76 ($r_p$ = 0.68 in the luminosity-luminosity plane) and a probability of no correlation of 0.00024\% (0.0070\% in the luminosity-luminosity plane), meaning that the correlation is even stronger in the flux-flux plane. Finally, we have a $r_p$ = 0.68 ($r_p$ = 0.63 in the luminosity-luminosity plane) and a probability of no correlation of 0.069\% (0.24\% in the luminosity-luminosity plane) for the relation between $P_{1.4\textrm{~GHz}}$ and $\textrm{BCG}_{\textrm{core}}$ in the flux-flux plane. Those values are similar and thus it confirms the correlations.

Finally, it is important to mention that the data used in this study is from various measuring methods and telescopes, and therefore this lack of homogeneity might affect the calculations of the correlations.

\section{Discussion}\label{Discussion}

\subsection{Detection of confirmed and new mini-halos}

\subsubsection{PKS~0745$-$191}

As introduced in Section~\ref{PKS}, PKS~0745-191 is an extremely strong radio source. In the right panel of Fig.~\ref{fig1}, we show our new VLA observations of the cluster at $1-2$~GHz. In addition to clearly detecting the central AGN, our observations also revealed the presence of an extended, very diffuse and faint component ($\mu$Jy level). As argued in Section~\ref{PKS}, we interpret this component to be a mini-halo, confirming its classification. It is highly elongated in the eastern to western direction (similar to the X-ray morphology of the cluster), but is also highly asymmetric and extends to roughly $\sim80~$kpc in eastern direction, while only $\sim45~$kpc in the western direction. Interestingly, \citet{Baum1991}, using their VLA observations of PKS~0745$-$191 in the L-band ($20~$cm $\sim1.5~$GHz) in B-configuration, found diffuse emission extending by $\sim40\arcsec \sim 76~$kpc to the north and the south of the nucleus at an angle of 15$\degr$ from a vertical line. We should be able to see this structure in the right panel of Fig.~\ref{fig1}, but there is an elongated artefact on each side of the nucleus exactly at this angle. This explains the lack of signal and the pinching of the contours at this position. The fact that there is an elongated artefact in our image at this position means that we are missing the potential diffuse emission that \citet{Baum1991} were seeing. More investigation with deeper radio observations are needed to confirm the nature of this emission. However, we are still able to see the two branches of emission at the south of the radio emission on either side of the elongated artefact. The increased sensitivity of our observations extends those observations of $\sim10\arcsec \sim19~$kpc. 

It is important to note that the images in \citet{Baum1991} were not able to show the mini-halo without a doubt as their observations were not deep enough. In the eastern direction, the emission now extend of $\sim 22\arcsec \sim 42~$kpc more than the emission in \citet{Baum1991} observations. However, it was included in the first samples of mini-halos assembled \citep{Gitti2004,Gitti2012,Doria2012} due to the presence of both a cool core cluster and a diffuse, amorphous radio emission with no direct association with the central radio source, even if its identification was uncertain. In those papers, it was already used to derive correlations between the radio and X-ray properties of clusters hosting a mini-halo. However, it was not included in the subsequent mini-halo samples \citep[e.g.][]{Cassano2008,Yuan2015,Bravi2016,Giacintucci2014,Giacintucci2017,Giacintucci2019}.

The mini-halo in PKS~0745$-$191 is the smallest mini-halo detected thus far (average radius of $\sim35~$kpc). Indeed, according to Table~\ref{tab:ClusterProperties}, the average radii of mini-halos usually range between 50 and 300~kpc. The mini-halo is contained in the multiple cold fronts (highlighted in Fig.~\ref{fig1}) found by \citet{Sanders2014}. Removing the contribution of the central AGN and jets, the diffuse component has a 1.4~GHz radio power of $P_{1.4\textrm{~GHz}}=(1.39\pm0.03)\times10^{24}~$W~Hz$^{-1}$, which is only $\sim$1\% of the total radio power of the integrated radio emission from this BCG.

Recently, \citet{Giacintucci2017} conducted a study based on a complete, mass-limited sample of 75 clusters from the Planck Sunyaev-Zeldovich (SZ) cluster catalog. These authors essentially found that the majority (12 out of 15, $80\%$) of massive ($M_{500}>6\times10^{14}M_{\rm \odot}$), strong cool core clusters ($K_0<20$ keV cm$^2$, where $K_0$ is the core entropy) host mini-halos. Given the properties of PKS~0745$-$191, being one of the most massive clusters in our sample and harboring a massive cool core, it is therefore not surprising to find a mini-halo in this cluster. As shown in the right and left panels of Fig.~\ref{fig:PowerM500}, the mini-halo in PKS~0745$-$191 falls at an average position in the plane of the mini-halo radio power ($P_{1.4\textrm{~GHz}}$) and cluster mass ($M_{500}$), as well as the well-known correlation between $P_{1.4\textrm{~GHz}}$ and cluster X-ray luminosity ($L_{\textrm {X}}$). It also follows the strong correlation newly found between $P_{1.4\textrm{~GHz}}$ and X-ray cavity power shown in the left panel of Fig.~\ref{fig:Cavity}. However, it does fall below the new correlation found in this manuscript between $P_{1.4\textrm{~GHz}}$ and the BCG component related to past AGN outbursts ($\textrm{BCG}_{\textrm{steep}}$). To further understand the properties of this intriguing mini-halo, additional deeper radio observations would be required to determine if the mini-halo extends beyond the inner cold front. Additional observations at other radio frequencies would also be needed to study the spectral index properties.

\subsubsection{MACS~J1447.4+0827}

MACS~J1447.4+0827, introduced in Section~\ref{MACS}, is an extremely massive cluster of galaxies ($M_{500}=7.46^{+0.80}_{-0.86}\times10^{14}M_{\rm \odot}$; \citealt{Planck2015}), that was identified as one of the strongest cool core clusters in the Massive Cluster Survey (MACS; \citealt{Julie2012}). A detailed analysis of the cluster, its X-ray and feedback properties are presented in \citet{Myriam2020}. Here, we focus on its radio properties and in particular, the new mini-halo discovered in Section \ref{MACS}.

In Section \ref{MACS}, we presented new $1-2$~GHz VLA observations that were obtained in 2016 for the cluster (see Fig. \ref{fig2}). While the A-configuration observations clearly reveal the presence of a central AGN, coincident with the BCG, as well as two collimated jets extending out to 20~kpc in radius, the B-configuration and C-configuration observations reveal the presence of an additional, faint and diffuse component extending out to $\sim 36.2\arcsec \sim160~$kpc in radius, based on the 3$\sigma_{\rm rms}$ radio contours. Removing the contribution of the central AGN and jets, the diffuse component has a $1.4~$GHz radio power of $P_{1.4\textrm{~GHz}}=(3.0\pm0.3)\times10^{24}$~W~Hz$^{-1}$. Using a temperature profile, \citet{Myriam2020} found evidence of two cold fronts located at radii $\sim 50~$kpc and $\sim 300~$kpc. The mini-halo with $R_{\rm MH} \sim 190~$kpc is therefore contained inside the outer cold front.

The structure appears to be elongated in the east-west direction, similar to the X-ray morphology of the cluster. No other deep radio observations are available on the source. Therefore, we can only provide a rough estimate of the spectral index of the mini-halo based on the $1-2~$GHz VLA observations. Using the Taylor coefficients images produced by the \textsc{clean} task in CASA, which describe the frequency structure of the source, we find that the diffuse component has a spectral index of $\alpha=-1.2\pm1.0$. This is typical of spectral indexes for mini-halos, and is usually interpreted as aging of the relativistic particles (e.g. \citealt{Ferrari2008}). 


Based on the study by \citet{Giacintucci2017}, given that MACS~J1447.4+0827 is a massive and strong cool core cluster, it is also not surprising to find a mini-halo in this cluster. In addition, the mini-halo in MACS~J1447.4+0827 falls directly onto the general trend found between the mini-halo radio power ($P_{1.4\textrm{~GHz}}$) and cluster mass ($M_{500}$), as well as the well-known correlation between $P_{1.4\textrm{~GHz}}$ and $L_{\textrm{X}}$. It also follows the strong correlations found between $P_{1.4\textrm{~GHz}}$ and X-ray cavity power shown in the left panel of Fig.~\ref{fig:Cavity}, and between $P_{1.4\textrm{~GHz}}$ and $\textrm{BCG}_{\textrm{steep}}$ shown in the left panel of Fig.~\ref{fig:Steep}. The average radius of the mini-halo is also similar to that seen in other clusters with mini-halos. Overall, MACS~J1447.4+0827 therefore appears to host a classical mini-halo given the properties of the cluster.

\subsection{Mini-halos and cluster scale properties}\label{sec:mHClusterProperties}

Even if the number of known mini-halos has more than doubled in the last decade, there are still only 28 clusters with confirmed detections so far, with five more candidate or uncertain detections. The understanding of mini-halos has therefore been mainly limited by their small numbers, and their origin remains controversial. 

Several authors have reported an interesting correspondence between cold fronts, caused by sloshing of the cool core in the event of a minor merger, and the boundaries of mini-halos (e.g. \citealt{Mazzotta2008}). This suggests that the turbulence being generated by the sloshing cores may be contributing to the reacceleration of the particles, ultimately producing the radio emission of mini-halos, with or without it being the dominant physical phenomenon. Other authors have also reported that mini-halo radio power scales with cooling flow power ($P_{\rm CF}={\dot M}kT/\mu m_{\rm p}$; \citealt{Gitti2004,Gitti2007,Gitti2012,Doria2012,Bravi2016}). This suggests that the thermal energy in cool cores must be connected at an even more fundamental level to the non-thermal energy of mini-halos. The cooling flow power is very similar to the X-ray luminosity of the cooling region of the cluster from this derivation \citep{Fabian1994,Gitti2004}. \citet{Ignesti2020} also found a super-linear scaling between the radio emission and the X-ray brightness in clusters when doing a point-to-point correlation, which also point towards a connection between the thermal and non-thermal particles.

Recently, \citet{Giacintucci2017} performed the first study based on a complete mass-limited sample of clusters. They found that mini-halos are exclusively found in cool cores (i.e. a cool core is required for the formation of mini-halos; with the exception of A1413) and that they may be rarer in lower-mass cool core clusters (i.e. a massive cluster provides a better environment to form a mini-halo). However, the latter could be caused by an observational bias, since if mini-halo radio power scales with cluster mass, than it would be more difficult to detect mini-halos in lower-mass clusters. 

Here, using the most up-to-date database of mini-halos (33 in total), including the confirmed and the new mini-halos reported in Section~\ref{OBS}, we confirm that there is a strong, statistically significant correlation between mini-halo radio power and cluster X-ray luminosity using a uniform way of calculating the X-ray luminosity, namely inside a radius of 600~kpc (Fig.~\ref{fig:PowerM500}, right panel; $r_p \approx 0.73$ and probability of no correlation of 0.00012$\%$ using the Pearson coefficients). We find a best-fitting relation, in agreement with previous relations (e.g. \citealt{Kale2013,Kale2015,Gitti2015b,Yuan2015}), such that: 

\begin{equation}\label{Eq:PvsLx}
\log(P_{1.4\textrm{~GHz}}) = (1.97 \pm 0.31) \times \log(L_{\rm X}) - (1.72 \pm 0.31).
\end{equation}

We therefore confirm that there appears to be a strong, intrinsic relation between the thermal and non-thermal properties of clusters, from the connection between the energy reservoir in cool core clusters with the non-thermal particles forming the mini-halos. 

In Table~\ref{tab:Lx}, we compare the correlations obtained depending on which radius was used to extract the X-ray luminosities. As expected, we found that using uniform radii gives stronger correlations, even if the probability of no correlation are on the same order of magnitude for all radii. The slope of all three methods intersect, but interestingly the one using $L_{\rm X}$ values from the literature is less similar to the uniform methods. Of the two methods using uniform radii, as predicted, the $L_{\rm X}$ found using a small radius is better correlated with the mini-halo radio power as cool core are centrally strongly peaked clusters. \citet{Giacintucci2019} also compared the mini-halo radio power to the bolometric $L_{\rm X}$ of the host clusters using three different radii representing: the central coolest core region ($r=70~$kpc), the core region ($r=0.15R_{500}$) and the whole cluster ($r=R_{500}$). All those relations show a strong correlation, the correlation becoming tighter the smaller the radius is. Interestingly, their correlation using $L_{\rm X, 70~kpc}$ is weaker ($r_s=0.67$ and a probability of no-correlation of $0.7\%$) than the one found in this work using $L_{\rm X, 600~kpc}$ ($r_s=0.71$ and $0.00031\%$). This is most probably due to the fact that \citet{Giacintucci2019} used the bolometric X-ray luminosity and not the X-ray luminosity in the $0.1-2.4~$keV band, suggesting that non-thermal particles have a stronger relation with lower energy thermal particles.

 \begin{table*}
  \caption{Best-fit correlation parameters using the BCES-orthogonal method without bootstrapping for the relations between mini-halo radio power ($P_{1.4\textrm{ {GHz}}}$) and cluster X-ray luminosity ($L_{\rm \textbf{X}}$) for three different methods to extract $L_{\rm \textbf{X}}$ of the clusters: the values found in the literature and the values measured inside a radius of 600~kpc and inside the radius $R_{\rm 500}$. The columns are: 1. Different methods to find the X-ray luminosity of the clusters; 2. \& 3. Slope and intercept of the power-law relation, following Equation~\eqref{Eq:Log}; 4. The Pearson rank correlation coefficient $r_p$; 5. The related probability of no correlation $\rho_p$; 6. The Spearman rank correlation coefficient $r_s$; 7. The related probability of no correlation $\rho_s$.}
  \label{tab:Lx}
  \begin{tabular}{lllllll}
    \hline
    \thead{Relation} & \thead{Slope (A$_{\rm Eq.~\ref{Eq:Log}}$)} & \thead{Intercept (B$_{\rm Eq.~\ref{Eq:Log}}$)} & \thead{$r_p$} & \thead{$\rho_p$} & \thead{$r_s$} & \thead{$\rho_s$}  \\
    \hline
    $\log(P_{1.4\textrm{ {GHz}}}) - \log(L_{\rm X, lit.})$ & 2.27 $\pm$ 0.47 & -2.26 $\pm$ 0.51 & 0.70 & 0.00079\% & 0.64 & 0.0080\% \\
    
    $\log(P_{1.4\textrm{ {GHz}}}) - \log(L_{\rm{X, 600~kpc}})$ & 1.97 $\pm$ 0.31 & -1.72 $\pm$ 0.31 & 0.73 & 0.00012\% & 0.71 & 0.00031\% \\
    
    $\log(P_{1.4\textrm{ {GHz}}}) - \log(L_{\rm{X, R_{500}}})$ & 1.91 $\pm$ 0.32 & -1.77 $\pm$ 0.35 & 0.72 & 0.00021\% & 0.71 & 0.00039\% \\
    
        \hline
  \end{tabular}
 \end{table*}
 

We also explored the relation between $P_{1.4\textrm{~GHz}}$ and $M_{500}$ using our sample. \citet{Giacintucci2014} initially explored this correlation based on the sample of 14 mini-halos known at the time. The authors found no evidence of a statistically significant correlation between these parameters ($r_s \approx 0.3$ and probability of no correlation of 10$\%$ using Spearman rank correlation coefficients), in contrast to the relation between radio halo power and cluster mass for giant radio halos \citep{Cassano2013, MartinezAviles2018}. \citet{Giacintucci2019}, using 23 mini-halos, also found a scattered distribution and no obvious correlation ($r_s \approx 0.06$ and probability of no correlation of 79$\%$), while \citet{Yuan2015} found, with 12 mini-halo, a marginal correlation with $r_s=0.59$. Interestingly, using a more up-to-date sample of 33 mini-halos, as well as more systematic estimates of $M_{500}$ from the Planck collaboration \citep{Planck2014}, we find evidence of a general trend, between $P_{1.4\textrm{~GHz}}$ and $M_{500}$ (Fig.~\ref{fig:PowerM500}, left panel; $r_p \approx 0.44$ and probability of no correlation of $0.95\%$ using the Pearson coefficients) such that: 

\begin{equation}\label{Eq:PvsM}
\log(P_{1.4\textrm{~GHz}}) = (8.38 \pm 3.75) \times \log(M_{500}) - (6.41 \pm 3.04).
\end{equation}

This relation was found using the BCES-orthogonal method. Interestingly, the values we found with this method are not coherent with the values found using the BCES-bisector method in this work (see Table~\ref{tab:Correlations}) and in \citet{Paul2019} where they found a slope of $3.26 \pm 0.62$, pointing towards more of a general trend.

\citet{Cassano2013} argued that the $P_{1.4\textrm{~GHz}}-M_{500}$ correlation for giant radio halos could be related to the available pool of energy in mergers, with mergers between the most massive clusters harboring a larger supply of energy and therefore turbulence that could drive more powerful giant radio halos (see also \citealt{Eckert2017}). Since mini-halos are found exclusively in cool core clusters, and therefore clusters that are considered as relaxed clusters, one could argue that such a correlation should not exist for mini-halos. Yet, the potential link between sloshing motions and mini-halos does indicate that mini-halos may be driven in part by mergers (although minor mergers).

Based on current observations and simulations, we believe that, in the cosmic web, clusters are at the intersection of cosmic filaments formed of galaxies. They form through gravitational collapse of primordial high peaks density perturbations and grow by an hierarchical sequence of mergers and accretion of smaller systems driven by gravity \citep[e.g.][]{Peebles1970,Press1974,Rosati2002,Voit2005,Kravtsov2012}. Thus, higher-mass clusters are in larger potential wells, and therefore they might have a higher frequency of (minor) mergers, resulting in more turbulence than in lower-mass clusters. This could explain why we find evidence of a weak trend between $P_{1.4\textrm{~GHz}}$ and $M_{500}$. Another possible explanation could be that more active AGN live in more massive clusters. A more detailed analysis, in particular using simulations, is needed before confirming such a trend.  


\subsection{Mini-halos and BCG properties}

The existence of mini-halos requires both the presence of magnetic fields, as well as a population of ultra-relativistic electrons. It has been proposed by several authors that the AGN in the BCG may provide the population of seed electrons needed for the reacceleration model (e.g. \citealt{Cassano2008}), given that mini-halos are found exclusively in cool core clusters (with the exception of A1413) and that cool core clusters almost always harbor a powerful central radio AGN \citep[e.g.][]{Hogan2015}. In addition, the jets that are being generated by the AGN in the BCGs may be driving turbulence into the ICM sufficient to offset cooling of the ICM \citep{Zhuravleva2014}, as well as provide the energy needed to reaccelerate the non-thermal particles \citep{Cassano2008,Hitomi2016,Gitti2015b,Bravi2016}. 

Interestingly, new high-dynamic range VLA images of the Perseus cluster revealed a previously unknown rich structure to the mini-halo \citep{Marie-Lou2017}. The shape of the mini-halo appears to be strongly influenced by the sloshing motions, with the mini-halo curving counterclockwise in the direction of sloshing. However, the mini-halo also appears to leak out beyond the inner cold fronts and is elongated in the same direction as the jet axis of the AGN in the BCG. Perseus is one of the rare clusters where evidence of multiple outbursts from the central AGN is detected \citep{Fabian2011} due to its closeness and brightness, allowing us to determine with accuracy the jet axis of the black hole over long time scales ($>10^8$ Gyrs). The association between the shape of the mini-halo and the jet axis of the AGN in the BCG therefore suggests that mini-halos may also be in part created by phenomena related to the BCG, in addition to those related to sloshing of the core. 


\subsubsection{New trends and correlations}

\paragraph{Mini-halos and BCG radio power}\label{sec:BCG}

\citet{Giacintucci2014} and \citet{Giacintucci2019} compared the mini-halo radio power to the radio power of the BCG, both at 1.4~GHz. These authors found evidence of a weak trend between these two quantities ($r_s\approx0.5$ and $r_s\approx0.43$ respectively, and a probability of no correlation of a few percent using the Spearman rank coefficients). \citet{Govoni2009} did the same study and also found a qualitative weak trend. \citet{Govoni2009} however argued that such a correlation was not expected, since the AGN in the centres of clusters undergo multiple cycles over the lifetime of a mini-halo (e.g. \citealt{Clarke2009}, \citealt{Randall2011}). Yet, recent studies have clearly shown that on average, feedback from the central AGN can offset cooling in clusters for Giga-year time-scales \citep{Julie2012,Julie2015}. 

An important aspect that was not considered in \citet{Giacintucci2014,Giacintucci2019} and \citet{Govoni2009} is the fact that the radio power of BCGs is complex, and that its radio SED often contains multiple components. In particular, a core component, originating from very near the AGN and reflecting active accretion, as well as a steeper component, originating from older AGN activity. They considered the radio power of the BCG at 1.4~GHz, without considering the fact that in some BCGs, the radio power at this frequency will be dominated by past activity (i.e. the steep component), whereas for other BCGs, it will be dominated by the current accretion (i.e. the core component). In this manuscript, we therefore explored for the first time the relation between mini-halos and BCG while taking into account the complex radio SED of BCGs. 

Figs.~\ref{fig:Steep} and \ref{fig:Core} show the new relations explored in this paper. We find evidence of a strong, statistically significant correlation between the mini-halo radio power ($P_{1.4\textrm{~GHz}}$) and the BCG component related to past AGN outbursts ($\textrm{BCG}_{\textrm{steep}}$), with $r_p \approx 0.68$ and probability of no correlation of $0.0070\%$, the second lowest probability of no correlation of all the relations we explored, such that:

\begin{equation}\label{Eq:steep}
\log(P_{1.4\textrm{~GHz}}) = (0.99 \pm 0.21) \times \log(\textrm{BCG}_{\textrm{steep}}) - (0.36 \pm 0.11).
\end{equation}

We note however that in the cases where this steep component has a particularly steep spectrum with $\alpha<-1.5$, such emission may be contaminated by other acceleration processes towards the centre of the galaxy cluster in which the BCG resides. In other words, in some clusters, part of the mini-halo or even other steep radio emissions may be contaminating the $\textrm{BCG}_{\textrm{steep}}$ component due to the resolution of the observations. Indeed, in \citet{Hogan2015}, the BCG radio powers at 1~GHz were found using L-band (1.4~GHz) observations that were extrapolated to 1~GHz using a spectral index of $\alpha \sim$~1. For all clusters, \citet{Hogan2015} used the NVSS (NRAO VLA Sky Survey) and SUMSS (Sydney University Molonglo Sky Survey) radio catalogues which have a poor spatial resolution, with a beam size of $\sim45\arcsec$, equivalent to $\sim 165$~kpc at $z=0.227$, the mean redshift of our clusters. As mini-halos have sizes between $\sim30$ and $\sim300$~kpc, this beam may be too large to separate completely the mini-halo emission from the BCG emission in certain clusters. However, for $\sim60$\% of the \citet{Hogan2015} clusters, the authors also used data from the FIRST (Radio Images of the Sky at Twenty-Centimeters) survey, which has a resolution of $\sim5\arcsec$, equivalent to $\sim 18$~kpc at $z=0.227$. Therefore, for the majority of the clusters, the radio SEDs of the BCGs should have sufficient resolution to isolate the emission of the BCG from the extended mini-halo emission, meaning that the beam is approximately the same size or a bit bigger than the emission from the central AGN. Therefore, for those clusters, only a small fraction of the mini-halo flux would be included in the flux from the BCG. This fraction should not be significant, as the radio emission from mini-halos is extremely faint compared to the radio emission from the central AGN. 

As a non-negligible part of the sample has only a resolution of $\sim45\arcsec$, we further explore this potential bias in Fig.~\ref{fig:Steep}. The dotted lines represent the location where the value of the mini-halo radio power ($P_{1.4\textrm{~GHz}}$) is equivalent to 1\% (purple), 10\% (blue), 20\% (cyan), 30\% (green), 40\% (light green), 50\% (orange) and 100\% (red) of the BCG steep radio power at 1.4~GHz. If the percentage is low for a data point, we can safely say that $\textrm{BCG}_{\textrm{steep}}$ is dominated by the power of the BCG and not by the power of the mini-halo. An extreme example is PKS~0745$-$191 where the emission from the mini-halo is only $\sim1\%$ of the emission from the $\textrm{BCG}_{\textrm{steep}}$. However, a number of clusters have values over the 30\% line. For those clusters, we still cannot interpret this by saying that the $\textrm{BCG}_{\textrm{steep}}$ comes almost entirely from the mini-halo, as it depends not only on the resolution of the data, but also on the extend of the mini-halo. It could be possible that a large fraction of the mini-halo emission comes from outside the $\textrm{BCG}_{\textrm{steep}}$ region. For example, the anomalously powerful mini-halo compare to its $\textrm{BCG}_{\textrm{steep}}$ in Fig.~\ref{fig:Steep} is the cluster RX~J1347.5$-$1145. The mini-halo in this cluster has a radius of 320~kpc. It is therefore understandable that the power of the mini-halo is higher that the power of the $\textrm{BCG}_{\textrm{steep}}$. 

Overall, this implies that although mini-halo and other radio emissions may be contaminating part of the $\textrm{BCG}_{\textrm{steep}}$ emission in some of our clusters, we estimate that it should not be a dominant factor in explaining the observed correlation. We note that this relation ($P_{1.4\textrm{~GHz}}$ versus $\textrm{BCG}_{\textrm{steep}}$) is one of the strongest thus known for mini-halos. However, we need to be cautious before stating that it indicates a connection between mini-halos and feedback processes in the BCG. Therefore, we considered another parameter in Section~\ref{sec:cavities} which supports, independently, the relation between mini-halos and AGN feedback.

Another option to consider is the possibility of a multivariable regression. In the left panel of Fig.~\ref{fig:Steep}, when $\textrm{BCG}_{\textrm{steep}} \sim 10^{25}~$W~Hz$^{-1}$, there is a range of two orders of magnitude in the mini-halo radio power with very small uncertainty on those values. Therefore, there could be a third parameter included in this relation, which would explain this discrepancy. The third parameter could be related to the age of the particles or to why, how or when the reacceleration of the particles is done. From the relations in this work however, this third parameter would not be related to the radius of mini-halos. This would be similar to the fundamental plane of black hole activity \citep[e.g.][]{Merloni2003} relating the radio and X-ray luminosity and the mass of black holes, where the correlations between each pair of parameters are highly significant but with a lot of scatter. A multivariable study has already been tried for mini-halos in \citet{Yuan2015}, where they looked at the radio power of mini-halos and clusters X-ray luminosity plane, and the dynamical parameters of the cluster were used as a third parameter. It did reduce significantly the data scatter, however a convincing fundamental plane for mini-halos was not found. With discoveries of more mini-halos, a multivariable study should be carried for $P_{1.4~{\rm~GHz}}$ in function of $\textrm{BCG}_{\textrm{steep}}$ and of $L_{\rm X}$, however this is beyond the scope of this paper.

Finally, we find evidence of a new correlation between $P_{1.4\textrm{~GHz}}$ and the radio component of the BCG that traces ongoing accretion ($\textrm{BCG}_{\textrm{core}}$) (see Fig.~\ref{fig:Core}; left panel) even if not comparable to the correlation between $P_{1.4\textrm{~GHz}}$ and the X-ray luminosity of the cluster, with a probability of 0.24\%. This is puzzling as we would not expect a clear correlation since AGNs vary in time. Biases should not be the cause of the correlation here as the mini-halo power should not contribute to the BCG radio power at 10~GHz as this is a much flatter emission and the resolution is good enough (typically 4" at C-band, equivalent to $14~$kpc at $z=0.227$) to isolate the emission of the BCG from the extended mini-halo emission. Yet, \citet{Hogan2015} found that the X-ray cavity power correlates with both the steep ($\textrm{BCG}_{\textrm{steep}}$) and core radio emission ($\textrm{BCG}_{\textrm{core}}$), suggestive of steady fueling of the AGN over bubble-rise time scales in clusters with X-ray cavities. Furthermore, if we look at Fig.~\ref{fig:Core}, the scatter is somewhat large, therefore it could be seen more as a general trend than a strong correlation.

\paragraph{Mini-halos and X-ray cavities}\label{sec:cavities}

The left panel of Fig.~\ref{fig:Cavity} shows the newly studied relation between the radio power of mini-halos and the X-ray cavity powers of BCGs. The latter were determined from X-ray observations and are therefore a priori independent from radio measurements. We found a strong correlation, with $r_p \approx 0.83$ and probability of no correlation of 0.022$\%$, such that:

\begin{equation}\label{Eq:cavity}
\log(P_{1.4\textrm{~GHz}}) = (0.82 \pm 0.11) \times \log(\textrm{Cavity Power}) - (2.12 \pm 0.30).
\end{equation}

If mini-halos are linked to AGN feedback processes, then we can expect mini-halos properties to be correlated to X-ray cavity properties. The strong relation seen in the left panel of Fig.~\ref{fig:Cavity} therefore provides an independent way, free of the biases mentioned in Section~\ref{sec:BCG}, to corroborate the relation between mini-halos and AGN feedback, especially considering the fact that X-ray cavity properties are measured from X-ray observations and not radio observations. It is to be noted however, that the strong relation in the left panel of Fig.~\ref{fig:Cavity} only includes 14 mini-halos (excluding A2204), as the other clusters hosting mini-halos do not have deep enough X-ray images to detect their X-ray cavities. Therefore, this relation should be studied further with a systematic study of the remaining systems with no reported X-ray cavities in a follow-up paper. However, since both relations in the left panel of Figs.~\ref{fig:Steep} and \ref{fig:Cavity} point towards the same conclusion with different and independent biases, we should study the implication of this conclusion.

Overall, our study strongly supports the fact that mini-halos must be connected to the feedback properties of BCGs. This link could arise if the particles forming mini-halos come from the central AGN and are reaccelerated, whether from sloshing motion, turbulence from the AGN or both. It could also exist simply if the particles of mini-halos are reaccelerated by turbulence generated by the jetted outflows of the central AGN. We will discuss this in more detail in Section~\ref{Origin}.

This link could exist if the particles forming mini-halos come from the central AGN and are reaccelerated, whether from sloshing motion, turbulence from the AGN or both. It could also exist simply if the particles of mini-halos are reaccelerated by turbulence from the central AGN.

In the right panel of Fig.~\ref{fig:Cavity}, we investigated the known relation between $\textrm{BCG}_{\textrm{steep}}$ and X-ray cavity power to study in more detail the composition of $\textrm{BCG}_{\textrm{steep}}$. Very interestingly, we find that this correlation is significantly weaker compared to the one between $P_{1.4\textrm{~GHz}}$ of the mini-halos and the X-ray cavity powers. Here, $r_p \approx 0.59$ and the probability of no correlation is 2.7$\%$. This points to the fact that, as mentioned earlier, the $\textrm{BCG}_{\textrm{steep}}$ parameter may include many different components, therefore it will not necessarily correlate directly with the jet emission. Note that in the right panel of Fig.~\ref{fig:Cavity}, we also included the 26 data points from fig.~13 in \citet{Hogan2015} in light grey. We clearly see that for clusters hosting a mini-halo, the $\textrm{BCG}_{\textrm{steep}}$ is weaker for a given cavity power. This may be indicating that clusters with mini-halos are older. Therefore, the emission from the steep component of the BCG would be fainter than when the cavities were first created. Mini-halos could be seen as forming only in clusters with older outbursts as they take time to form maybe due to the time necessary for the turbulence to reach the outer regions.

\paragraph{Mini-halos sizes}

\begin{figure}
\centering
\subfloat{\includegraphics[width=0.48\textwidth]{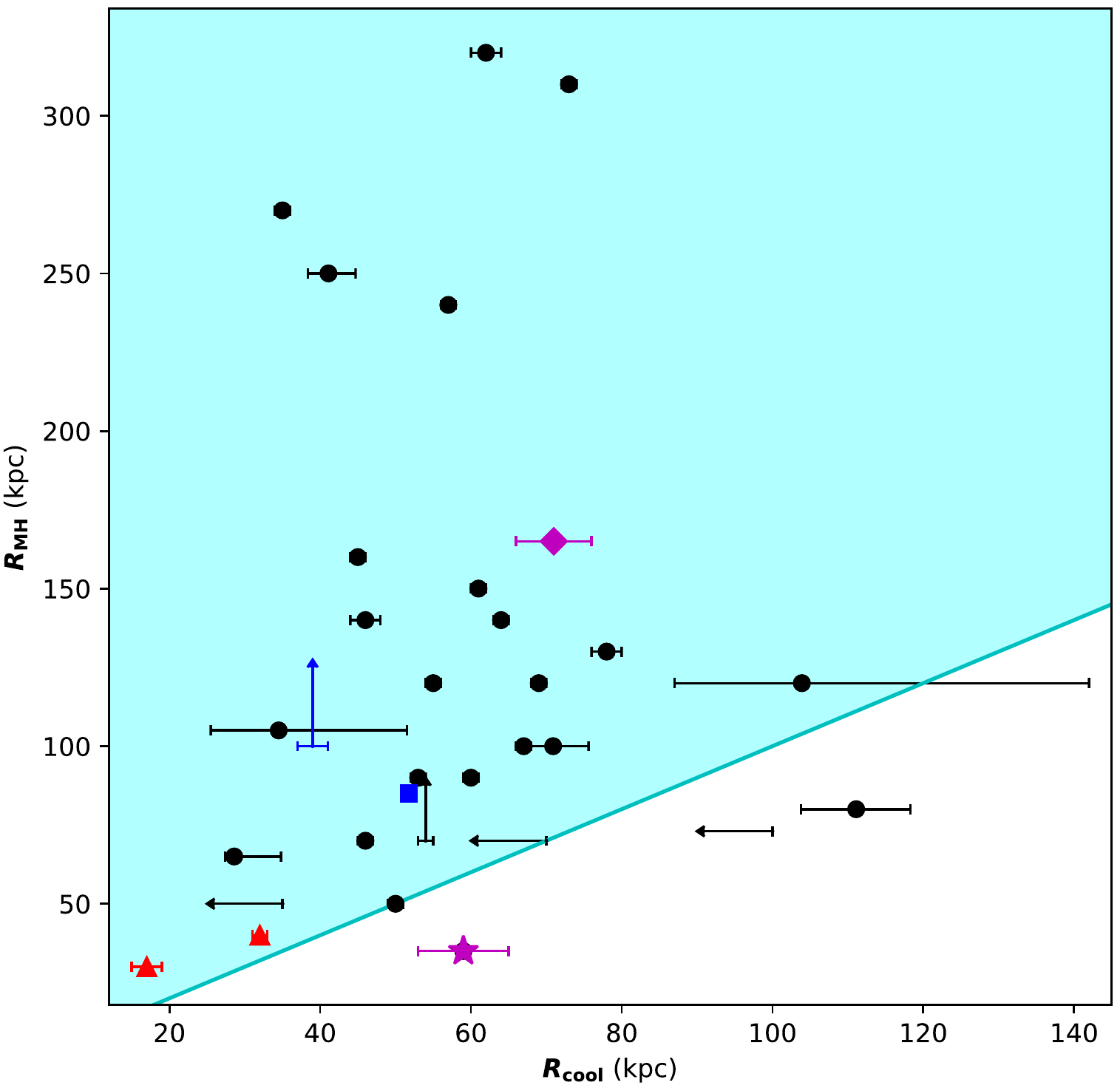}}
\caption[]{Average radius of the mini-halo ($R_{\textrm{MH}}$) as a function of the cooling radius of the cluster ($R_{\textrm{cool}}$), for 27 mini-halos including five upper/lower limits (arrows). The candidate (blue squares) and uncertain (red triangles) mini-halos are shown. The solid cyan line illustrates the one-to-one line, while the cyan region represents where the average radius of the mini-halos is bigger than the cooling radius of the cluster. The magenta star represents PKS~0745$-$191, while the magenta diamond represents MACS~J1447.4+0827.}
\label{fig:Rcool}
\end{figure}

In Fig.~\ref{fig:Rcool} we illustrate the average mini-halo radius as a function of the cooling radius. Interestingly, in general, mini-halos appear to be systematically larger than the cool cores. Here, we define the cooling radius as the radius at which the cooling time is equal to 3 Gyrs \citep{Bravi2016}, such that this time represents the average time since the last major collision of the galaxy cluster. Often, the cooling radius is also defined as the radius at which the cooling time is equal to the $z = 1$ look-back time ($\approx7.7$ Gyrs). This corresponds to the time that a cooling flow should have had the time to establish itself since many clusters at $z = 1$ appear to have similar properties to present-day ones. Given that the cooling time profiles of strong cool cores, such as those that harbor mini-halos, are very similar \citep[e.g.][]{Voigt2004,Julie2012,Panagoulia2014}, such a definition would have resulted in the cooling radii being on average 2 times larger. The size of mini-halos would therefore appears to be the same order of magnitude as the cool cores size, implying yet again a connection between cool cores and mini-halos.

However, we note that there appears to be no strong correlation between the size of the mini-halos and the radio properties of BCGs, their X-ray cavity power or the mini-halo's radio power at 1.4~GHz (see right panels of Figs.~\ref{fig:Steep} and \ref{fig:Core}, and both panels of Fig.~\ref{fig:PvsR}). 

\subsection{Implications for the origin of mini-halos}\label{Origin}


Mini-halos arise from synchrotron emission that is produced by relativistic particles and magnetic fields. Magnetic fields are known to be embedded within the ICM, but the relativistic particles need to either be created or reaccelerated in-situ. The relativistic particles are thought to come, at least in part, from the central AGN of the cluster. If this is the case, reacceleration processes need to occur to explain the extended nature of mini-halos. Among the possibilities regarding the origin of mini-halos discussed in the literature and in Section~\ref{intro}, the preferred one relates them to sloshing motions of cool cores in clusters, since many mini-halos are bounded by cold fronts \citep[e.g.][]{Mazzotta2003,Markevitch2007}. This sloshing motion, arising from minor mergers, is thought to cause turbulence in the ICM which can then reaccelerate the particles of mini-halos \citep[e.g.][]{Mazzotta2008,ZuHone2013,Giacintucci2014}.

However, based on the strong correlations found in this paper linking mini-halo properties to AGN feedback properties (BCG radio emission and X-ray cavity power), the main conclusion of this paper is that mini-halos and AGN feedback processes must be connected at a fundamental level. As proposed earlier in the literature, the mini-halo's relativistic particles may originate from the central AGN and they are afterwards reaccelerated by a source such as sloshing. However, another possibility, suggested by the morphology of the mini-halo in the Perseus cluster \citep{Marie-Lou2017} and in the Phoenix cluster \citep{McDonald2019,Raja2020}, and by earlier studies \cite[e.g.][]{Cassano2008,Gitti2015,Gitti2015b,Bravi2016,Gitti2018}, is that the jets from the central AGN generate enough turbulence to reaccelerate the non-thermal particles of mini-halos. Those particles could come from the central AGN or from other sources, for example from AGNs in other galaxies of the cluster \citep{Marie-Lou2020}. \citet{Bravi2016} found a dependence between the mini-halo radio power and the cooling flow power in clusters, which hints that mini-halos should be powered by the same turbulence than the one offsetting the cooling flow, linking the thermal and the non-thermal particles. \citep[e.g.][]{McNamara2007,Fabian2012,Gitti2012}. In this paper, we found for the first time quantitative relations between mini-halos and AGN feedback properties, supporting the hypothesis of reacceleration of the non-thermal particles of mini-halos by the AGN feedback. 

Therefore, sloshing motions due to minor mergers may not be the dominant driver of the formation of mini-halos. The main driver may instead be AGN feedback creating turbulence in the ICM. This is also supported by the fact that we are starting to realize that with good enough X-ray images, we find cold fronts (the signature of sloshing motions) in almost all cool core clusters, yet only $\sim$30 mini-halos are known so far. Furthermore, we did not find a strong correlation between the power of mini-halos and the $M_{500}$ of the clusters. This correlation in giant radio halo is thought to be related to the available pool of energy in mergers. Thus, the fact that we found only a general trend here might indicate that even if mini-halos may be driven in part by minor mergers, these mergers may not be the dominant factor for the reacceleration of the particles and thus the origin of mini-halos. 

The sloshing motion creating the cold fronts could instead be responsable for driving the overall shape of the mini-halos in the sense that these motions could distribute the relativistic electrons throughout the cluster while creating physical boundaries at the location of the cold fronts \citep[e.g.][]{ZuHone2013}. This could explain why some mini-halos are now found with emission spilling past their cold fronts (e.g. the mini-halo in the Perseus cluster, \citealt{Marie-Lou2017}; in the Phoenix cluster, \citealt{Raja2020}; in RXC~J1504.1$-$0248, \citealt{Giacintucci2011}; in Z3146, \citealt{Giacintucci2014}). This theory could also explain the cold front between the mini-halo and the giant radio halo in PSZ1~G139.61+24.20 \citep{Savini2019}.

Hence, here we argue that sloshing may only be contributing to shaping the overall shape of the mini-halos, but the underlying engine lighting up the mini-halos may be AGN feedback. Such a scenario should be tested with simulations and would imply that AGN feedback not only plays a fundamental role in energizing the thermal gas in clusters of galaxies (preventing massive cooling flows from occurring), but it may also play a crucial role in energizing the non-thermal particles. 

It is to be noted that this conclusion is not in disagreement with the recent result found in \citet{Giacintucci2019} claiming that the origin of mini-halos should be closely related to the properties of the cool core gas, as the hypothesis suggested in our work claims that the AGN feedback could be the driving phenomena to create the turbulence in the cool core gas, which would in turn reaccelerate particles to create mini-halos. Therefore mini-halos would still be related to the properties of the cool core gas. This way, the central AGN has an impact via the AGN feedback not only on the thermal particles like it was believe earlier, but also on the relativistic particles, the non-thermal particles. The impact of the central AGN is global and explains the link between thermal and non-thermal particles.



\section{Summary and future prospects}\label{Summary}

In this study, we have confirmed one mini-halo and identified one previously unknown mini-halo using new, high-dynamic range VLA images. These mini-halos reside in the massive, strong cool core clusters PKS~0745$-$191 and MACS~J1447.4+0827. Combining these new detections to the known mini-halos in the literature, we have explored several new correlations between mini-halo properties and cluster properties. 

In addition to confirming the known correlation between mini-halo radio power ($P_{1.4\textrm{~GHz}}$) and cluster X-ray luminosity using an uniform way of measuring this last parameter, namely inside a radius of 600~kpc, we have also found evidence of a trend between $P_{1.4\textrm{~GHz}}$ and the cluster mass ($M_{500}$) using a homogeneous database for the cluster mass based on the Planck database, such that the most massive clusters of galaxies host the most luminous mini-halos. However, we did not find any relation between the size of mini-halos and the properties of the clusters.

We also explored for the first time in detail the relation between radio mini-halos and AGN feedback processes in clusters of galaxies using new measurements and parameters. By decomposing the radio emission of the BCG into a component associated with on-going accretion (core emission) and another associated with past AGN activity (steep component), we find evidence of a strong correlation between the mini-halo radio power and the BCG steep radio power component, link to the jets of the AGN, as well as between the mini-halo radio power and the X-ray cavity power, created by the jets of the AGN. This strong correlation between the mini-halo radio power and the X-ray cavity power is found only if A2204, the cluster with the largest bubble heating power known, is excluded due to the incapacity to isolate its most recent outburst and therefore its inner cavities, which were used for every other cluster for uniformity.

A possible contamination of the BCG steep radio power relation was studied as well as the small number of data in the mini-halo radio power and the X-ray cavity power plane. However, as those two relations point towards the same conclusion with different biases, from these results, our study suggests that there must be a connection between the feedback processes of the central AGN and the mini-halo. This discovery completely supports the hypothesis proposed by e.g. \citet{Bravi2016} after finding that mini-halos and gas heating in cool core clusters might have a common origin, that particles from mini-halos are reaccelerated by AGN feedback. However, exactly how the connection occurs will require a more in depth study using simulations. We suggest that the main driver for the creation of mini-halos could be AGN feedback creating turbulence in the ICM, while sloshing motions would drive the overall shape of the mini-halos. Similar trends are emerging for other diffuse radio sources in clusters. In particular, \citet{vanWeeren2017} reported the discovery of a direct connection between a radio relic and a radio galaxy in A3411-3412. Overall, they find that radio AGNs play an important role in determining the non-thermal properties of the ICM in clusters. Therefore, it may not be surprising to find a similar trend for mini-halos. 


With the advent of new radio telescopes in the near future, the study of diffuse radio structures in galaxy clusters will improve dramatically. LOFAR will enable us to discover the full extent of mini-halos (e.g. \citealt{Savini2018,Savini2019}) and its deep low-frequency surveys, like the LOFAR Two-Metre Sky Survey (LoTSS), could discover up to $\sim1,400$ new mini-halos \citep{Gitti2018}. On a longer time-scale, the Square Kilometer Array (SKA1 and SKA2) is predicted to detect hundreds to thousands of new mini-halos up to $z=1$, enabling detailed statistical analyses of mini-halos to be performed for the first time \citep{Kale2016b,Iqbal2017,Gitti2018}.

\section*{Acknowledgements}

We would like to thank the editors and the anonymous referee for providing constructive comments and raising interesting points that helped in improving the quality and presentation of the manuscript. ARL was supported for part of this work (during her Master's degree at Universit\'{e} de Montr\'{e}al) by NSERC (Natural Sciences and Engineering Research Council of Canada) through the NSERC Alexander-Graham-Bell Canada Graduate Scholarships-Master's Program (CGS M) and by FRQNT (Fonds de recherche du Qu\'{e}bec - Nature et technologies) through the FRQNT Graduate Studies Research Scholarship - M. Sc. level under grant \#209839. ARL is now supported by the Gates Cambridge Scholarship, by the St John's College Benefactors' Scholarships, by NSERC through the Postgraduate Scholarship-Doctoral Program (PGS D) under grant PGSD3-535124-2019 and by FRQNT through the FRQNT Graduate Studies Research Scholarship - Doctoral level under grant \#274532.  JHL is supported by NSERC through the discovery grant and Canada Research Chair programs, as well as FRQNT. RN acknowledges support by FAPESP (Funda\c{c}\~ao de Amparo \`{a} Pesquisa do Estado de S\~ao Paulo) under grant 2017/01461-2. ACF acknowledges ERC Advanced Grant 340442.

\section*{Data availability}

The data underlying this article are available in the article. The \textit{Chandra} and VLA data used in this article to create Figs.~\ref{fig:Xray}, \ref{fig1} and \ref{fig2} are available through the respective archives. The observation identification numbers for the \textit{Chandra} observations are 2427 \& 12881 for PKS~0745-191 and 10481, 17233 \& 18825 for MACS~J1447.4+0827. The project codes for the VLA observations are SC0344 for PKS~0745-191 and SG0848 for MACS~J1447.4+0827. The reduced images in Figs.~\ref{fig:Xray}, \ref{fig1} and \ref{fig2} generated for this research will be shared on reasonable request to the corresponding author.




\bibliographystyle{mnras}
\bibliography{PHY3030} 




\appendix

\section{Complementary tables}


 \begin{table*}
  \caption{Mini-halos and BCG fluxes, and spectral indexes for the 33 galaxy clusters with mini-halos. The columns are: 1. Cluster Name; 2. 1.4~GHz flux of the mini-halo ($S_{1.4\textrm{~GHz}}$); 3. Reference for $S_{1.4\textrm{~GHz}}$; 4. Steep BCG flux measured at 1~GHz (Flux $\textrm{BCG}_{\textrm{steep}}$), from \citet{Hogan2015} unless specified otherwise; 5. Spectral index at 1~GHz ($\alpha_{\rm steep}$), from \citet{Hogan2015} unless specified otherwise. If no value was available, $\alpha_{\rm steep}= 1.0$; 6. Core BCG flux at 10~GHz (Flux $\textrm{BCG}_{\textrm{core}}$), from \citet{Hogan2015} unless specified otherwise; 7. Spectral index at 10~GHz ($\alpha_{\rm core}$), from \citet{Hogan2015} unless specified otherwise. If no value was available, $\alpha_{\rm core}= 0.2$. \textbf{Reference code:} (1) \citet{Giacintucci2014}, (2) \citet{Knowles2019}, (3) \citet{Giacintucci2019}, (4) \citet{Govoni2009}, (5) \citet{Gitti2013}, (6) \citet{Kale2015}, (7) This work, (8) \citet{Sijbring1993}, (9) \citet{Raja2020}, (10) \citet{Kale2015b}, (11) \citet{Giacintucci2011}, (12) \citet{Giacintucci2014b}, (13) \citet{Yuan2015}.} 
  \label{tab:flux}
  \begin{tabular}{lllllll}
    \hline
   \thead{Name} & \thead{$S_{1.4\textrm{~GHz}}$} & \thead{Ref.} &  \thead{Flux BCG steep} &  \thead{$\alpha_{\rm steep}$} & \thead{Flux BCG core} & \thead{$\alpha_{\rm core}$} \\
      &  \thead{[mJ]} &  & \thead{[mJ]} &  & \thead{[mJ]} &  \\
    \hline
2A~0335+096 &	21 $\pm$	2	&	(1) &	62.5	$\pm$	0.8	&	1.48	$\pm$	0.01	&	$<3.3$			&	0.2			\\
ACT$-$CL~J0022.2$-$0036 &	2.5  $\pm$	0.8	$^a$ &	(2) &	NA	$^c$		&	NA			&	NA			&	NA			\\
A478 &	17 $\pm$	3	&	(1) &	34	$\pm$	4	&	1.0 	$\pm$	0.2	&	4.8	$\pm$	0.3	&	0.59	$\pm$	0.07	\\
A907 &	14  $\pm$	3	&	(3) &	134	$\pm$	8	&	0.88 	$\pm$	0.03	&	$<46$			&	0.2			\\
A1068 &	3.4 $\pm$	1.1	&	(4) &	13	$\pm$	8	&	1.09 	$\pm$	0.13	&	$<1.2$			&	0.2			\\
A1413 &	1.9 $\pm$	0.7	&	(4) &	$<2$			$^d$ &	1			$^d$ &	$<1$			$^d$ &	0.2			$^d$ \\
A1795 &	85 $\pm$	5	&	(1) &	1295	$\pm$	20	&	0.94 	$\pm$	0.03	&	27	$\pm$	9	&	0.51	$\pm$	0.08	\\
A1835 &	6.1 $\pm$	1.3	&	(1) &	48	$\pm$	6	&	0.8 	$\pm$	0.2	&	$<3.6$			&	0.2			\\
A2029 &	20 $\pm$	3	&	(1) &	757	$\pm$	7	&	1.23 	$\pm$	0.04	&	2.70	$\pm$	0.10	&	0.2	$\pm$	0.2	\\
A2204 &	8.6 $\pm$	0.9	&	(1) &	77	$\pm$	24	&	1.16 	$\pm$	0.10	&	10	$\pm$	3	&	0.32	$\pm$	0.10	\\
A2626 &	18.0 $\pm$	1.8	&	(5) &	92	$\pm$	7	&	2.09 	$\pm$	0.06	&	5.4	$\pm$	0.5	&	0.28	$\pm$	0.03	\\
A2667 &	7.1 $\pm$	0.6	&	(3) &	$<28.1$ 			&	1.0 			&	8	$\pm$	2	&	0.20	$\pm$	0.17	\\
A3444 &	12.1 $\pm$	0.9	&	(3) &	13.2			&	1.0	$\pm$	0.2	&	$<0.26$			&	0.2			\\
AS780 &	11 $\pm$	3	&	(6) &	45			$^d$ &	1.1 			$^d$ &	150			$^d$ &	0.2			$^d$ \\
MACS~J0159.8$-$0849 &	2.4 $\pm$	0.2	&	(1) &	5			$^d$ &	1.2 			$^d$ &	70			$^d$ &	0.2			$^d$ \\
MACS~J0329.6$-$0211 &	3.8 $\pm$	0.4	&	(1) &	6			$^d$ &	1.2 			$^d$ & 	1			$^d$ &	0.2			$^d$ \\
MACS~J1447.4+0827 &	5.7 $\pm$	0.5	&	(7) &	8			$^d$ &	1.0 			$^d$ &	10			$^d$ &	0.2			$^d$ \\
MACS~J1931.8$-$2634 &	48 $\pm$	3	&	(1) &	160			$^d$ &	2.0 			$^d$ &	60			$^d$ &	0.2			$^d$ \\
MS~1455.0+2232 &	8.5 $\pm$	1.1	&	(1) &	21	$\pm$	2	&	1.09 	$\pm$	0.10	&	1.5	$\pm$	0.6	&	0.5	$\pm$	0.3	\\
Ophiucus &	62 $\pm$	9	&	(3) &	30			$^d$ &	0.8			$^d$ &	$<5$			$^d$ &	0.2			$^d$ \\
Perseus &	3020 $\pm$	153	&	(8) &	4000	$\pm$	400	&	1.0 			&	11000	 $\substack{+6600  \\ -5600}$ 		&	0.2			\\
Phoenix &	9.9 $\pm$	1.4	$^b$ &	(9) &	50			$^d$ &	1.2 			$^d$ &	2.5			$^d$ &	0.2			$^d$ \\
PKS~0745$-$191&	50.2 $\pm$	0.7	&	(7) &	3738	$\pm$	57	&	1.0 			&	$<75$			&	0.2			\\
PSZ1~G139.61+24.20 &	0.6 $\pm$	0.1	&	(3) &	NA			&	NA			&	NA			&	NA			\\
RBS~797 &	5.2 $\pm$	0.6	&	(1) &	20			$^d$ &	1.2 			$^d$ &	3			$^d$ &	0.2			$^d$ \\
RXC~J1115.8+0129 &	17.90		&	(10) &	NA			&	NA			&	NA			&	NA			\\
RX~J1347.5$-$1145 &	36  $\pm$	3	&	(3) &	20			$^d$ &	1.5 			$^d$ &	20			$^d$ &	0.2			$^d$ \\
RXC~J1504.1$-$0248 &	20 $\pm$	1	&	(11) &	55			$^d$ &	1.0 			$^d$ &	30			$^d$ &	0.2			$^d$ \\
RX~J1532.9+3021 &	7.5 $\pm$	0.4	&	(1) &	27	$\pm$	3	&	0.66 	$\pm$	0.08	&	$<4$			&	0.2			\\
RXJ1720+2638 &	72 $\pm$	4	&	(12) &	101	$\pm$	9	&	1.08 	$\pm$	0.03	&	3.4	$\pm$	1.8	&	0.19	$\pm$	0.12	\\
RX~J2129.6+0005 &	2.4 $\pm$	0.4	&	(6,13) &	26	$\pm$	4	&	1.0 	$\pm$	0.2	&	3.7	$\pm$	1.0	&	0.44	$\pm$	0.11	\\
Z3146 &	5.2 $\pm$	0.8	&	(1,13) &	8.3	$\pm$	1.1	&	1.0 	$\pm$	0.2	&	0.6	$\pm$	0.3	&	0.4	$\pm$	0.3	\\
ZwCl~1742.1+3306 &	13.8 $\pm$	0.8	&	(1) &	46	$\pm$	9	&	1.45 	$\pm$	0.07	&	59	$\pm$	12	&	0.21	$\pm$	0.06	\\

    \hline
\multicolumn{7}{l}{$^a$ Found flux at 1.4~GHz using $\alpha=1.15 \pm 0.15$ from the flux at 610~MHz in \citet{Knowles2019}.}\\
\multicolumn{7}{l}{$^b$ Found flux using \citet{Raja2020} power at 1.4~GHz and their spectral index between 610~MHz and 1.52~GHz}\\
\multicolumn{7}{l}{\quad of $\alpha = -0.98 \pm 0.16$.}\\
\multicolumn{4}{l}{$^c$ NA means that the value is not available.}\\
\multicolumn{7}{l}{$^d$ From this work.}\\
\end{tabular}
 \end{table*}



 \begin{table*}
  \caption{The columns are: 1. Cluster Name; 2. Observation IDs (ObsID(s)) used from the \textit{Chandra} archive; 3. Column density ($n_H$); 4. Cluster X-ray luminosity ($L_{\rm X, lit.}$) from the literature, derived from observations in the $0.1-2.4$ keV band, corrected for our cosmology; 5. Reference for $L_{\rm X, lit}$; 6. Cluster X-ray luminosity ($L_{\rm X, 600~kpc}$) derived from observations in the $0.1-2.4$ keV band inside a circular region of 600~kpc of radius; 7.  Cluster X-ray luminosity ($L_{{\rm X}, R_{500}}$) derived from observations in the $0.1-2.4$ keV band inside a circular region of $R_{500}$ of radius; 8. Radius ($R_{500}$) for which the density is 500 times the critical density of the universe at this redshift, calculated from the $M_{500}$ given in Table~\ref{tab:ClusterProperties}. \textbf{Reference code:} (1) \citet{Yuan2015}, (2) \citet{Menanteau2013}, (3) \citet{Mulroy2019}, (4) \citet{Ebeling1998}, (5) \citet{Walker2014}, (6) \citet{Cassano2008}, (7) \citet{Mantz2016}, (8) \citet{Ebeling2010}, (9) This work, (10) \citet{McDonald2012}, (11) \citet{Birzan2019}, (12) \citet{Ettori2013}.}
  \label{tab:LxMethods}
  \begin{tabular}{llllllll}
    \hline
    \thead{Name} & \thead{ObsID(s)} & \thead{$n_H$} & \thead{$L_{\rm X, lit.}$} & \thead{Ref.} & \thead{$L_{\rm X, 600~kpc}$} & \thead{$L_{{\rm X}, R_{500}}$} & \thead{$R_{500}$}  \\
    
    \thead{} & \thead{} & \thead{[$e^{22}~$cm$^2$]} & \thead{[$10^{44}~$erg~s$^{-1}$]} &  \thead{} & \thead{[$10^{44}~$erg~s$^{-1}$]} & \thead{[$10^{44}~$erg~s$^{-1}$]} & \thead{[Mpc]}  \\
    
    \hline
2A~0335+096 &	919 &	0.176 &	2.27 $\pm$ 0.05 &	(1) &	1.176 $\pm$ 0.011 &	1.111 $\pm$ 0.011 &	0.94 $\substack{+0.03  \\ -0.04}$ 	\\
ACT$-$CL~J0022.2$-$0036 &	16226 &	0.026 &	6 $\pm$ 4 &	(2) &	11.3 $\substack{+2.0  \\ -1.5}$ &	14.0 $\substack{+0.8   \\ -1.3}$ &	0.96 $\substack{+0.12  \\ -0.13}$ 	\\
A478 &	1669 &	0.148 &	7.5 $\pm$ 0.3 &	(1) &	5.89  $\pm$ 0.12 &	5.89 $\pm$ 0.12  &	1.35 $\substack{+0.02  \\ -0.03}$ 	\\
A907 &	3185, 3205 &	0.0569 &	6.0 $\pm$ 0.5 &	(3) &	3.940 $\substack{+0.065  \\ 0.045}$ &	4.805 $\substack{+0.097   \\ -0.058}$ &	1.19 $\substack{+0.03  \\ -0.04}$ 	\\
A1068 &	1652 &	0.137 &	4.76 &	(4) &	4.073 $\substack{+0.056  \\ -0.051}$ &	5.17 $\substack{+0.20   \\ -0.16}$ &	1.09 $\substack{+0.04  \\ -0.05}$ 	\\
A1413 &	537 &	0.143 &	8.62 &	(4) &	5.78 $\substack{+0.22  \\ -0.25}$ &	6.6 $\pm$ 0.2  &	1.25 $\pm$ 0.03 	\\
A1795 &	493 &	0.062 &	4.99 &	(5) &	4.552 $\substack{+0.047  \\ -0.056}$ &	4.337 $\substack{+0.028   \\ -0.038}$ &	1.17 $\substack{+0.01  \\ -0.02}$ 	\\
A1835 &	495 &	0.253 &	25 $\pm$ 3 &	(1) &	20.1 $\substack{+0.6  \\ -0.3}$ &	21.5 $\pm$ 0.4  &	1.35 $\pm$0.03  	\\
A2029 &	4877 &	0.076 &	9 $\pm$ 3 &	(1) &	7.366 $\substack{+0.029  \\ -0.044}$ &	8.18 $\pm$ 0.04  &	1.33 $\substack{+0.02  \\ -0.01}$ 	\\
A2204 &	499 &	0.057 &	14.0 $\pm$ 0.6 &	(1) &	12.21 $\substack{+0.19  \\ -0.25}$ &	14.1 $\substack{+2.5   \\ -0.3}$ &	1.37 $\substack{+0.03  \\ -0.02}$ 	\\
A2626 &	3192 &	0.043 &	0.90 $\pm$ 0.10 &	(6) &	0.825 $\substack{+0.029  \\ -0.037}$ &	0.868 $\substack{+0.037   \\ -0.029}$ &	0.95 $\substack{+0.06  \\ -0.07}$ 	\\
A2667 &	2214 &	0.0163 &	13.4 $\pm$ 0.3 &	(7) &	11.19 $\substack{+0.34  \\ -0.37}$ &	13.9 $\pm$ 0.5  &	1.27 $\substack{+0.03  \\ -0.04}$ 	\\
A3444 &	9400 &	0.054 &	14 $\pm$ 1.3 &	(1) &	11.2 $\substack{+3.0  \\ -1.8}$ &	14.74 $\substack{+0.38   \\ -0.32}$ &	1.30 $\pm$ 0.03 	\\
AS780 &	9428 &	0.078 &	16 $\pm$ 3 &	(1) &	7.36  $\pm$ 0.15  &	10.76 $\substack{+0.27   \\ -0.25}$ &	1.32 $\substack{+0.03  \\ -0.04}$ 	\\
MACS~J0159.8$-$0849 &	3265 &	0.02 &	20.0 $\pm$ 0.7 &	(8) &	15.12 $\substack{+0.65  \\ -0.59}$ &	18.1 $\substack{+1.4   \\ -1.0}$ &	1.19 $\substack{+0.05  \\ -0.06}$ 	\\
MACS~J0329.6$-$0211 &	3257, 3582 &	0.0588 &	NA $^a$ &	 &	11.3  $\pm$ 0.7  &	13.5 $\pm$ 0.7  &	1.040 $\substack{+0.050  \\ -0.049}$ 	\\
MACS~J1447.4+0827 &	17233, 18825 &	0.0209 &	12.40 $\pm$ 0.10 &	(9) &	24.0  $\pm$ 0.4  &	25.6 $\pm$ 0.4  &	1.20 $\substack{+0.04  \\ -0.05}$ 	\\
MACS~J1931.8$-$2634 &	9382 &	0.0893 &	24.1 $\pm$ 1.3 &	(8) &	15.0  $\pm$ 0.2  &	17.51 $\substack{+0.26   \\ -0.21}$ &	1.17 $\pm$0.05  	\\
MS~1455.0+2232 &	4192 &	0.0301 &	9 $\pm$ 2 &	(1) &	9.18 $\substack{+0.15  \\ -0.10}$ &	9.70 $\pm$ 0.17  &	1.000 $\substack{+0.040  \\ -0.036}$ 	\\
Ophiucus &	3200 &	0.138 &	5.30 $\pm$ 0.12 &	(1) &	2.035 $\substack{+0.055  \\ -0.036}$ &	1.872 $\substack{+0.091   \\ -0.084}$ &	1.65 $\substack{+0.06  \\ -0.05}$ 	\\
Perseus &	3209, 4289 &	0.139 &	7.89 $\pm$ 0.18 &	(1) &	1.9599 $\substack{+0.0024  \\ -0.0025}$ &	1.979 $\pm$ 0.003  &	1.31 $\pm$ 0.04  	\\
Phoenix &	16135 &	0.015 &	84  $\substack{+1 \\ -2}$ $^b$ &	(10) &	59.7 $\substack{+1.8  \\ -2.0}$ &	70 $\substack{+8   \\ -12}$ &	1.37 $\substack{+0.07  \\ -0.06}$ 	\\
PKS~0745$-$191 &	6103, 7694 &	0.407 &	12.47  $\substack{+0.03 \\ -0.04}$ &	(9) &	6.27 $\substack{+0.08  \\ -0.11}$ &	7.22 $\substack{+0.14   \\ -0.11}$ &	1.35 $\pm$ 0.05  	\\
PSZ1~G139.61+24.20 &	15139 &	0.267 &	9.2 $\pm$ 0.2 &	(11) &	6.94 $\substack{+0.31  \\ -0.29}$ &	10.1 $\substack{+1.2   \\ -1.3}$ &	1.35 $\pm$ 0.04  	\\
RBS~797 &	2202, 7902 &	0.021 &	20.8 $\pm$ 1.0 &	(1) &	18.3 $\substack{+0.6  \\ -0.5}$ &	19.8 $\substack{+1.2   \\ -1.0}$ &	1.22 $\pm$ 0.05  	\\
RXC~J1115.8+0129 &	3275 &	0.0443 &	17.1 $\pm$ 0.6 &	(8) &	12.1  $\pm$ 0.5  &	14.4 $\substack{+0.9  \\ -0.8}$ &	1.19 $\substack{+0.04  \\ -0.05}$ 	\\
RX~J1347.5$-$1145 &	13516, 13999, 14407 &	0.0488 &	46 $\pm$ 5 &	(1) &	40.1 $\substack{+0.6  \\ -0.5}$ &	35.1 $\substack{+0.5   \\ -0.4}$ &	1.35 $\substack{+0.03  \\ -0.05}$ 	\\
RXC~J1504.1$-$0248 &	5793 &	0.0604 &	28.6 $\pm$ 1.3 &	(1) &	22.5 $\substack{+0.3  \\ -0.1}$ &	24.93 $\substack{+0.41   \\ -0.14}$ &	1.28 $\substack{+0.04  \\ -0.03}$ 	\\
RX~J1532.9+3021 &	14009 &	0.0217 &	16.9 $\pm$ 0.8 &	(1) &	18.71 $\substack{+0.18  \\ -0.23}$ &	20.8 $\pm$ 0.3  &	1.06 $\substack{+0.05  \\ -0.04}$ 	\\
RXJ1720+2638 &	3224, 4361 &	0.0402 &	7.5 $\pm$ 0.5 &	(1) &	7.17 $\substack{+0.12  \\ -0.11}$ &	8.08 $\pm$ 0.15  &	1.26 $\pm$ 0.03 	\\
RX~J2129.6+0005 &	552 &	0.0429 &	12 $\pm$ 4 &	(1) &	7.42 $\substack{+0.40  \\ -0.39}$ &	8.91 $\substack{+0.42   \\ -0.44}$ &	1.08 $\substack{+0.04  \\ -0.05}$ 	\\
Z3146 &	9371 &	0.0294 &	19.8 $\pm$ 1.8 &	(1) &	17.67 $\substack{+0.33  \\ -0.30}$ &	18.9 $\substack{+0.7   \\ -0.6}$ &	1.23 $\pm$ 0.05  	\\
ZwCl~1742.1+3306 &	11708 &	0.0383 &	2.39 &	(12) &	2.2  $\pm$ 0.3 &	2.14 $\pm$ 0.05  &	0.97 $\substack{+0.03  \\ -0.04}$ 	\\
    
\hline
        
\multicolumn{4}{l}{$^a$ NA means that the value is not available.}\\
\multicolumn{4}{l}{$^b$ From observations in the $2.0-10.0$ keV band.}\\

  \end{tabular}
 \end{table*}
 
 

 \begin{table*}
  \caption{Observation IDs and parameters used to find the cooling profile of the 11 galaxy clusters not in \citet{Bravi2016}, \citet{Fabian2007}, \citet{Sanders2014} or \citet{Myriam2020}. The columns are: 1. Cluster Name; 2. Redshift ($z$); 3. Column density ($n_H$); 4. Observation IDs (ObsID(s)) used from the \textit{Chandra} archive.}
  \label{tab:ObsID}
  \begin{tabular}{lllllll}
    \hline
    \thead{Name} & \thead{$z$} & \thead{$n_H$ [$e^{22}~$cm$^2$]} & \thead{ObsID(s)} \\
    \hline
    ACT$-$CL~J0022.2$-$0036 & 0.805 & 0.026 & 16226  \\
    A907 & 0.153 & 0.0569 & 3185, 3205 \\
    A1068 & 0.137  &  0.137 & 1652  \\
    A1413 & 0.143 & 0.022 & 537, 1661, 5002, 5003, 7697, 12194, 12195, 12196, 13128 \\
    A2667 &  0.230 & 0.0163 & 2214  \\
    A3444 & 0.254 & 0.057 & 9400 \\
    AS780 & 0.236 & 0.073 & 9428 \\
    Ophiuchus & 0.028 & 0.2 & 3200, 16142, 16143, 16464, 16626, 16627, 16645 \\
    PSZ1~G139.61+24.20 & 0.267  & 0.267 & 15139  \\
    RXC~J1115.8+0129 & 0.350 & 0.044 & 3275 \\
    RX~J2129.6+0005 & 0.235 & 0.043 & 552 \\
        \hline
  \end{tabular}
 \end{table*}
 

\section{Statistical Analysis}\label{appendix:Stats}
 
\subsection{Fitting methods and upper limits study}\label{appendix:BCES}

The Bivariate Correlated Errors and intrinsic Scatter (BCES) fitting method is used for astronomical data analysis \citep[e.g.][]{Brunetti2009,Cassano2013,Zhao2013,Kale2015,Yuan2015,Bravi2016,Giacintucci2019,Paul2019}, as, amongst other things, it allows to perform linear regression to study correlations of astronomical datasets with measurement uncertainties on both variables and it takes into account the intrinsic scatter of the data. The BCES-bisector and BCES-orthogonal methods treat the variables symmetrically, thus it is useful when the independent variable is unknown. The BCES-bisector linear regression represents the line that bisects the two fits if X is used as the independent variable for the first fit, and Y is used as the independent variable for the second one, whereas the BCES-orthogonal method consists of minimizing the orthogonal distances of the points to the line. \citet{Isobe1990} performed Monte Carlo simulations and found that the BCES-bisector method was more accurate with a small amount of data. However, it was found that the BCES-bisector method is self inconsistent \citep{Hogg2010}, as well as being difficult to represent by a justifiable likelihood function. For these reasons, we chose to have the BCES-orthogonal algorithm as our reference method. Nevertheless, we chose to also use the BCES-bisector method for comparison purposes between both algorithms and with previous work since almost all papers studying mini-halos or giant radio halos correlations with the BCES fitting method used this algorithm as their main method \citep[e.g.][]{Cassano2013,Kale2015,Yuan2015,Bravi2016,Paul2019}. Therefore, on each figure, the best-fit using the BCES-orthogonal method and its $95\%$ confidence regions are shown, as well as the linear fit using the BCES-bisector\footnote{The linear regressions and confidence bands were found using the script on \url{https://github.com/rsnemmen/BCES}}. The confidence region represent the area that has a 95\% chance of containing the true regression line. Each fit is done using every mini-halo, including the candidate and uncertain mini-halos. The upper limits are not included as they should not have a considerable impact since they mostly follow the relations (see left panels of Fig.~\ref{fig:Steep} and \ref{fig:Core}). In fact, using the Bayesian linear regression method \textsc{linmix\_err} package that takes into account both measurement errors and non-detections \citep{Kelly2007}, we were able to find the linear regressions taking into account the upper limits for the relations of the left panels of Fig.~\ref{fig:Steep} and \ref{fig:Core}. The fits not considering the upper limits are consistent within one standard deviation with those that take them into account, and in this paper we therefore choose to only quote fits using the BCES methods without upper limits. 

For each linear regression and each algorithm, $10,000$ bootstrap resamples were also done. Bootstrapping generates a large number of shuffled samples of the original dataset, performs the fit of each realization, and reports the mean and standard deviation of the results. Bootstrapping is very useful for small samples, like in this paper, as it tries to get the empirical distribution function of a N mini-halo sample (here $\textrm{N}=10,000$) using only the small sample. Therefore, this technique gives more conservative uncertainties on the fits especially for small samples and not so clear correlations.
 
\subsection{Pearson and Spearman tests}\label{appendix:tests}
 
When studying the relation between the cooling flow power and the mini-halo-integrated radio power, \citet{Bravi2016} used the Spearman test to evaluate the strength of the correlation. This test studies the monotonic relationship between two variables, and is often used in astronomical data analysis (e.g. \citealt{Cassano2013}, \citealt{Giacintucci2014}, \citealt{Kale2015b}, \citealt{Yuan2015}, \citealt{Bravi2016}). Instead, we decided to use a Pearson test which is related to the Spearman test, to the difference that it looks only at the linear relationship between two parameters. As we are only looking at relations between two parameters to see if they have a linear correlation, the Pearson test is best suited for our study. We still provide the coefficients of both tests in Table~\ref{tab:Correlations} for comparison. Again, every mini-halo is used for each test. The coefficients of the Pearson test are $r_p$ and $\rho_p$ and the strength of the correlation is confirmed if $r_p$ is close to 1 or -1 and $\rho_p$ close to 0. $r_p$ represents the statistical dependence of two variables, with $r_p$ = 1/-1 associated with a positive/negative linear correlation and $r_p$ = 0 with no linear correlation. $\rho_p$ represents the two-sided significance level of deviation from zero, which can be understood as the probability for $r_p$ to be the same for an uncorrelated system or the probability of no correlation. The relation is considered strong if $r_p > 0.60$, moderate if $0.40 < r_p < 0.59$ and weak if $\rho_p > r_p$ (see \citealt{Press1992}, p. 634). Similarly for the coefficients $r_s$ and $\rho_s$ of the Spearman test.

\bsp	
\label{lastpage}
\end{document}